\documentclass[11pt]{amsart}

\usepackage{amsmath}
\usepackage{amssymb}
\usepackage{amsaddr}
\usepackage{wasysym}
\usepackage{amsfonts}
\usepackage{mathrsfs}
\usepackage{mathtools}
\usepackage{color}
\usepackage{nicefrac}
\usepackage{graphicx}
\usepackage[labelfont=rm]{subcaption}
\usepackage{array}
\usepackage{tikz}
\usetikzlibrary{quantikz}

\makeatletter
\renewcommand{\email}[2][]{%
  \ifx\emails\@empty\relax\else{\noindent\g@addto@macro\emails{,\space}}\fi%
  \@ifnotempty{#1}{\noindent\g@addto@macro\emails{\textrm{(#1)}\space}}%
  \g@addto@macro\emails{#2}%
}
\makeatother

\title{Quantum Error Correction Scheme \\
for Fully-Correlated Noise}

\author{Chi-Kwong Li$^{*}$, Yuqiao Li$^{\dag}$, Diane Christine Pelejo$^{\ddag}$, Sage Stanish$^{\S}$}
\address{Department of Mathematics, College of William and Mary,\\ Williamsburg, Virginia, 23185, USA}

\email{$^{*}$ckli@math.wm.edu,$^{\dag}$yuqiaoli@uw.edu, $^{\ddag}$dcpelejo@gmail.com, $^{\S}$sagestanish@posteo.net}

\date{}

\def\IC{{\mathbb C}}
\def\ra{{\rangle}}
\def\la{{\langle}}
\def\cE{{\mathcal E}}

\begin{document}

\begin{abstract}

This paper investigates quantum error correction schemes for fully-correlated noise channels on an $n$-qubit system, where error operators take the form $W^{\otimes n}$, with $W$ being an arbitrary $2\times 2$ unitary operator. In previous literature, a recursive quantum error correction scheme can be used to protect $k$ qubits using $(k+1)$-qubit ancilla. We implement this scheme on 3-qubit and 5-qubit channels using the IBM quantum computers, where we uncover an error in the previous paper related to the decomposition of the encoding/decoding operator into elementary quantum gates. 

Here, we present a modified encoding/decoding operator that can be efficiently decomposed into (a) standard gates available in the  \texttt{qiskit} library and (b) basic gates comprised of single-qubit gates and CNOT gates. Since IBM quantum computers perform relatively better with fewer basic gates, a more efficient decomposition gives more accurate results. Our experiments highlight the importance of an efficient decomposition for the encoding/decoding operators and demonstrate the effectiveness of our proposed schemes in correcting quantum errors.

Furthermore, we explore a special type of channel with error operators of the form $\sigma_x^{\otimes n}, \sigma_y^{\otimes n}$ and 
$\sigma_z^{\otimes n}$, where $\sigma_x, \sigma_y, \sigma_z$ are the Pauli matrices. For these channels, we implement a hybrid quantum error correction scheme that protects both quantum and classical information using IBM's quantum computers. We conduct experiments for $n = 3, 4, 5$ and show significant improvements compared to recent work.

\end{abstract}

\maketitle

\medskip

 \medskip

\section{Introduction}

Quantum information science concerns the use of quantum systems as computational resources to store, communicate, and process information. One of the obstacles for building quantum computers is decoherence, which is a process caused by the coupling between a quantum system and its environment. A pure state, to be used as a computational resource, becomes a dirty mixed state due to decoherence, which makes the computational outcome unreliable. There are different strategies to fight against decoherence and quantum error correcting codes (QECC) is one of them. The main idea of QECC is to embed quantum information into a higher dimensional Hilbert space so that either 
\begin{itemize}
\item[(i)] the error acting on the physical qubits may be identified by introducing the error syndrome measurement qubits, so that the initial quantum information is recovered after applying appropriate corrections, or 
\item[(ii)] the error operator acts only on a part of the Hilbert space, keeping the initial quantum information intact. 
\end{itemize}
The second QECC scheme is often called the “error-avoiding” coding scheme. Decoherence free subspace (DFS) and noiseless subsystem (NS) are two popular examples of error-avoiding QECC schemes [1-14].


In this paper, we consider the second approach to deal with quantum channels in which all physical qubits involved in coding suffer from the same error operators with a certain probability distribution. Such a channel is what we refer to as a channel with fully-correlated noise. Mathematically, if we let $U(2)$ be the group of $2\times 2$ unitary matrices, and $\mu$ be a probability measure on U(2), then a channel with fully correlated noise can be represented as an operator $\Phi$ that transforms an $n$-qubit state $\hat{\rho}$ into 
\begin{equation}\label{c-channel}
\Phi(\hat{\rho}) = \int W^{\otimes n}\hat{\rho} (W^{\otimes n})^{\dagger} d \mu(W).
\end{equation}
As mentioned in \cite{Utkan}, there are two relevant cases in which such error operators are in action; (i) when the size of a quantum computer is much smaller than the wavelength of the external disturbances, and  (ii) when photonic qubits are sent one by one through an optical fiber with a fixed imperfection. In both cases, the qubits suffer from the same errors leading to decoherence. Another instance in which such encoding is useful is when Alice sends quantum information to Bob, possibly billions of light years away, without knowing which basis  vectors Bob employs. In this case, mismatching of the basis vectors is common for all qubits and such mismatching is regarded as collective noise.

In \cite{Tomita}, an explicit recursive implementation of encoding/decoding circuits for an arbitrary number $n$ of physical qubits was presented. Remarkably, the scheme depends only on the algebra generated by the error operators $W^{\otimes n}$ with $W \in U(2)$, but not affected by the probability measure $\mu(W)$. The study demonstrates that for an $n$-qubit channel with $n = 2k+1$, the scheme can protect data encoded in $k$ logical qubits using the other $k+1$ qubits. This leads to the asymptotic encoding rate of $1/2$. In \cite{[17]},  the maximum dimension $2^k$ of the error correction code corresponding to the subspace in $\IC^{2^n}$ immune to collective noise operators of the form $W^{\otimes n}$ was determined. The study shows that the encoding rate $k/n$ approaches 1 as the positive integer $n$ gets much larger. Nevertheless, as mentioned in \cite{Utkan}, the recursive scheme in \cite{Tomita} is more practical compared to other schemes;  for example, see \cite{[11],Yang}.

In this paper, we implement the recursive QECC scheme proposed in \cite{[15]} for channels with fully-correlated noise 
using the IBM quantum computers and study the mathematical issues associated with the process. During our investigation, we identify an error in the decomposition of the encoding/decoding operator into elementary  quantum gates as illustrated in the quantum circuits in \cite{[15]};  see also \cite{Utkan}. 
We fix the error by finding a new encoding operator $U$ for the fully-correlated channels on 3 qubits, which is the base case for the recursive scheme. In particular, the matrix $U$  can be decomposed into a product of three CNOT gates, a single-qubit gate and two controlled-gates, which are {\bf standard
gates} available in \texttt{qiskit}---the Python library used to interface with the IBM quantum computers. We then implement the scheme using the IBM quantum computers for the fully-correlated channels on 3 qubits and 5 
qubits. However, the numerical results vary among different quantum computers. 
It turns out that the IBM quantum computers would decompose the 
standard gates into more {\bf basic gates} to run the program, and the 
decomposition varies on different machines on different runs. In response, we further decompose our encoding operator $U$ into a 
product of 6 CNOT gates and 8 single qubit gates before feeding the 
circuit into the IBM quantum computers. Intriguingly, qiskit has an internal algorithm that further adjusts the 
decomposition based on the specific machine being utilized. 

To deepen our understanding of the errors in the implementation 
process, we compare our numerical results on different quantum 
computers and their associated factors such as qubit connections, gate errors, decoherence time, 
etc. In Section 2, we will present and analyze these results. 

In Section 3, we consider the special case of fully-correlated channels that use only the error operators 
$W^{\otimes n}$ for $W = I_2, \sigma_x, \sigma_y, \sigma_z$, 
where $\sigma_x, \sigma_y, \sigma_z$ are the Pauli matrices and $I_2$ is the $2\times 2$ identity matrix. 
As mentioned before, the general recursive scheme is good for
fully correlated quantum channels with any probability 
distribution $D\mu(U)$. In this special case, the probabilities
are nonzero only for the four special choices of $W$, and thus it simplifies the error structure significantly.
As a result, we can have a much more efficient
quantum error correction scheme. In \cite{LLP},
a hybrid quantum error correction scheme was proposed
for this special case, where a
1-qubit ancilla  can be used to protect $2k$ qubits of quantum information,  
and a 2-qubit ancilla can be used to protect $2k$ qubits of quantum information and 2 bits of classical information. While the scheme was implemented on IBM quantum computers in \cite{LLP}, the results were not satisfactory for $4$-qubit and $5$-qubit channels. In our study, we present an improved implementation of the hybrid scheme with significantly better computational results. Furthermore, we analyze the error patterns of different quantum computers with  different 
qubit connections, gate errors, decoherence time, etc. 

Finally, in Section 4, we conclude this paper with a short summary of our work and a discussion of future research topics.

\section{The recurrence scheme and the correction of previous error}

Denote the space of all $N\times N$ complex matrices by $M_N$ and consider the fully-correlated channel $\Phi: M_{2^n} \rightarrow M_{2^n}$ defined in (\ref{c-channel}) with  $n = 2k+1$. In \cite{Tomita} (see also \cite{Utkan}), a recursive quantum error correction scheme was presented for protecting $k$-qubits. Suppose we wish to protect $k$-qubits of data, realized as a density matrix $\rho$ in $M_{2^k}$. First, we will embed the information into a higher-dimensional Hilbert space having initial state $\rho\otimes \sigma$ where $\sigma = |u \ra \la u|$ and $|u\ra = |0\ra^{\otimes k}\otimes |v_0\ra$ is the product state of $k$ copies of the pure state $|0\ra$ 
and one arbitrary qubit $|v_0\ra$.  
Then, we apply a carefully-chosen encoding operator $\cE: M_{2^n} \rightarrow M_{2^n}$ such that the encoded state is equal to $\cE(\rho\otimes \sigma)$  and the decoded state of the system after going through the noisy channel is 
 \[\cE^{-1}\circ \Phi\circ \cE(\rho\otimes \sigma)= \rho \otimes \Big(|0\ra\la 0|\Big)^{\otimes k} \otimes |v_1\ra \la v_1|.\]
Finally, we can take a partial trace to recover the data state $\rho$.

It is worth noting that in some scenarios, error correction may need to be done multiple times, such as periodically when the data state $\rho$ is attacked by the correlation error regularly before the computation or transmission process is done.  In such cases, there is no need to do the encoding and decoding multiple times, as doing so may cause additional errors. One only needs to let the encoded state $\cE(\rho \otimes \sigma)$ stay in the environment, going through $m$ rounds of error attack, and then apply the decoding scheme $\cE^{-1}$ to obtain the final decoded state
\[(\cE^{-1}\Phi^m \cE)(\rho\otimes \sigma) = (\cE^{-1}\Phi \cE)^m(\rho\otimes \sigma) = \rho \otimes \Big(|0\ra\la 0|\Big)^{\otimes k} \otimes |v_m\ra \la v_m|.\]

In the following, we will use the IBM quantum computers to implement the recursive quantum error correction  schemes for the fully correlated channel for 3-qubits ($k=1$) and 5-qubits ($k=2$).

\subsection{Three-qubit case}

For convenience, we will reorder the positions of the data qubits and the ancilla qubits. The basic case is when $k = 1$, and the encoding operation is done by 
\[\hat{\cE}(|0\ra \la 0| \otimes \rho \otimes |v_0\ra \la v_0| ) = U\Big(|0\ra \la 0| \otimes \rho \otimes |v_0\ra \la v_0|\Big)U^{\dagger},\]
where $\rho = |\psi\ra \la \psi|$ is the data qubit to be transmitted, $|v_0\ra$ is an arbitrary qubit, and the unitary matrix $U$ is chosen such that for any $W\in U(2)$, we have
\begin{equation}\label{encodmat}
U^{\dagger}(W\otimes W \otimes W) U = \mu_W\Big((I_2 \otimes W) \oplus F_W\Big)=\mu_W\Big(|0\rangle\langle 0|\otimes (I_2 \otimes W) + |1\rangle\langle 1|\otimes F_W\Big),
\end{equation}
for some complex unit $\mu_W$ and some $F_W \in M_4$. 
The existence of such a $U$ is guaranteed by a result in representation theory (see \cite{[5]} and \cite{Tomita}). Thus, for any qubit $|\psi\ra$, we have
\begin{equation}\label{WWWscheme}
U^{\dagger}(W\otimes W \otimes W)U \Big(|0\ra |\psi\ra |v_0\rangle\Big) = \mu_W\Big[(I_2 \otimes W) \oplus F_W\Big]\Big(|0\ra |\psi \ra |v_0\ra\Big)= |0\ra|\psi \ra (\mu_W W| v_0\ra).
\end{equation}
In other words, one can use an arbitrary qubit $|v_0\ra$ and the qubit $|0\ra$ to protect a given quantum bit $|\psi\ra$. This effect is illustrated in the circuit diagram in Figure \ref{OldvNew}(\subref{Ueff}). This scheme was utilized in \cite{Utkan} with the unitary matrix $U$ given in Figure \ref{OldvNew}(\subref{oldU}). 

\begin{figure}[!ht]
\begin{subfigure}{0.4\textwidth}
\quad\scalebox{0.85}{\begin{quantikz}[row sep=0.5cm,column sep=0.3cm]
\lstick{$|v_0\rangle$} 	& \qw & \gate[3]{U} & \gate{W} & \gate[3]{U^{\dagger}} & \qw & \qw & \rstick{$\mu_W W|v_0\rangle$}\\ 
\lstick{$|\psi\rangle$}  & \qw & \qw      & \gate{W} & \qw 			 & \qw & \qw & \rstick{$|\psi\rangle$}\\
\lstick{$|0\rangle$}		& \qw & 	\qw 	    & \gate{W} & 	\qw 			 & \qw & \qw & \rstick{$|0\rangle$}
\end{quantikz}}\vspace{0.6cm}
\caption{The circuit diagram of (\ref{WWWscheme}).}
\label{Ueff}
\end{subfigure}
\begin{subfigure}{0.55\textwidth}
\footnotesize
$\begin{pmatrix}
0 & 0 & 0 & 0 &
	1 & 0 & 0 & 0 \\
\sqrt{\nicefrac{2}{3}} & 0 & 0 & 0 &
	0 & \sqrt{\nicefrac{1}{3}} & 0 & 0 \\
-\sqrt{\nicefrac{1}{6}} & 0 & \sqrt{\nicefrac{1}{2}} & 0 &
 	0 & \sqrt{\nicefrac{1}{3}} & 0 & 0 \\
0 & \sqrt{\nicefrac{1}{6}} & 0 & \sqrt{\nicefrac{1}{2}} &
 	0 & 0 & \sqrt{\nicefrac{1}{3}} & 0 \\
-\sqrt{\nicefrac{1}{6}} & 0 & -\sqrt{\nicefrac{1}{2}} & 0 &
 	0 & \sqrt{\nicefrac{1}{3}} & 0 & 0 \\
0 & \sqrt{\nicefrac{1}{6}} & 0 & -\sqrt{\nicefrac{1}{2}} &
 	0 & 0 & \sqrt{\nicefrac{1}{3}} & 0 \\
0 & -\sqrt{\nicefrac{2}{3}} & 0 & 0 &
 	0 & 0 & \sqrt{\nicefrac{1}{3}} & 0 \\
0 & 0 & 0 & 0 &
 	0 & 0 & 0 & 1 
\end{pmatrix}$\normalsize
\caption{A unitary matrix $U$ satisfying (\ref{encodmat}).}
\label{oldU}
\end{subfigure}
\caption{}
\label{OldvNew}
\end{figure}

\begin{figure}[!ht]
\begin{center}
\scalebox{0.9}{\begin{quantikz}[row sep=0.1cm,column sep=0.35cm]
 \lstick{$|v\rangle$}& \qw[above]{0}  &\qw\gategroup[3,steps=7,style={inner sep=7pt, line width=0.5}, background]{{}} & \qw & \gate{\sigma_z}\gategroup[3,steps=2,style={dashed, rounded corners,inner ysep=0pt, inner xsep=0pt, yshift=2pt}, background]{{}} & \ctrl{2} &\qw &  \targ & \qw & \qw &\qw & \qw\\
   \lstick{$|\psi\rangle$} & \qw[above]{1}  & \ctrl{1} & \gate{R_y(\tfrac{-\pi}{2})}  & \qw &\qw & \qw & \octrl{-1} & \targ  & \qw &\qw & \qw\\ 
  \lstick{$|0\rangle$} & \qw[above]{2} & \gate{R_y(\theta)} & \octrl{-1} & \qw  & \targ & \qw &  \qw & \qw &  \ctrl{-1} & \qw & \qw 
\end{quantikz}}\quad  
\begin{minipage}{0.35\textwidth}
$\begin{array}{lcl}
R_y(\alpha) & = & \mbox{exp}({-i\frac{\alpha}{2}\sigma_y}) \medskip\\
& = & \begin{pmatrix}
\cos(\frac{\alpha}{2}) & -\sin(\frac{\alpha}{2}) \medskip\\
\sin(\frac{\alpha}{2})  & \cos(\frac{\alpha}{2})
\end{pmatrix}
\end{array}$

\medskip
\ \ Here $\theta$ 
satisfies $\sin(\frac{\theta}{2})=-\sqrt{\nicefrac{2}{3}}$.
\end{minipage}
\caption{The (erroneous) circuit diagram presented 
in \cite{Tomita} for the decomposition of the matrix 
in Figure \ref{OldvNew}(\subref{oldU}).
}
\label{Udec}
\end{center}
\end{figure}
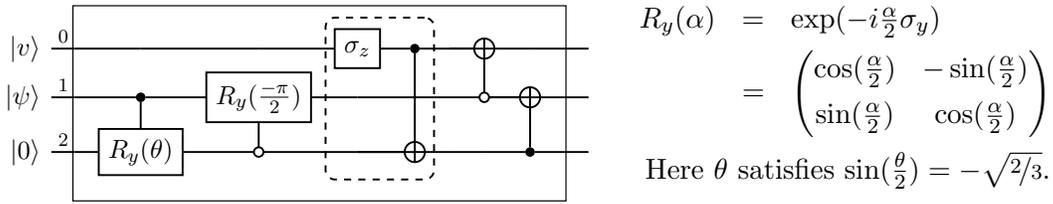

To implement the quantum error correction scheme, it is necessary to decompose the encoding operator $U$ into elementary gates/operations that the IBM quantum computers can carry out. In \cite{Tomita}, the matrix $U$ in Figure \ref{OldvNew}(\subref{oldU}) was supposed to have a simple realization by the circuit diagram in Figure \ref{Udec}, where the part surrounded by the broken line is not needed if $|v\ra = |0\ra$, as claimed in the paper. However, we found that the product of the 6 elementary gates do not actually produce the matrix $U$ as asserted (see Appendix 1). With some effort,
we identified a different decomposition of the unitary matrix in 
Figure \ref{OldvNew}(\subref{oldU}) to construct a circuit diagram for implementation. However, it requires 9 standard gates in the IBM quantum computer library including the use of a Toffoli gate, and the implementation of the scheme using the IBM quantum computers did not yield satisfactory results.

To get around the problem, we modify the encoding matrix $U$, satisfying equation (\ref{encodmat}), given by \small
\begin{equation}\label{newU}
U=\left(\begin{array}{cccccccc}
0 & 0 & 0 & 0 & 0 & 0 & 0 & -1\\
\sqrt{\nicefrac{2}{3}}  & 0 & 0 & 0& \sqrt{\nicefrac{1}{3}} & 0 & 0 & 0\\
-\sqrt{\nicefrac{1}{6}} & 0 & \sqrt{\nicefrac{1}{2}} & 0 &\sqrt{\nicefrac{1}{3}} & 0 & 0 & 0\\
0 & \sqrt{\nicefrac{1}{6}} & 0 & \sqrt{\nicefrac{1}{2}}& 0 & -\sqrt{\nicefrac{1}{3}} & 0 & 0\\
-\sqrt{\nicefrac{1}{6}} & 0 & -\sqrt{\nicefrac{1}{2}} & 0 & \sqrt{\nicefrac{1}{3}} & 0 & 0 & 0\\
0 & \sqrt{\nicefrac{1}{6}} & 0 & -\sqrt{\nicefrac{1}{2}}& 0 & -\sqrt{\nicefrac{1}{3}} & 0 & 0\\
0 & -\sqrt{\nicefrac{2}{3}} & 0 & 0 & 0 & -\sqrt{\nicefrac{1}{3}} & 0 & 0\\
0 & 0 & 0 & 0 & 0 & 0 & 1 & 0
\end{array}\right).
\end{equation}\normalsize
It is worth noting that the modification of the unitary matrix was based on the fact that any unitary matrix $U$ with the first four columns equal to that of the unitary matrix in Figure \ref{OldvNew}(\subref{oldU}) will satisfy equation (\ref{encodmat}). 

Now, for the matrix $U$ in (\ref{newU}), we have $U=P_1P_2P_3Q_1Q_2Q_3$, where $Q_1$ is the single qubit gate $Q_1=\sigma_z\otimes I_4$; while $Q_2$ and $Q_3$ are the controlled gates 
\[Q_2=\frac{1}{\sqrt{2}}\begin{pmatrix}
I_2 & -I_2 \\I_2 & I_2
\end{pmatrix}\oplus I_4\quad Q_3=\begin{pmatrix}
-\sqrt{\frac{1}{3}} & 0 & \sqrt{\frac{2}{3}} & 0\\
0 & 1 & 0 & 0\\
\sqrt{\nicefrac{2}{3}} & 0 & \sqrt{\nicefrac{1}{3}} & 0\\
0 & 0 & 0 & 1
\end{pmatrix}\otimes I_2;\]
and $P_1,P_2$ and $P_3$ are the CNOT gates
\[P_1=\begin{pmatrix}
0 & 0 & I_2 & 0\\
0 & I_2 & 0 & 0\\
I_2 & 0 & 0 & 0\\
0 & 0 & 0 & I_2
\end{pmatrix},\qquad P_2=I_4\oplus (I_2\otimes \sigma_x), \qquad P_3=I_2\otimes \begin{pmatrix}
1 & 0 & 0 & 0 \\0 & 0 & 0 & 1\\ 0 & 0 & 1 & 0\\ 0& 1 & 0 & 0
\end{pmatrix}.\]
This decomposition is illustrated in the circuit diagram in Figure \ref{newUdec}. (See Appendix 2 for a Matlab script
to verify this decomposition.) Moving forward, we will refer to this as the \textit{standard gates decomposition of $U$.}

\begin{figure}[!ht]
\begin{center}
\scalebox{.75}{\begin{quantikz}[row sep=0.25cm,column sep=0.25cm]
\lstick{$|q_0\rangle$}  & & \gate[3]{U} & \qw &  & &  \lstick{$|q_0\rangle$} & 
\qw & \qw & \qw & \gate{\sigma_z} & \ctrl{1} & \targ{} & \qw  & \qw\\
\lstick{$|q_1\rangle$} & & & \qw & \push{\rule{0.1em}{0em}=\rule{1em}{0em}} & & \lstick{$|q_1\rangle$} & 
\qw  & \octrl{1} & \gate{R_y(\frac{\pi}{2})}  & \qw  & \targ{} & \qw & \octrl{1} & \qw \\
 \lstick{$|q_2\rangle$} & & & \qw & & & \lstick{$|q_2\rangle$} &
 \qw  &\gate{R_y(\alpha)\sigma_x} & \octrl{-1} & \qw & \qw &  \ctrl{-2}  & \targ{}  & \qw \\
\end{quantikz}}\vspace{-0.5cm}
\end{center}
\caption{Here, $U$ is the matrix in $(\ref{newU})$ and $\alpha=2\arcsin(\sqrt{\nicefrac{1}{3}})$.}
\label{newUdec}
\end{figure}
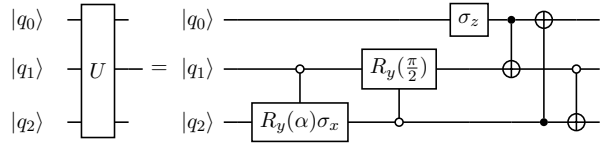

\begin{figure}[!ht]
\begin{subfigure}{0.47\textwidth}
\includegraphics[scale=0.34]{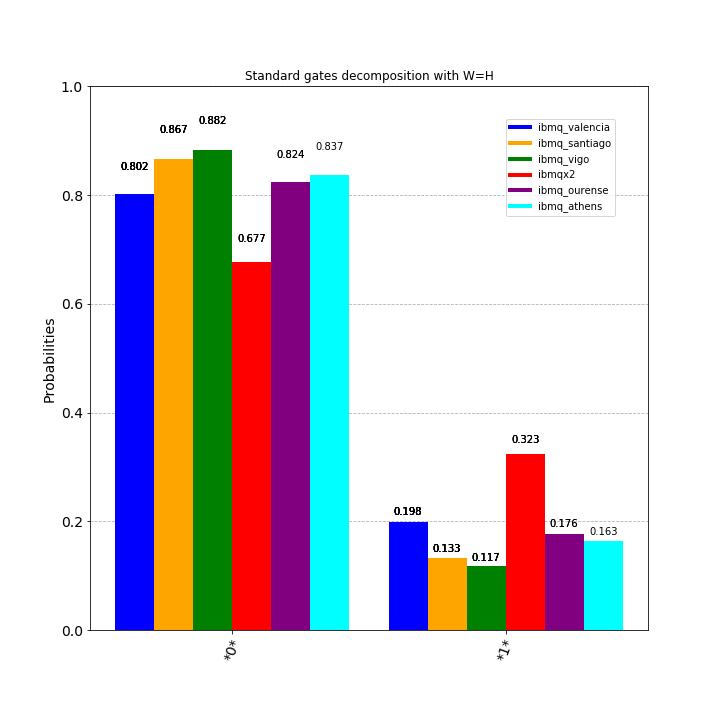}\vspace{-0.65cm}
\caption{}
\label{3sh}
\end{subfigure}
\begin{subfigure}{0.47\textwidth}
\includegraphics[scale=0.34]{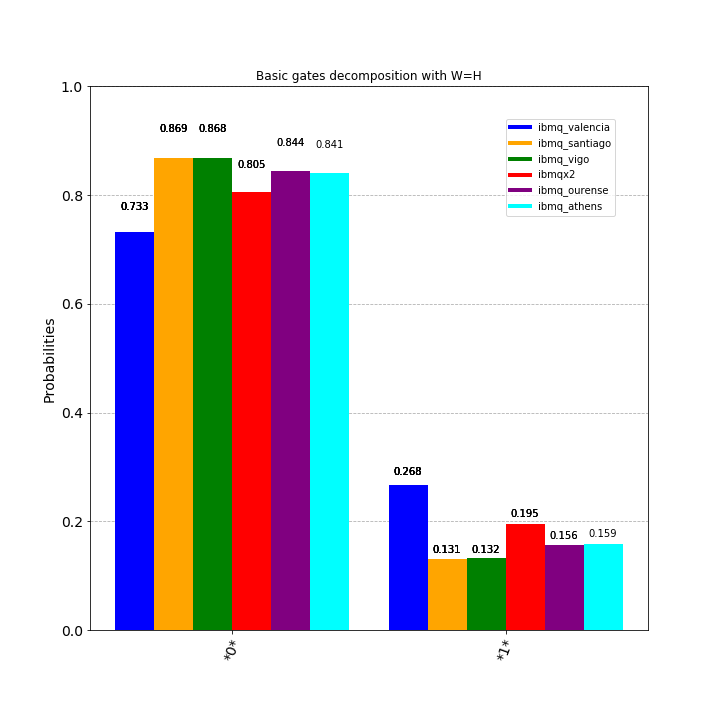}\vspace{-0.65cm}
\caption{}
\label{3bh}
\end{subfigure}
\caption{}
\label{3h}
\end{figure}

Using this decomposition, we implement the error correction scheme illustrated in Figure \ref{OldvNew}(\subref{Ueff}) using six IBM quantum computers: \texttt{ibmq\_valencia}, \texttt{ibmq\_santiago},  \texttt{ibmq\_vigo}, \texttt{ibmq\_5\_yorktown},   \texttt{ibmq\_ourense} and \texttt{ibmq\_athens}. The results are shown in Figure \ref{3h}(\subref{3sh}). Here, we set $W$ to be the Hadamard gate $H$ defined as:
\[H=\frac{1}{\sqrt{2}}\begin{pmatrix}
1 & 1 \\1 & -1
\end{pmatrix}\] 
and we measure only the data qubit (middle qubit). The results are quite satisfactory as shown in Figure \ref{3h}(\subref{3sh}). Specifically, the protected qubit is set to $|0\rangle$ and most outputs show a measurement of $|0\rangle$ higher than 80\%. Furthermore, we also performed experiments using alternative choices of $W$ and obtained comparable results.

It turns out the IBM quantum computers will further decompose the standard gates into basic gates when they execute the encoding and decoding scheme. These decompositions will usually create many CNOT gates. (For example, see the circuit diagrams produced by \texttt{ibmq\_valencia} in Figure \ref{app3}(\subref{IBMCQ1sh}) of Appendix 3). In view of this,  we try to improve the accuracy by decomposing $U$ into a product of 6 CNOT gates and 8 single qubit gates as shown in Figure \ref{newUbasic}. (See Appendix 2 for a Matlab script to verify this decomposition.) From here on out, we refer to this decomposition of $U$ as the \textit{basic gates 
decomposition of $U$.}

\begin{figure}[!ht]
\begin{center}
\scalebox{0.75}{\begin{quantikz}[row sep=0.25cm,column sep=0.25cm]
\lstick{$|q_0\rangle$}  & & \gate[3]{U} & \qw &  & &  \lstick{$|q_0\rangle$} & 
\qw & \qw & \qw & \qw & \qw & \qw & \gate{\sigma_z} & \ctrl{1} & \targ{} & \qw & \qw & \qw \\
\lstick{$|q_1\rangle$} & & & \qw & \push{\rule{0.1em}{0em}=\rule{1em}{0em}} & & \lstick{$|q_1\rangle$} &  
\gate{\sigma_x} & \ctrl{1}& \qw & \targ{} & \gate{R_y(\frac{\pi}{4})} & \targ{} & \gate{R_y(-\frac{\pi}{4})} & \targ{} & \qw  & \ctrl{1} & \gate{\sigma_x} & \qw\\
 \lstick{$|q_2\rangle$} & & & \qw & & & \lstick{$|q_2\rangle$} & 
\gate{R_y(-\frac{\alpha}{2})} &\targ{} & \gate{\sigma_xR_y(\frac{\alpha}{2})} &\ctrl{-1} & \qw & \ctrl{-1}& \gate{\sigma_x} & \qw & \ctrl{-2}  &\targ{} & \qw & \qw \\
\end{quantikz}}\vspace{-0.5cm}
\end{center}
\caption{Decomposition of the matrix $U$ in $(\ref{newU})$ as 
a product of 6 CNOT gates and 8 single qubit gates.}
\label{newUbasic}
\end{figure}
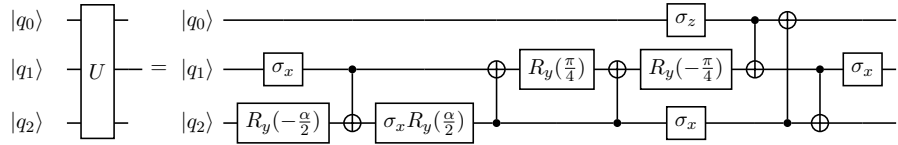

As shown in Figure \ref{3h}(\subref{3bh}), we obtain improvements in all
the machines except for \texttt{ibmq\_valencia} when we use the basic gates decomposition. It is also worth noting that the IBM quantum computers will still make small changes before carrying out the user-specified decomposition of the scheme. (See Figure \ref{app3}(\subref{IBMCQ1bh}) in Appendix 3.)

\subsection{Five-qubit case}
\label{5qWWW}
\begin{figure}[!ht]
\begin{center}
\scalebox{0.6}{\begin{quantikz}[row sep=0.4cm,column sep=0.1cm]
\lstick{$|0\rangle$} & \qw & \gate[3]{\ \hat{U}\ } & \qw & \qw & \qw & \gate[5]{\ \Phi\ } & \qw & \qw & \qw & \gate[3]{\ \hat{U}^{\dagger}} & \qw &  \rstick{$|0\rangle$} \\
\lstick{$|\psi_1\rangle$} & \qw & \qw & \qw & \qw & \qw & \qw & \qw &\qw & \qw & \qw & \qw  \rstick{$|\psi_1\rangle$}\\
\lstick{$|v\rangle$} &\qw & \qw & \qw & \gate[3]{\ U\ } & \qw & \qw & \qw & \gate[3]{\ U^{\dagger}} & \qw & \qw & \qw & \rstick{$|\hat{v}\rangle$}  \\
\lstick{$|\psi_2\rangle$} & \qw & \qw & \qw & \qw & \qw & \qw & \qw &\qw & \qw & \qw & \qw & \rstick{$|\psi_2\rangle$}  \\
\lstick{$|0\rangle$} & \qw & \qw & \qw & \qw & \qw & \qw & \qw &\qw & \qw & \qw & \qw & \rstick{$|0\rangle$}  
\end{quantikz}}\quad 
\scalebox{0.6}{\begin{quantikz}[row sep=0.15cm,column sep=0.1cm]
\lstick{$|0\rangle$} &
 \qw  &\gate{R_y(\theta)\sigma_x} & \octrl{1} & \qw & \qw  &  \ctrl{2} & \targ{} & \qw & \qw &\qw & \gate{W}
 & \qw & \qw & \qw & \targ{} & \ctrl{2} & \qw & \qw & \octrl{1} & \gate{\sigma_x R_y(-\theta)} & \qw &  \rstick{$|0\rangle$}  \\
\lstick{$|\psi_1\rangle$} & 
\qw  & \octrl{-1} & \gate{R_y(\frac{\pi}{2})}  & \qw &\targ{} &\qw & \octrl{-1}  &  \qw &\qw & \qw &\gate{W}  
& \qw & \qw & \qw & \octrl{-1} & \qw & \targ{} & \qw & \gate{R_y(-\frac{\pi}{2})} & \octrl{-1} &\qw  & \rstick{$|\psi_1\rangle$}   \\
\lstick{$|v\rangle$} & 
\qw & \gate{\sigma_z} & \qw & \qw & \ctrl{-1} & \targ{} & \gate{\sigma_z} & \ctrl{1} & \targ{} & \qw & \gate{W} &  \qw & \targ{} & \ctrl{1} & \gate{\sigma_z} & \targ{} & \ctrl{-1} & \qw & \qw & \gate{\sigma_z} & \qw & \rstick{$|\hat{v}\rangle$} \\
\lstick{$|\psi_2\rangle$} & 
\qw &\octrl{1} &\gate{R_y(\frac{\pi}{2})} &\qw &\qw  & \qw & \qw  &  \targ{} & \qw &\octrl{1} & \gate{W} & 
\octrl{1} & \qw & \targ{} & \qw & \qw & \qw & \qw & \gate{R_y(-\frac{\pi}{2})} & \octrl{1} & \qw  & \rstick{$|\psi_2\rangle$} \\
\lstick{$|0\rangle$} &
\qw & \gate{R_y(\theta)\sigma_x} & \octrl{-1} &\qw &\qw & \qw & \qw & \qw & \ctrl{-2} & \targ{}  &\gate{W} &
\targ{} & \ctrl{-2} & \qw & \qw & \qw & \qw & \qw & \octrl{-1} & \gate{\sigma_x R_y(-\theta)} & \qw & \rstick{$|0\rangle$}  
\end{quantikz}}
\end{center}
\caption{The circuit for the recursive coding scheme for $2$ protected data qubits 
$|\psi_1\ra$ and $|\psi_2\ra$.}
\label{5qbit}
\end{figure}
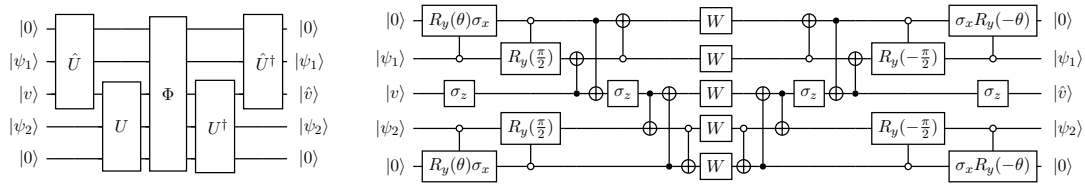

As shown in \cite{Tomita} and \cite{Utkan}, one can then apply the error correction scheme recursively to protect  two qubits of data using a 3-qubit ancilla having a product state $|0\ra|0\ra|v\ra$, where $|v\ra$ is an arbitrary qubit. The circuit diagram used in the error 
correction scheme is shown in Figure \ref{5qbit}. Recursively, we 
can protect $k$ data qubits using $k+1$ ancillary qubits having a product state $|0\ra^{\otimes k}|v\ra$, where $|v\ra$ is an arbitrary qubit. 

\begin{figure}[!ht]
\includegraphics[scale=0.2]{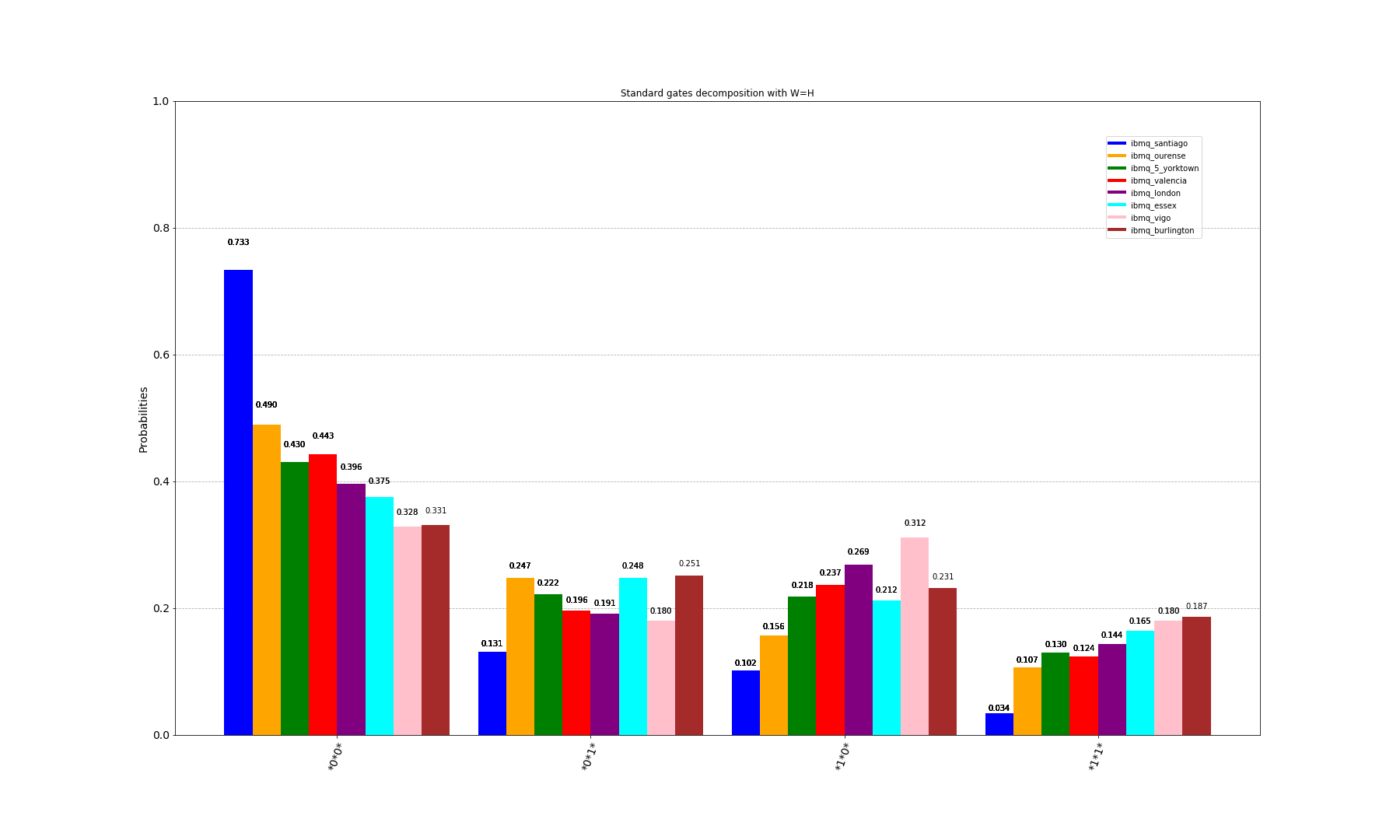}\vspace{-0.65cm}
\caption{5-qubit QECC with circuit diagram illustrated in Figure \ref{5qbit}}
\label{5sh1}
\end{figure}

\begin{figure}[!ht]
\includegraphics[scale=0.2]{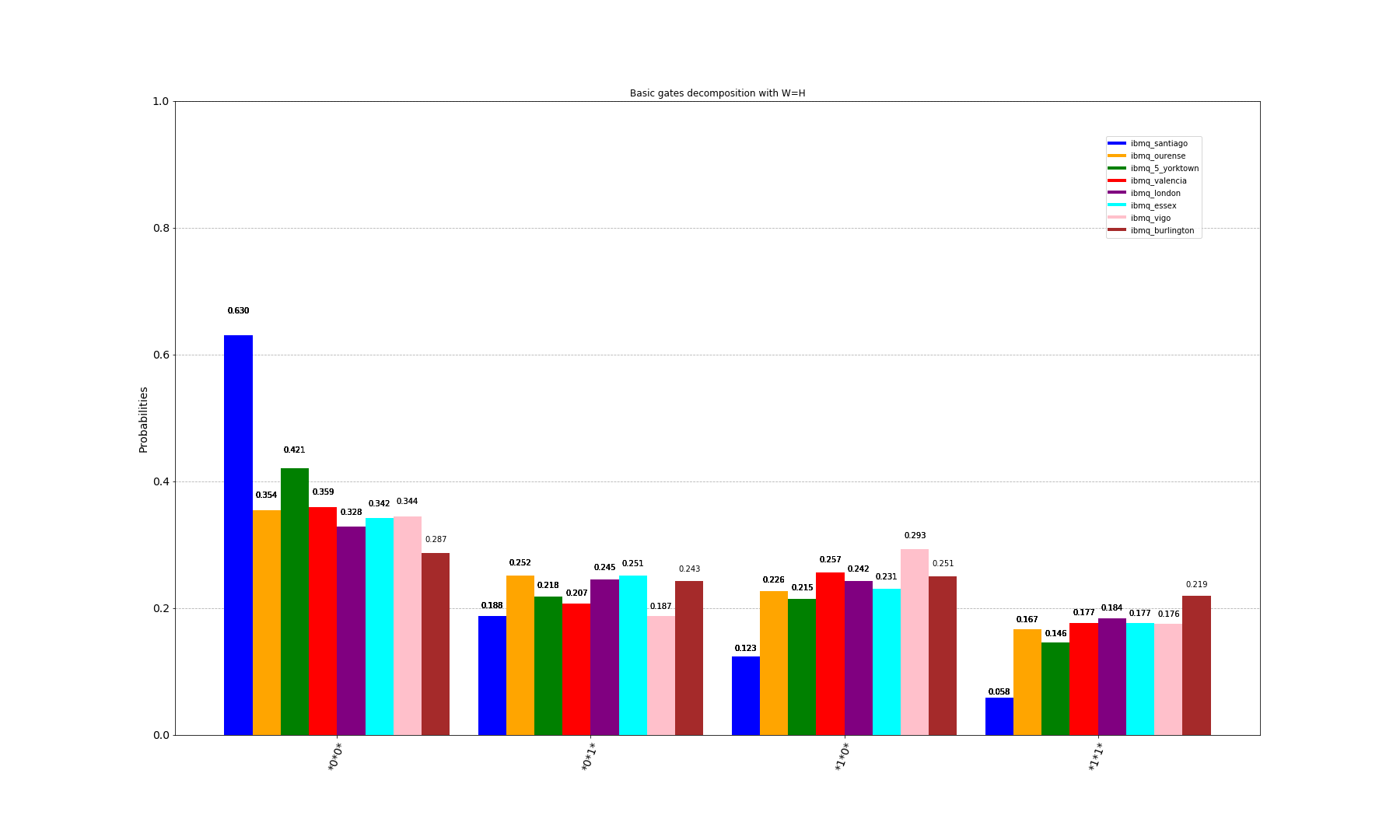}\vspace{-0.65cm}
\caption{5-qubit QECC using the basic gates decomposition of $U$.}
\label{5bh1}
\end{figure}

Setting $W=H, |\psi_1\rangle=|\psi_2\rangle=|0\rangle$, and using the decomposition in Figure \ref{5qbit}, we get the results shown in Figure \ref{5sh1} upon measurement of the data qubits (second and fourth qubit positions). On the other hand, if we utilize the decomposition of $U$ given in Figure \ref{newUbasic}, we obtain the results shown in Figure \ref{5bh1}. We observe that the results using the basic gates decomposition of $U$ in Figure \ref{newUbasic} are worse than those  using the standard gates decomposition of $U$ in Figure \ref{newUdec}. 
This suggests that the internal algorithm used by the IBM computers to decompose standard gates into basic gates work well with the machines even though more CNOT gates are used.  


We conducted the same experiments a few more times and  used a variety of error operators; readers can view the results of these additional experiments in Appendix 4. The Jupyter notebooks used to run these experiments are also available in the following Github repository:\medskip
\begin{center}
\texttt{https://github.com/dcpelejo/QECC/tree/main/WWW\%20experiments}\medskip
\end{center}

To further understand the performance of
our scheme on different IBM quantum computers, we compare the results in relation to factors such as  gate errors, qubit connections, decoherence time of the machines.

\begin{table}[htbp]\includegraphics[scale=0.7]{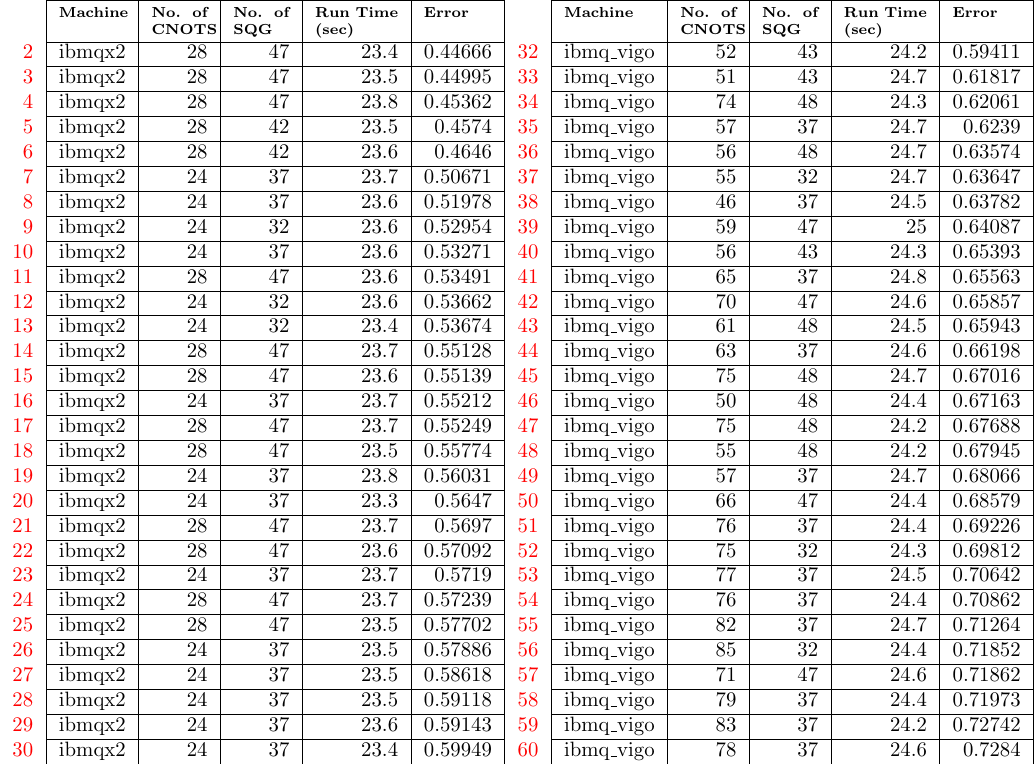}\medskip\\
\caption{}
\label{tab1}
\end{table}
\begin{table}[htbp]\includegraphics[scale=0.63]{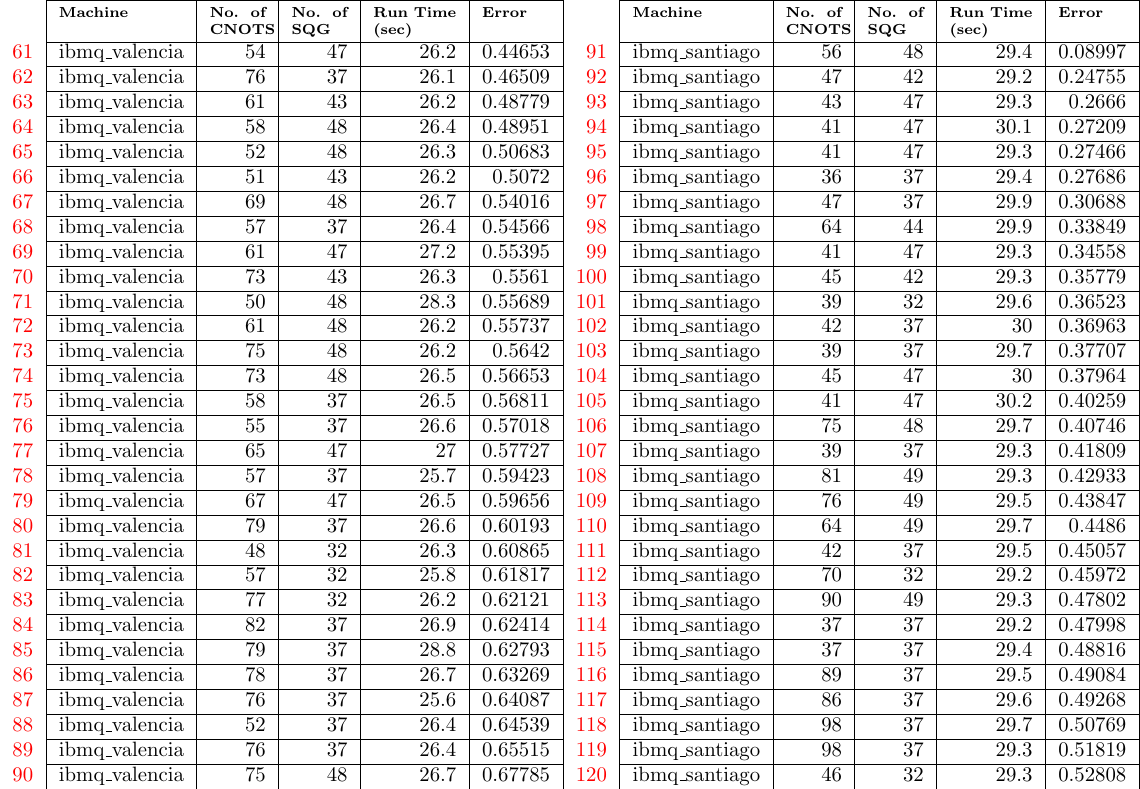}\medskip\\
\caption{}
\label{tab2}
\end{table}
\begin{table}[htbp]\includegraphics[scale=0.65]{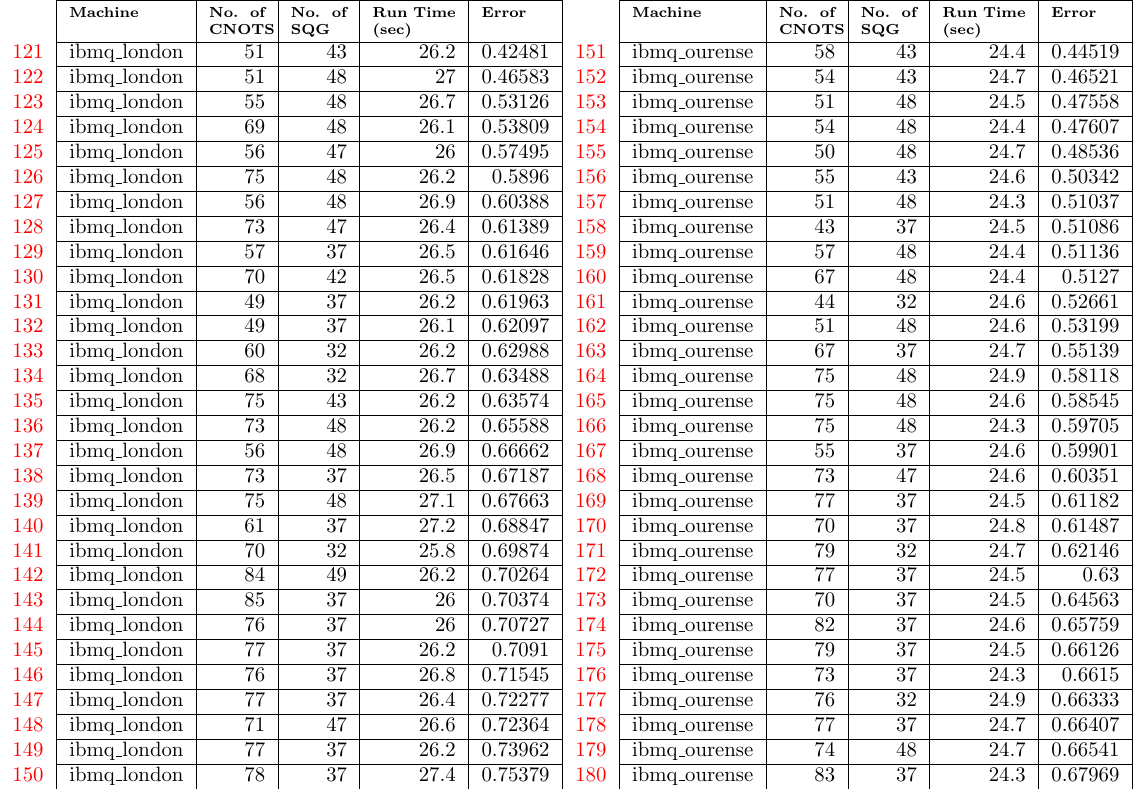}\medskip\\
\caption{}
\label{tab3}
\end{table}
\begin{table}[htbp]\includegraphics[scale=0.67]{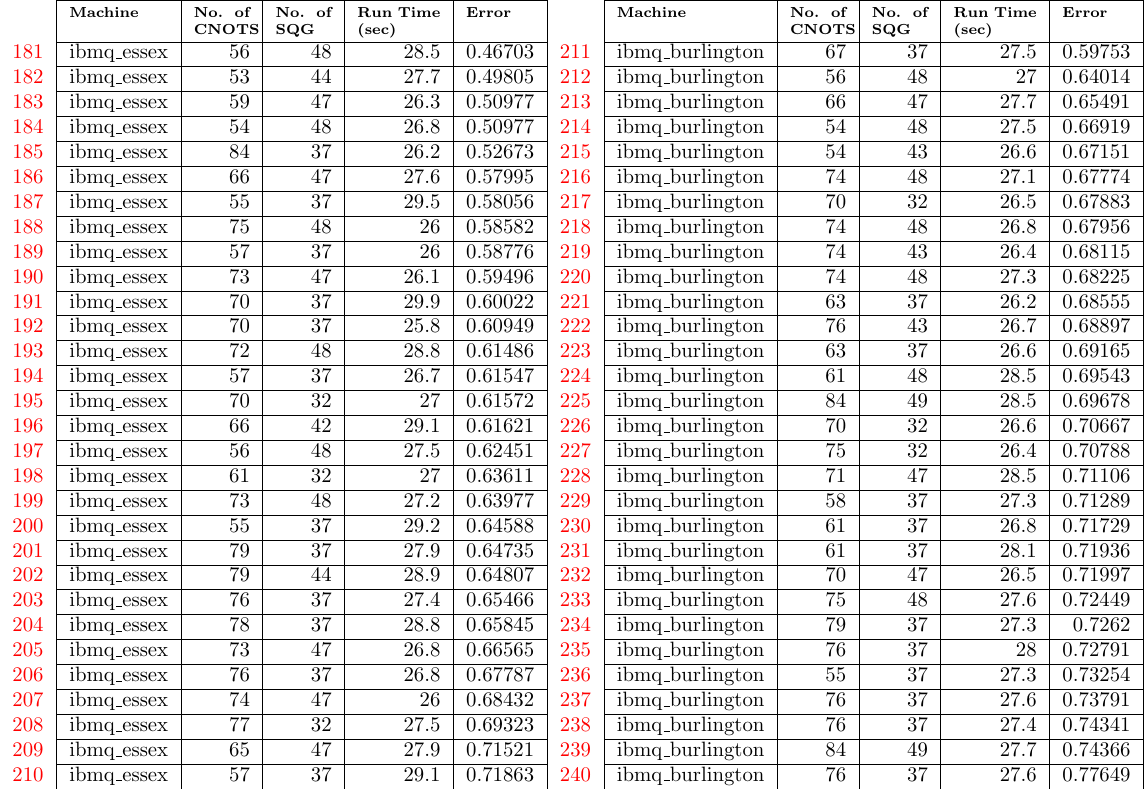}\medskip\\
\caption{}
\label{tab4}
\end{table}

%
%
%

\newpage

\section{Correlated channels with error operators
$\sigma_x^{\otimes n}, \sigma_y^{\otimes n}, \sigma_z^{\otimes n}$}

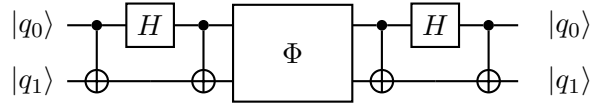
\begin{figure}[!ht]
\begin{quantikz}[row sep=0.2cm,column sep=0.23cm]
\lstick{$|q_0\rangle$}& \ctrl{1} & \gate{H} & \ctrl{1} &  \gate[2]{\ \quad\Phi\ \quad} & \ctrl{1} & \gate{H} & \ctrl{1} & \qw & \rstick{$|q_0\rangle$}\\
\lstick{$|q_1\rangle$} & \targ \qw   &      \qw & \targ  \qw  &  & \targ \qw &         \qw      & \targ \qw   & \qw & \rstick{$|q_1\rangle$}
\end{quantikz}
\caption{Hybrid QECC when $n = 2$ and $|q_1q_0\ra \in \{|00\ra, |01\ra, |10\ra, |11\ra\}$}
\label{xyz2}
\end{figure}

In \cite{LLP}, a hybrid quantum error correction scheme for fully-correlated channels with  error operators $\sigma_x^{\otimes n}, \sigma_y^{\otimes n}, \sigma_z^{\otimes n}$ was implemented using the IBM quantum computers.
Theoretically, this scheme allows the use of a single arbitrary ancilla to protect
$n-1$ data qubits if $n$ is odd, and use two classical bits to protect
$n-2$ data qubits and yet preserving the two classical bits. However, their numerical experiments using the IBM quantum computers failed to produce good results for 
$n=4$ and $n=5$. Here, we conducted additional experiments and discovered that we can indeed obtain reasonable results as shown in Appendix 4. Conceivably, the IBM quantum computers may have undergone improvements since our initial implementation. For ease of reference, we provide the circuit diagrams for the recursive scheme in Figures \ref{xyz2}-\ref{xyzeven}.
\begin{figure}[!ht]

\begin{quantikz}[row sep=0.2cm,column sep=0.3cm]
  \lstick{$|q_0\rangle$} & \qw & \ctrl{2} & \targ \qw & \gate[3]{\ \quad\Phi\ \quad} & \targ \qw & \ctrl{2} & \qw & \qw & \rstick{$|q_0\rangle$}\\
  \lstick{$|q_1\rangle$} & \targ \qw & \qw & \ctrl{-1} & & \ctrl{-1} & \qw & \targ \qw & \qw &\rstick{$|q_1\rangle$}\\
  \lstick{$|q_2\rangle$} & \ctrl{-1} & \targ \qw & \qw &  & \qw & \targ \qw & \ctrl{-1} & \qw &\rstick{$|\hat{q}_2\rangle$}
\end{quantikz} 
\caption{Hybrid QECC when $n = 3$, where $|q_2\ra$ can be any qubit state}
\label{xyz3}
\end{figure}
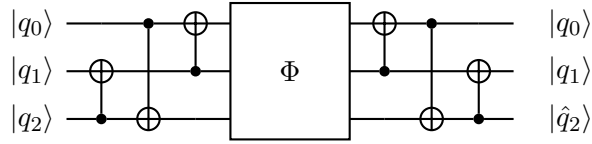
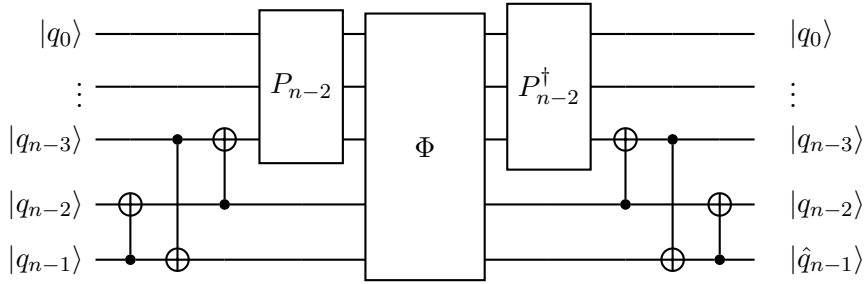
\begin{figure}[!ht]
\begin{quantikz}[row sep=0.3cm,column sep=0.3cm]
\lstick{$|q_0\rangle$}   & \qw & \qw      & \qw      &\gate[3]{P_{n-2}}       &\gate[5]{\ \quad \Phi \ \quad } & \gate[3]{P^\dag_{n-2}}      & \qw      &\qw       &\qw &\qw &\rstick{$|q_0\rangle$}\\
\lstick{\vdots}       &\qw          &\qw       & \qw      & & &  & \qw      &\qw     &\qw & \qw &\rstick{\vdots}\\
\lstick{$|q_{n-3}\rangle$} &\qw & \ctrl{2} & \targ\qw  & & & & \targ\qw & \ctrl{2} & \qw &\qw &\rstick{$|q_{n-3}\rangle$}\\
\lstick{$|q_{n-2}\rangle$} & \targ \qw & \qw & \ctrl{-1} & \qw & & \qw & \ctrl{-1} & \qw & \targ\qw & \qw &\rstick{$|q_{n-2}\rangle$} \\
\lstick{$|q_{n-1}\rangle$} & \ctrl{-1} & \targ\qw & \qw & \qw & &\qw & \qw & \targ\qw & \ctrl{-1} & \qw &\rstick{$|\hat{q}_{n-1}\rangle$}
\end{quantikz}
\caption{Hybrid QECC for odd $n$, where $|q_{n-1}\ra$ can be any qubit state}
\label{xyzodd}
\end{figure}
\begin{figure}[!ht]
\begin{quantikz}[row sep=0.3cm,column sep=0.3cm]
 \lstick{$|q_0\rangle$}    & \qw      & \qw      &\qw       & \gate[3]{P_{n-1}} &\gate[4]{\ \quad \Phi\ \quad } & \gate[3]{P^\dag_{n-1}}&\qw      & \qw      &\qw       & \qw & \rstick{$|q_0\rangle$}\\
 \lstick{\vdots}               &\qw       & \qw      &\qw        & & & & \qw      & \qw      &\qw     &\qw &  \rstick{\vdots}\\
    \lstick{$|q_{n-2}\rangle$}   & \ctrl{1} & \gate{H} & \ctrl{1}  & & & & \ctrl{1} & \gate{H} & \ctrl{1}    &\qw & \rstick{$|q_{n-2}\rangle$}\\
    \lstick{$|q_{n-1}\rangle$} & \targ \qw   &      \qw & \targ \qw & \qw   &  & \qw   & \targ \qw   & \qw      & \targ \qw   &\qw   &\rstick{$|q_{n-1}\rangle$}
\end{quantikz}
\caption{Hybrid QECC for even $n$ and $|q_{n-1}q_{n-2}\ra \in \{ |00\ra, |10\ra, |01 \ra, |11\ra\}$ }
\label{xyzeven}
\end{figure}
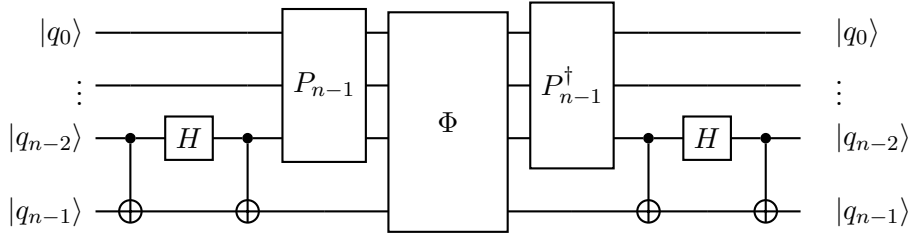

Here $P_n$ denotes the encoding matrix for the $n$-qubit case. That is, 
$P_2=\frac{1}{\sqrt{2}}(I_2\otimes \sigma_z+\sigma_x\otimes \sigma_x)$ and $P_3$ is the permutation matrix such that for any $a,b,c\in \{0,1\}$, $P_3|abc\rangle=|a\oplus c\rangle|a\oplus b\rangle|a\oplus b\oplus c\rangle$ or \footnotesize
\[
P_3\begin{bmatrix}
v_{000} & v_{001} & v_{010} & v_{011} & v_{100} & v_{101} & v_{110} & v_{111} 
\end{bmatrix}^T= \begin{bmatrix}
v_{000} & v_{111} & v_{101} & v_{010} & v_{110} & v_{001} & v_{011} & v_{100} 
\end{bmatrix}^T
\]
and for $k\geq 2$,
\[P_{2k}=(I_2\otimes P_{2k-1})(P_2\otimes I_{2^{2k-2}}) \mbox{ and } P_{2k+1}=(I_4\otimes P_{2k-1})(P_3\otimes I_{2^{2k-2}})\]
\normalsize
\subsection{Experimental results using Qiskit}
The Jupyter notebooks used to run these experiments are also available in the following Github repository:\medskip
\begin{center}
\texttt{https://github.com/dcpelejo/QECC/tree/main/XYZ\%20hybrid\%20experiments}\medskip
\end{center}

\subsubsection{Pure state, arbitrary state, and improvements}\label{pure_vs_arbi}
\begin{figure}[!ht]
\begin{subfigure}{0.49\textwidth}
\includegraphics[scale=0.18]{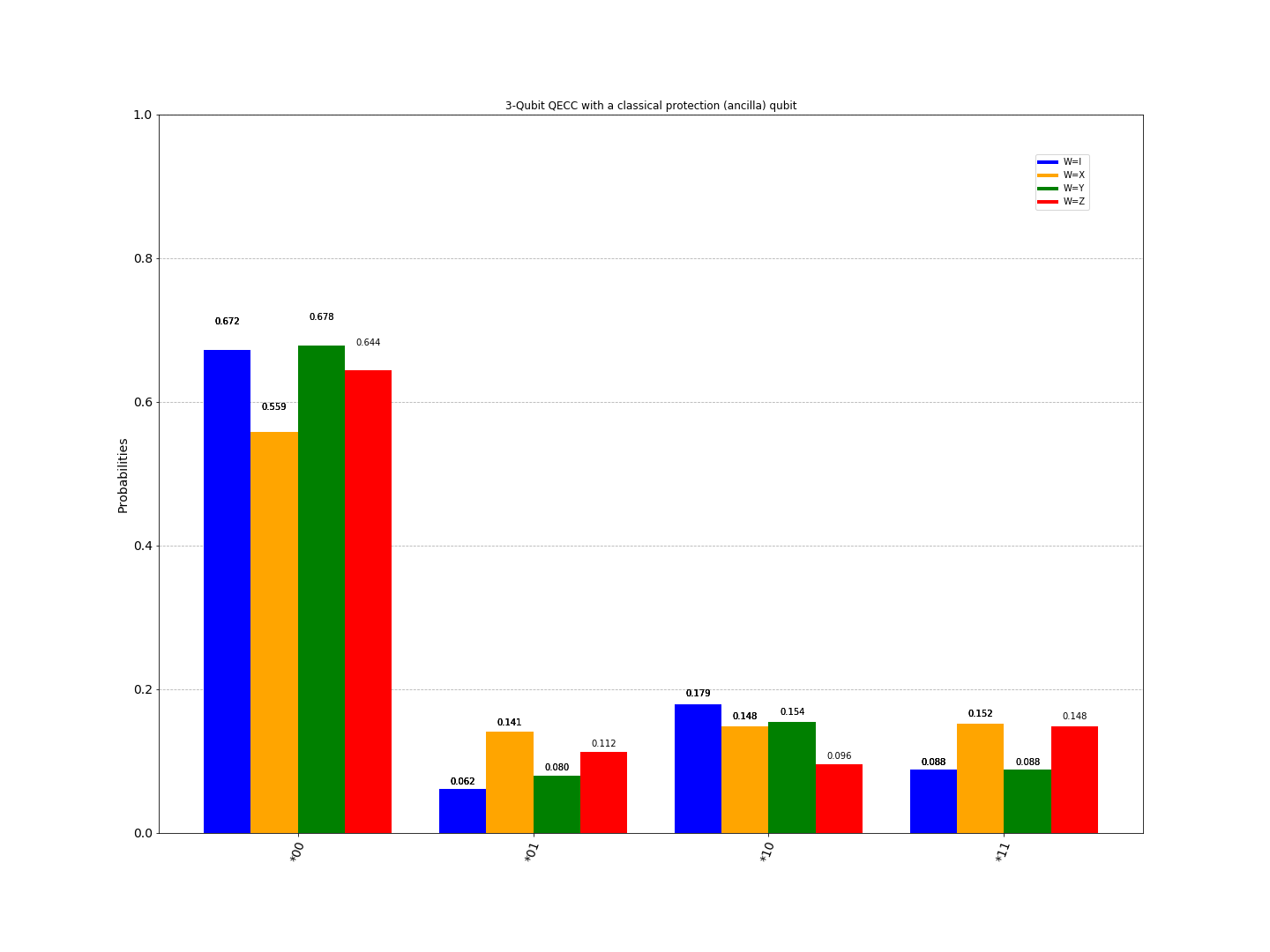}  
\caption{$|q_2q_1q_0\rangle=|000\rangle$}
\label{fig:3-pure}
\end{subfigure}
\begin{subfigure}{0.49\textwidth}
\includegraphics[scale=0.18]{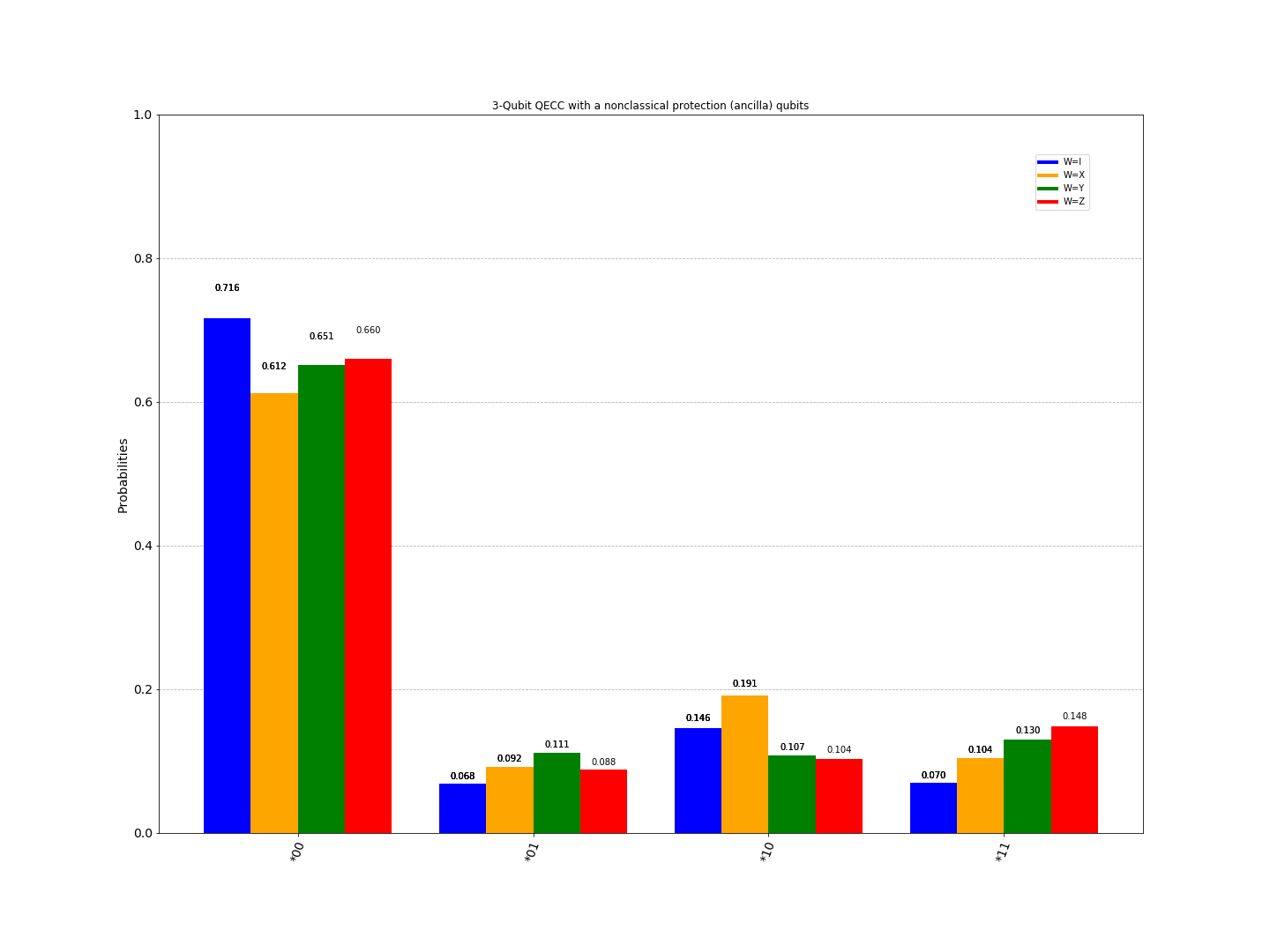}  
\caption{$|q_2q_1q_0\rangle=R_y(\frac{3\pi}{4})|0
\rangle\otimes |00\rangle$}
\label{fig:3-arbi}
\end{subfigure}
\caption{3-qubit QECC with circuit diagram illustrated in (\ref{xyz3})}
\label{fig:3-pa}
\end{figure}

First, we implemented the QECC scheme for $n = 3$ as illustrated in Figure \ref{xyz3}.
The results improved those in \cite{LLP} as shown in Figure~\ref{fig:3-pa}(\subref{fig:3-pure}). Note that experiments on this section were done on \texttt{ibmq\_burlington}. 

Besides using the pure state $|0\ra$ as the protection qubit, we may use any other qubit state. From Figure~\ref{fig:3-pa}(\subref{fig:3-arbi}), we observe that the results using the protection qubit $|q_2\rangle=|0\rangle$ and using $|q_2\rangle=R_y(\frac{3\pi}{4})|0\rangle$ are fairly similar. Since the quantum channel only has error operator
$W^{\otimes n}$ for $W = \{I_2, \sigma_x, \sigma_y, \sigma_z\}$, 
it is not surprising that we have 
a more effective scheme compared to the one in Section 2. 
Moreover,  we only need 
to use one ancilla qubit to protect two qubits, and the experimental results are better than that of the general scheme which uses three ancilla qubits to  protect two qubits as shown in Section 2.2.

\begin{figure}[!ht]
\begin{subfigure}{0.49\textwidth}
\includegraphics[scale=0.18]{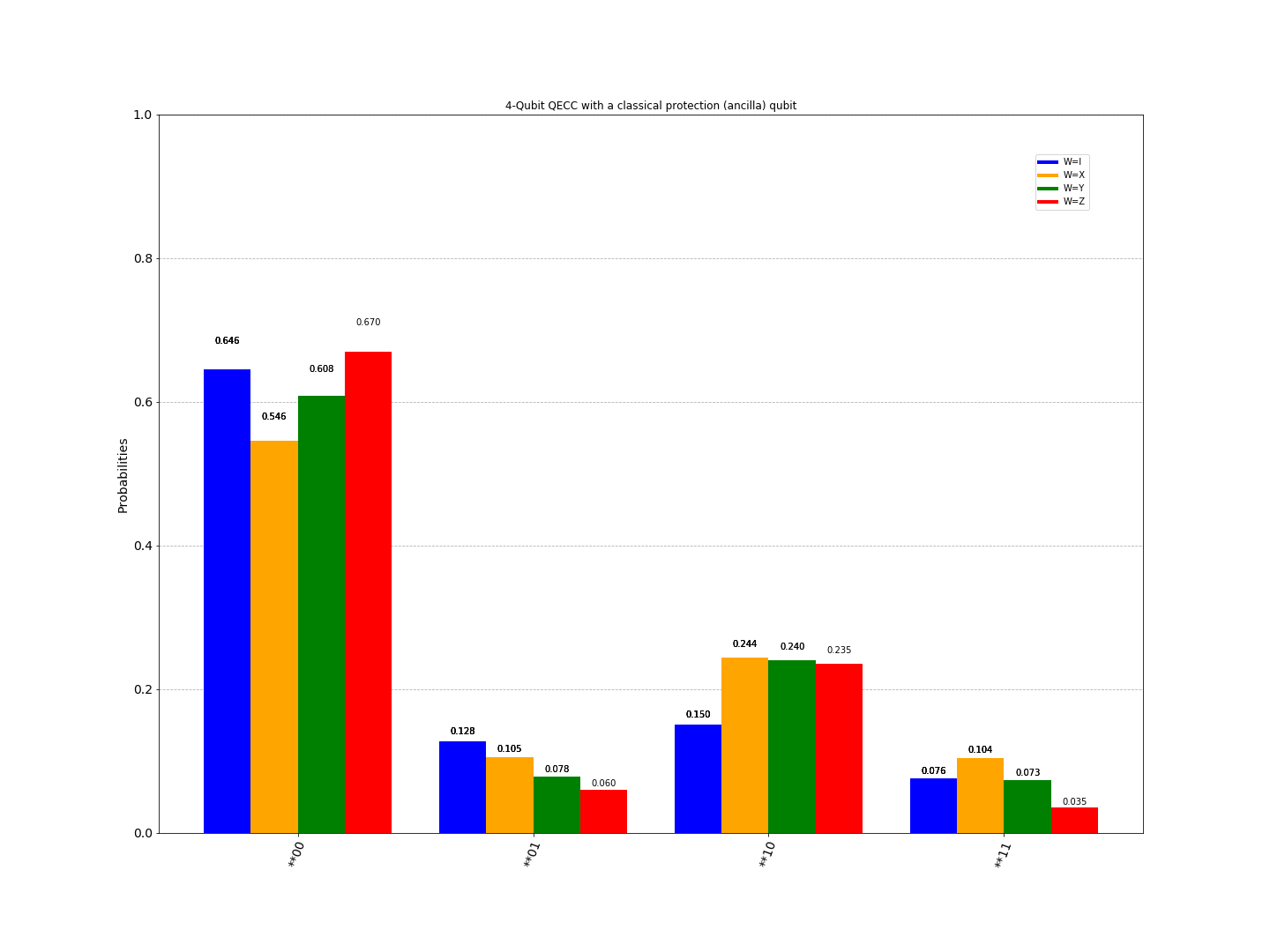}  
\caption{$|q_3q_2q_1q_0\rangle=|0000\rangle$}
\label{fig:4-pure}
\end{subfigure}
\begin{subfigure}{0.49\textwidth}
\includegraphics[scale=0.18]{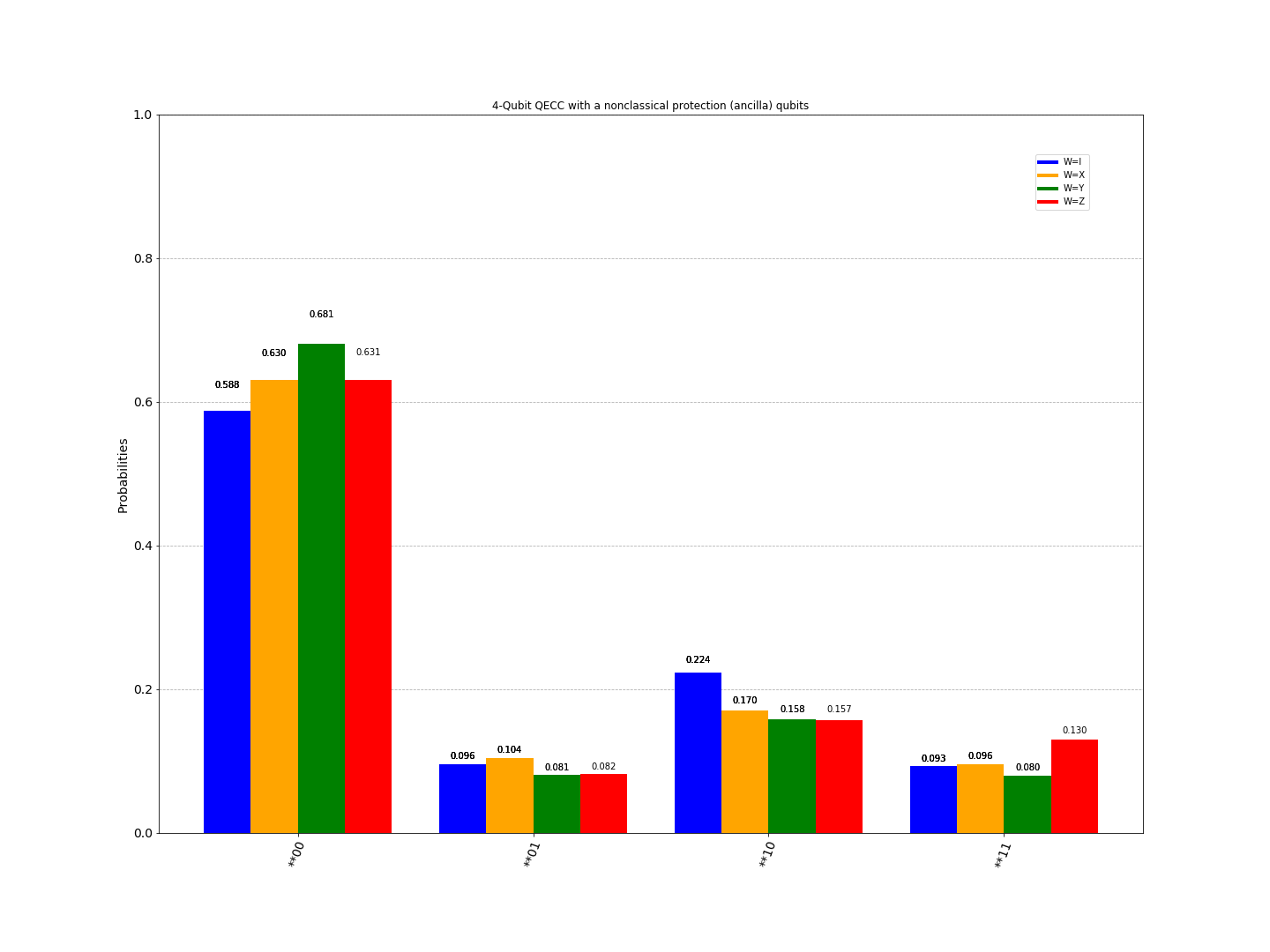}  
\caption{$|q_3q_2q_1q_0\rangle=R_y(\frac{3\pi}{4})|0
\rangle\otimes R_y(\frac{\pi}{4})|0
\rangle\otimes |00\rangle$}
\label{fig:4-arbi}
\end{subfigure}
\caption{4-qubit QECC with circuit diagram illustrated in (\ref{xyzeven})}
\end{figure}

Next, we implemented the scheme for $n$ qubits with $n = 4, 5$. Now, for the 4-qubit channel, we may use two pure classical ancillas or two arbitrary ancillas to protect two qubits. The experimental results are similar as shown in \ref{fig:4-pure} and \ref{fig:4-arbi}. Note that if we want to protect two classical bits of information encoded in the ancillas as well, then we must use two pure classical ancillas. The result of experiments where the protection ancillas $|q_3q_2\rangle$ are set to be $|01\rangle, |10\rangle$ or $|11\rangle$ can be found in Appendix 5. 

\begin{figure}[!ht]
\begin{subfigure}{1\textwidth}
\includegraphics[width=1\linewidth]{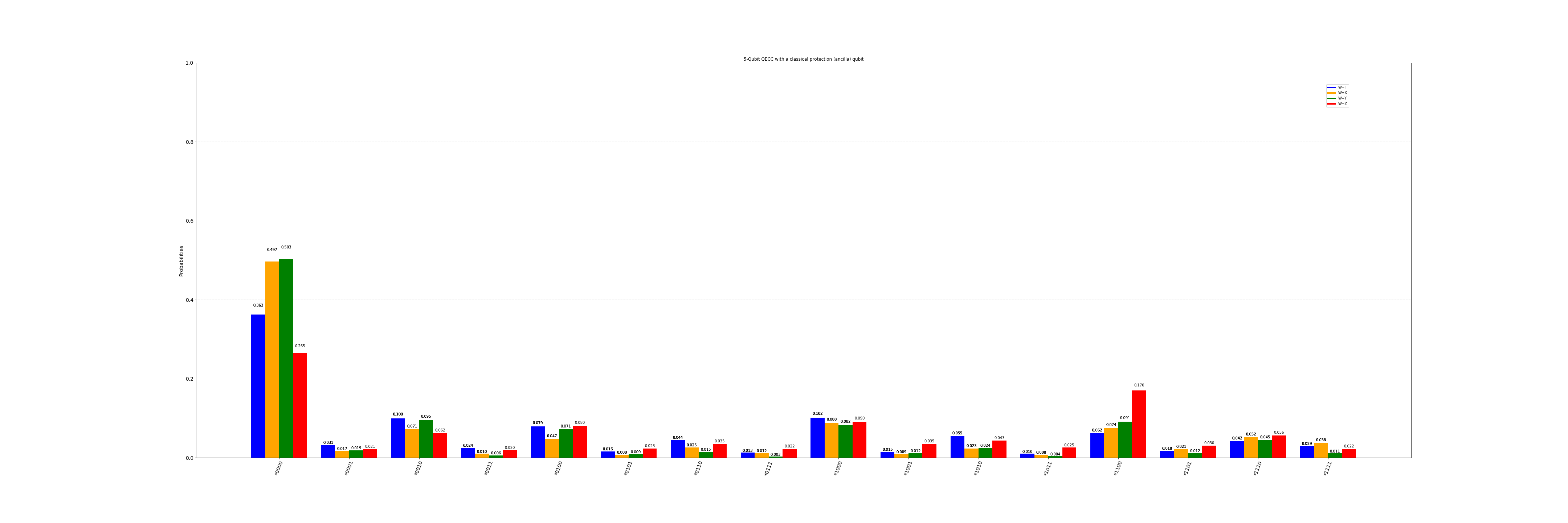}  
\caption{$|q_4q_3q_2q_1q_0\rangle=|00000\rangle$}
\label{fig:5-pure}
\end{subfigure}
\begin{subfigure}{1\textwidth}
\includegraphics[width=1\linewidth]{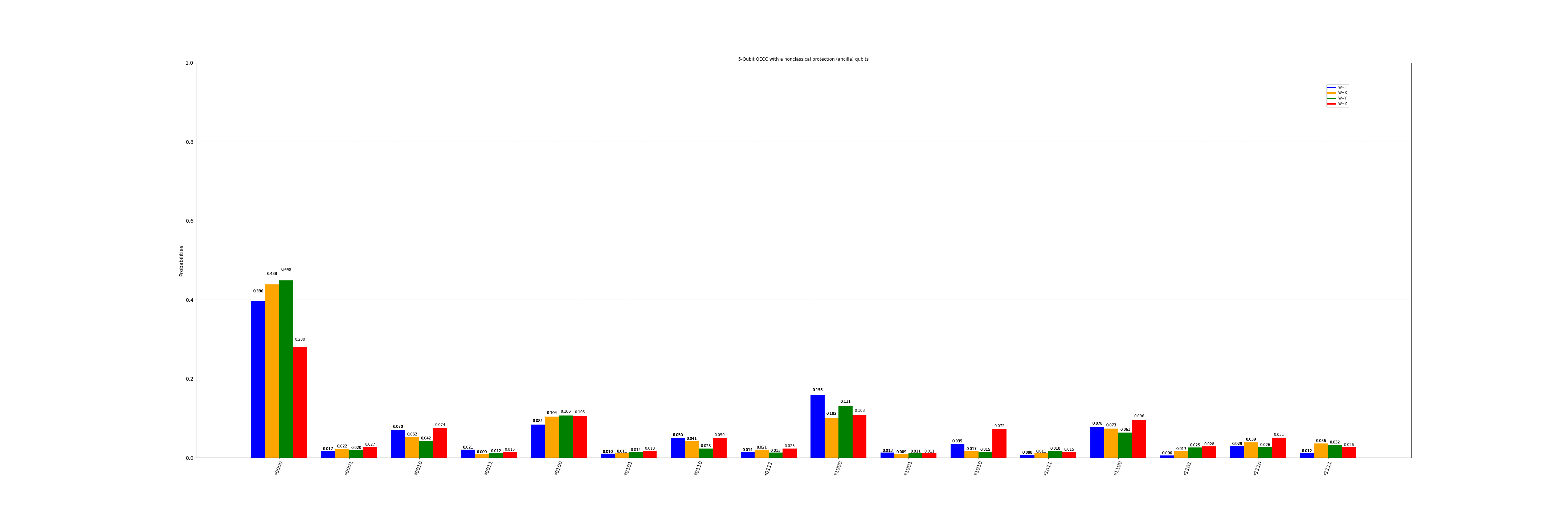}  
\caption{$|q_4q_3q_2q_1q_0\rangle=R_y(\frac{3\pi}{4})|0
\rangle\otimes |0000\rangle$}
\label{fig:5-arbi}
\end{subfigure}
\caption{5-qubit QECC with circuit diagram illustrated in (\ref{xyzodd})}
\label{fig:5-pa}
\end{figure}

Figure~\ref{fig:5-pa}(\subref{fig:5-pure})-(\subref{fig:5-arbi}) show the 5-qubit scheme with pure classical and arbitrary state protection qubits respectively. In \cite{LLP}, the experimental results for 4 and 5-qubit schemes were not satisfactory. However, our new experiments produced significantly improved outcomes, as demonstrated in our current findings.

\subsubsection{Errors across different IBM machines}

We note that our implementation uses only CNOT gate and Hadamard
gates, which are basic gates in the IBM quantum computers. We posited that the choice of whether to use arbitrary or pure classical state protection qubits will not impact our results. Therefore, in this subsection, we will focus on letting the protection qubits be any arbitrary state since, in practice, an arbitrary qubit is less  expensive to prepare than pure classical ones. 

For our numerical experiments in this subsection, we use 5 different IBM machines: \texttt{ibmq\_santiago}, \texttt{ibmq\_vigo},\texttt{ibmq\_valencia}, \texttt{ibmq\_ourense}, and \texttt{ibmq\_yorktown}. In \cite{LLP}, \texttt{ibmq\_yorktown} does not yield satisfactory results in 4 and 5-qubit experiments. However, we can see the improvement of this machine now, as it produces reasonable results using our implementation. 

\begin{figure}[!ht]
\begin{subfigure}{.32\textwidth}
  \includegraphics[scale=0.1]{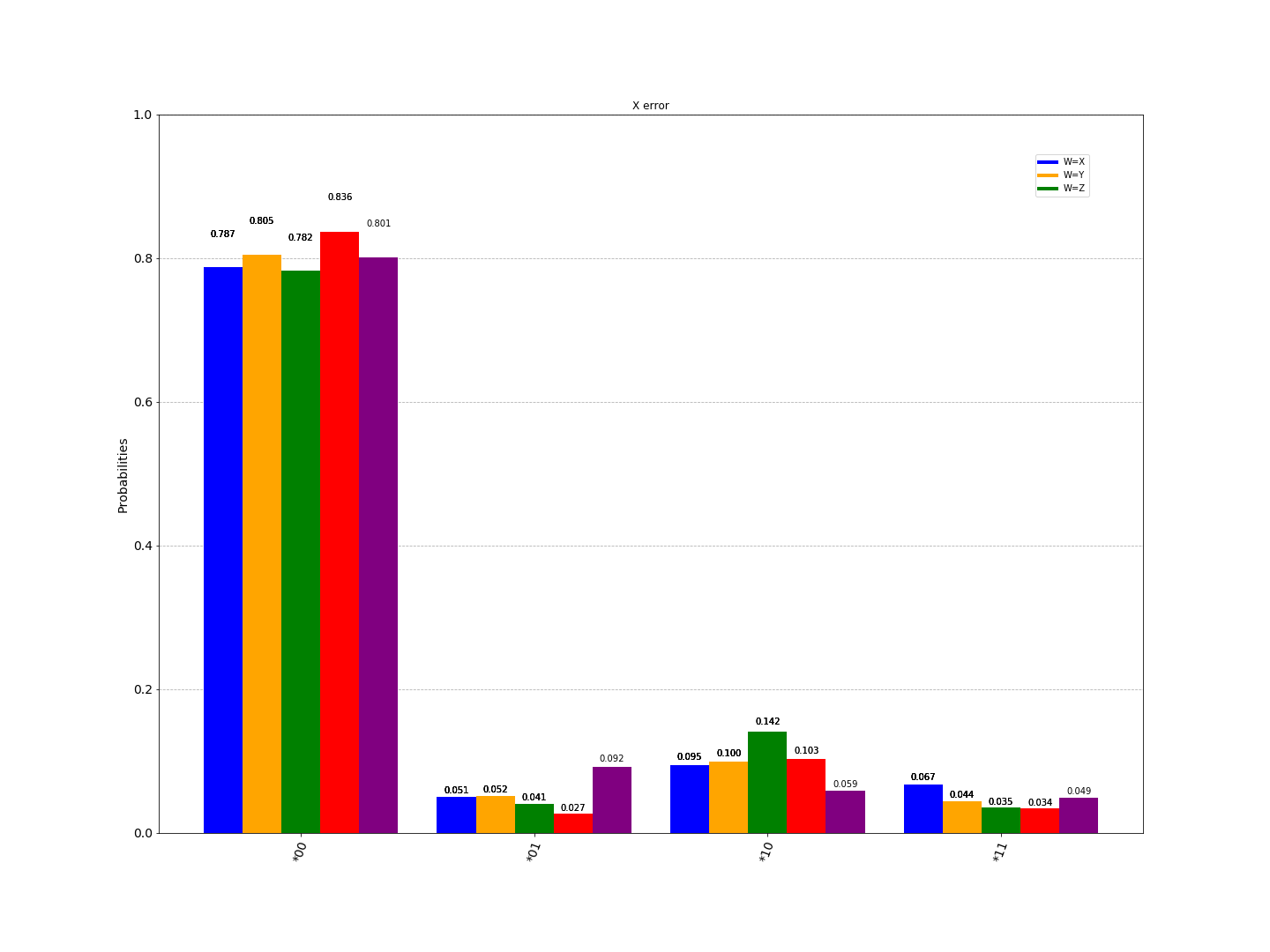}  
  \caption{$\sigma_x^{\otimes 3}$ error}
\end{subfigure}
\begin{subfigure}{.32\textwidth}
  \includegraphics[scale=0.1]{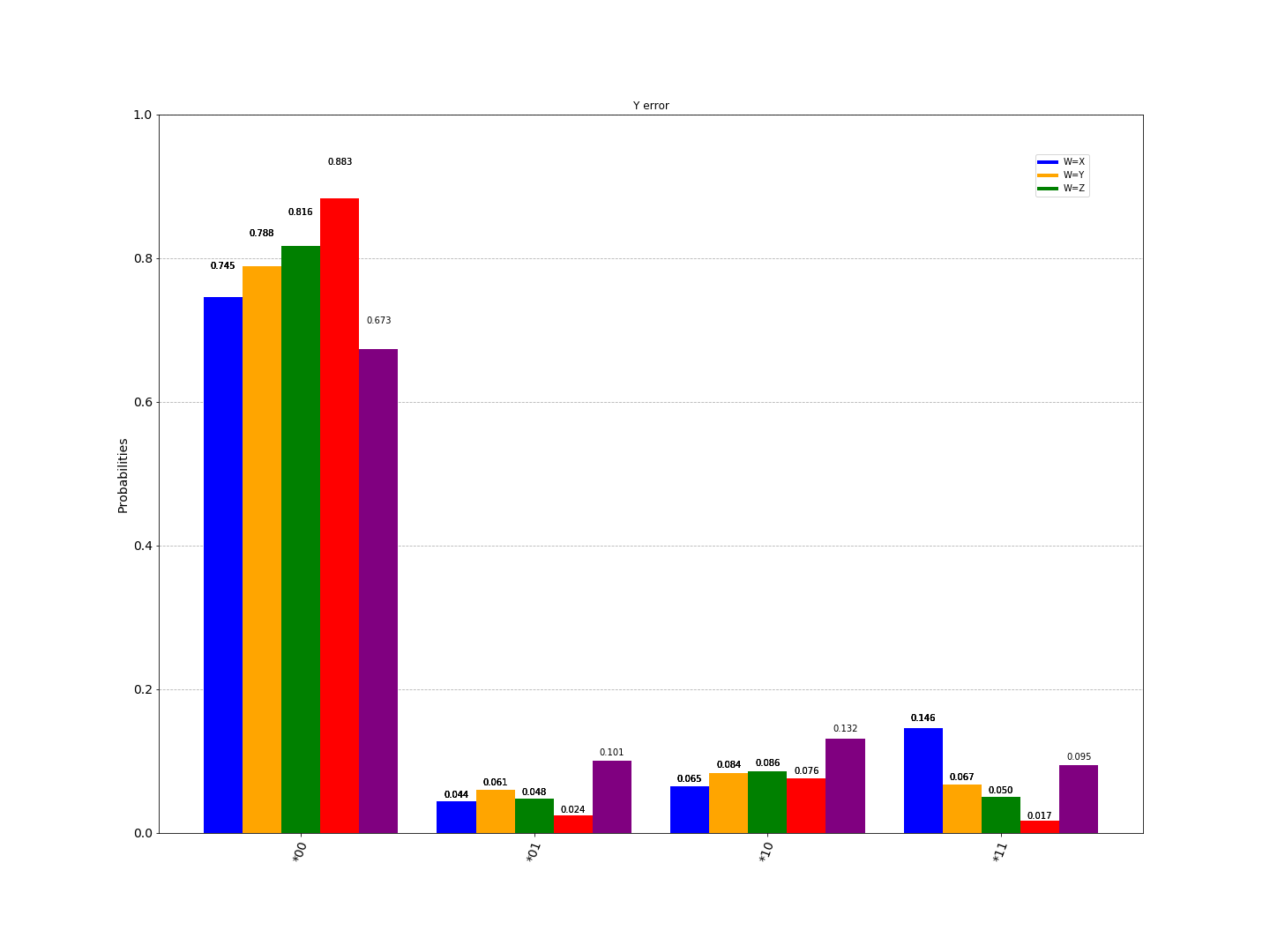}  
  \caption{$\sigma_y^{\otimes 3}$ error}
\end{subfigure}
\begin{subfigure}{.32\textwidth}
  \includegraphics[scale=0.1]{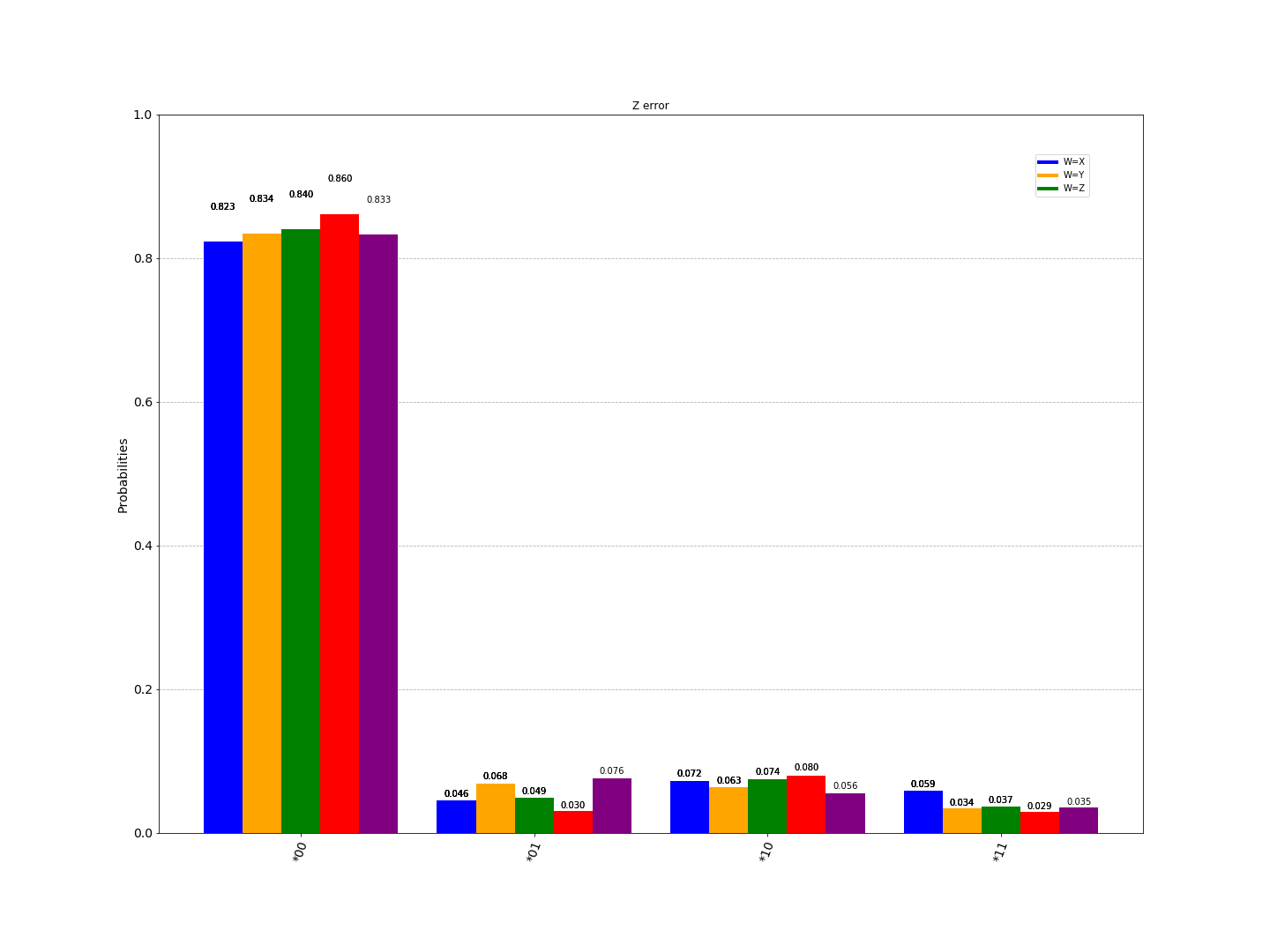}   
  \caption{$\sigma_z^{\otimes 3}$ error}
\end{subfigure}
\caption{Results of $\sigma_x^{\otimes 3}, \sigma_y^{\otimes 3}, \sigma_z^{\otimes 3}$ errors with arbitrary state protection in 5 IBM machines}
\label{fig:3-mult}
\end{figure}

We would like to note that since the two protection qubits are corrupted in
the 4-qubit scheme, only the other two qubits are measured. 

\begin{figure}[!ht]

\begin{subfigure}{.32\textwidth}
 \includegraphics[scale=0.1]{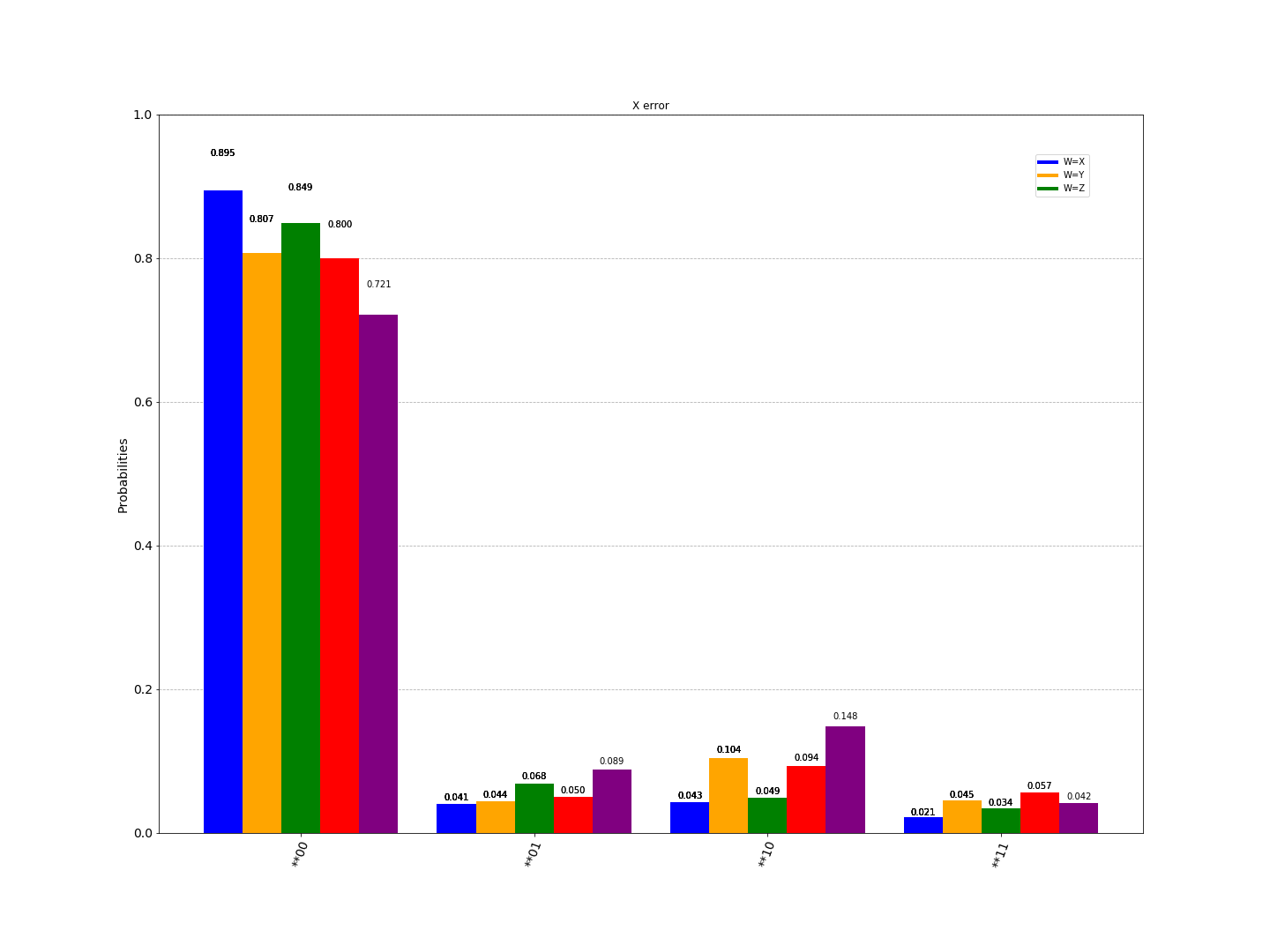}   
\caption{$\sigma_x^{\otimes 4}$ error}
\end{subfigure}
\begin{subfigure}{.32\textwidth}
 \includegraphics[scale=0.1]{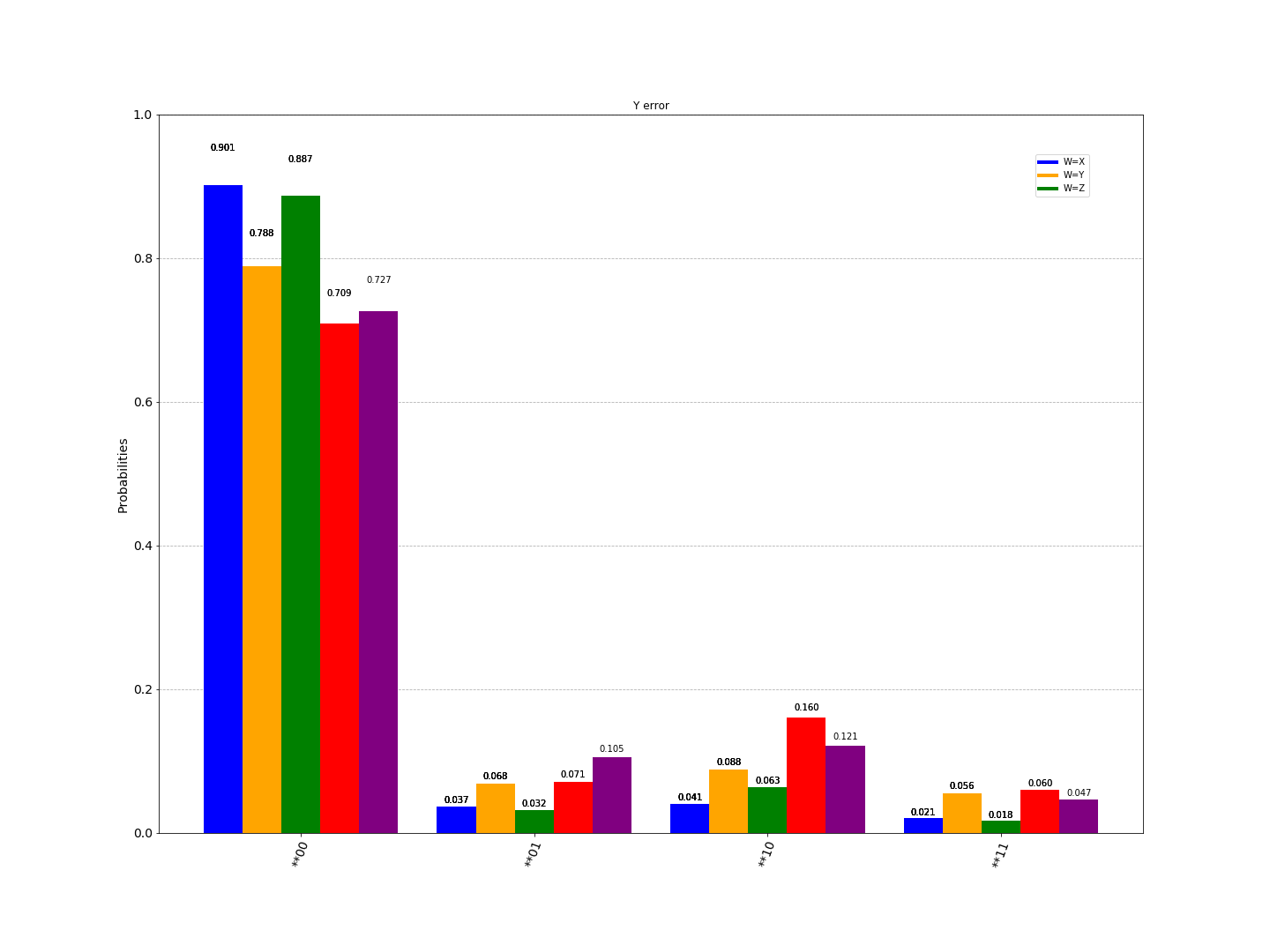}     
\caption{$\sigma_y^{\otimes 4}$ error}
\end{subfigure}
\begin{subfigure}{.32\textwidth}
 \includegraphics[scale=0.1]{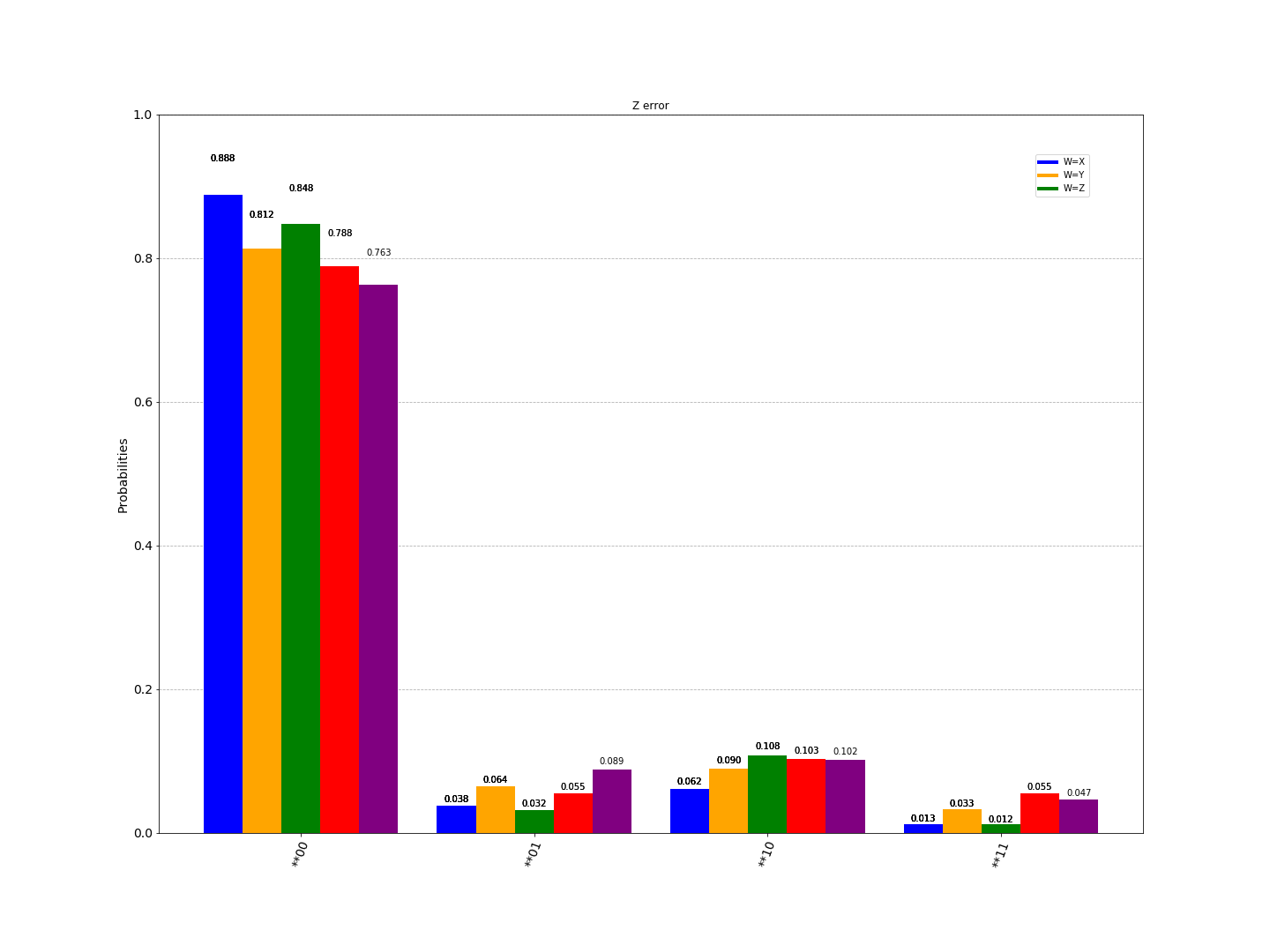}   
\caption{$\sigma_z^{\otimes 4}$ error}
\end{subfigure}
\caption{Results of $\sigma_x^{\otimes 4}, \sigma_y^{\otimes 4}, \sigma_z^{\otimes 4}$ errors with arbitrary state protection in 5 IBM machines}
\label{fig:4-mult}
\end{figure}

\begin{figure}[!ht]
\begin{subfigure}{1\textwidth}
\includegraphics[width=1\linewidth]{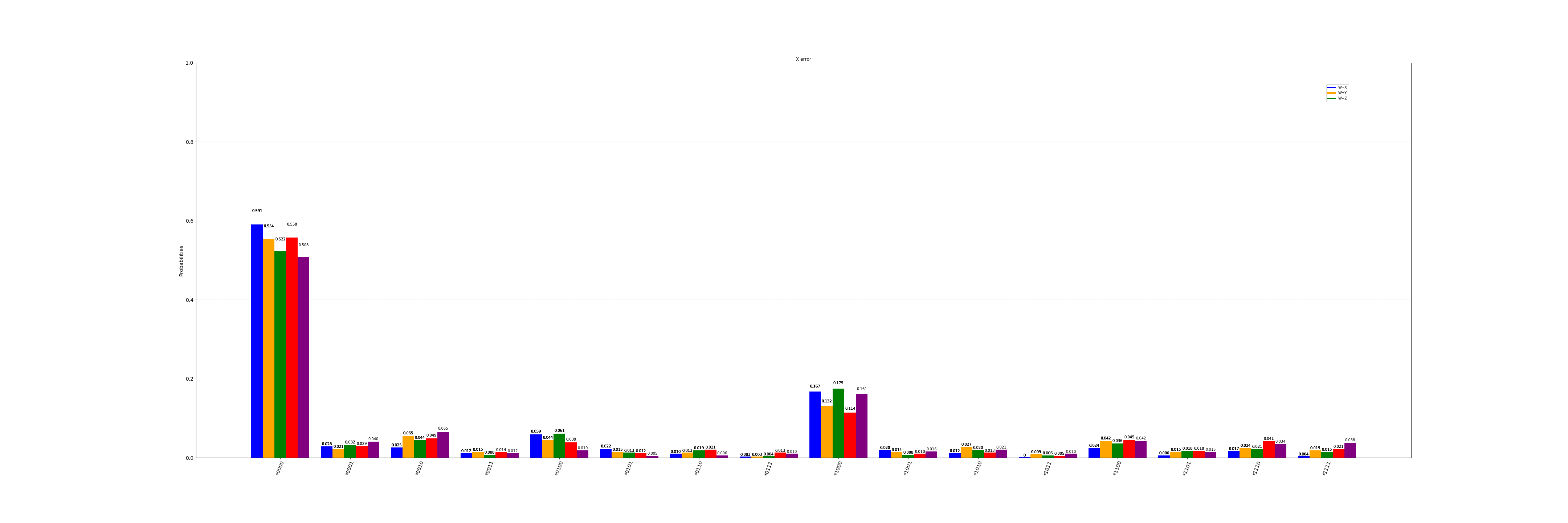}  
\caption{$\sigma_x^{\otimes 5}$ error}
\end{subfigure}
\begin{subfigure}{1\textwidth}
\includegraphics[width=1\linewidth]{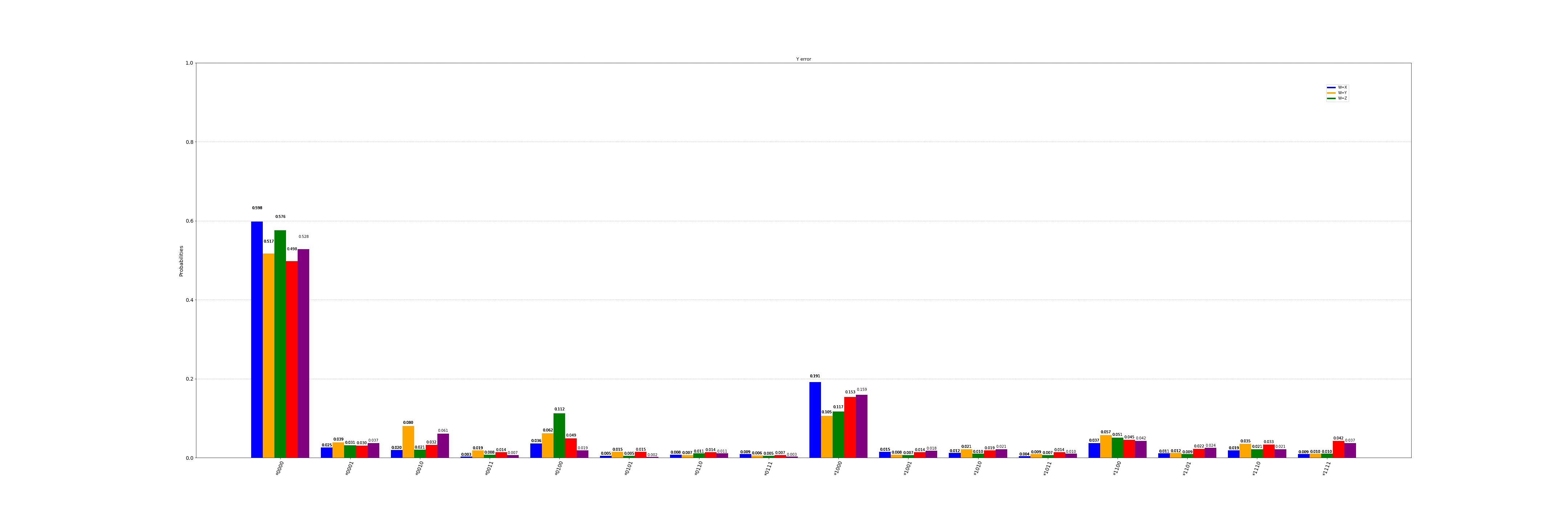}  
\caption{$\sigma_y^{\otimes 5}$ error}
\end{subfigure}
\begin{subfigure}{1\textwidth}
\includegraphics[width=1\linewidth]{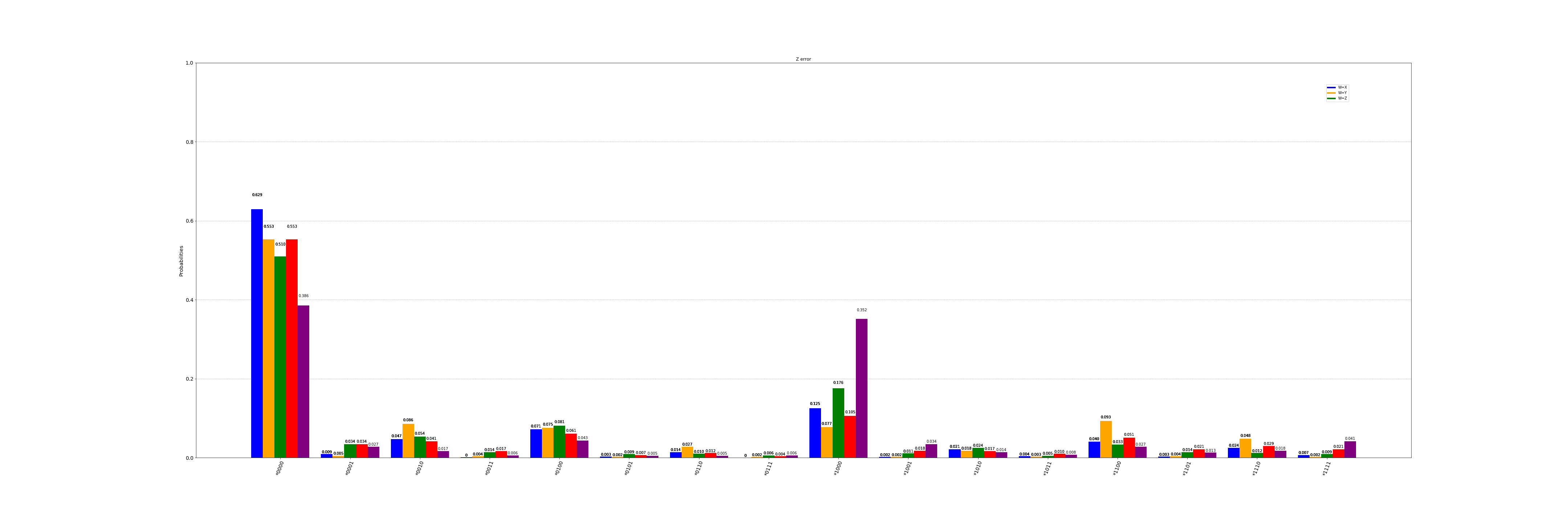}  
\caption{$\sigma_z^{\otimes 5}$ error}
\end{subfigure}
\caption{Results of $\sigma_x^{\otimes 5}, \sigma_y^{\otimes 5}, \sigma_z^{\otimes 5}$ errors with arbitrary state protection in 5 IBM machines}
\label{fig:5-mult}
\end{figure}

From the figures above, we can conclude that \texttt{ibmq\_ourense} is the best machine for 3-qubit schemes and \texttt{ibmq\_santiago} is best for 4 and 5 qubit schemes. Also, results from all machines match our predictions.

\section{Conclusion and further research}
In our study, we implemented a general recursive quantum error correction scheme for fully-correlated channels on $n$-qubits with error operators of the form $W^{\otimes n}$ using different IBM quantum computers. This scheme was proposed in earlier papers, where an erroneous decomposition of the encoding operator was given for the $3$-qubit channels. We modified the encoding operator so that it can be decomposed as the product of simple standard quantum gates which the IBM quantum computers can readily implement. We compared the errors on different IBM quantum computers, and tried to find out the key factors that will affect the accuracy of the results.  Furthermore,
we decomposed the encoding matrix as the product of  basic gates,
namely, CNOT gates and single unitary gates, and improved the
results in all but one of the IBM quantum computers we used. 
Then, we implement the recurrence scheme for 5-qubit channels.
It was somewhat surprising that better results were 
obtained by the standard gates 
decomposition instead of the basic gate decomposition 
despite the fact that much more CNOT gates were used in the former decomposition. 

We also implemented a hybrid quantum error correction scheme for the subclass of fully-correlated channels
where the error operators has the special form
$\sigma_x^{\otimes n}, \sigma_y^{\otimes n},
\sigma_z^{\otimes n}$ for the $n$-qubit channels when 
$n=4,5$. The scheme was implemented and good results were obtained, which covered cases a previous paper failed to handle.

There are a number of future research directions worth pursuing and we suggest a few of them in the following.

\begin{enumerate}
\item Implement the QECC schemes in our study for
$n$-qubit channels for higher $n$ by 
improving our schemes or finding better quantum computers.

\item Implement other quantum error correction schemes on IBM or other quantum computers.

\item In our study, we performed numerical experiments using different IBM quantum computers aiming to test how the architecture (e.g. network connections between the qubit nodes, range of approximate gate or measurement errors) of the quantum computers may perform differently, and to examine if some general beliefs in quantum computing are valid in practice. For instance, in general, one would believe that the use of more CNOT gates will cause more errors. In our case, we manually found a decomposition of our encoding and decoding operators using as few CNOT gates as possible. However, different IBM quantum computers  may provide different decompositions, often with more CNOT gates. Yet, the accuracy is comparable or even better than. So, it is of interest to examine the issues of how to optimize the performance of numerical implementations of quantum algorithms in connection to the hardware. 

\item While our study focused on comparing different IBM quantum computers, as pointed out by the referee, it would be interesting to perform experiments of our scheme on other quantum computing platforms such as Origin Quantum.
\end{enumerate}
\bigskip

\noindent{\bf Acknowledgments and Data Availability}

\medskip
The datasets generated during and/or analysed during the current study are available in the GitHub repository \texttt{https://github.com/dcpelejo/QECC}. We acknowledge the use of IBM Quantum services for this work. The views expressed are those of the authors, and do not reflect the official policy or position of IBM or the IBM Quantum team.

We thank Wenxuan Ding, Mikio Nakahara, Yiu-Tung Poon, and Yiyang Zhang for some discussion. We also thank the referee for many helpful suggestions. 

CK Li is an affiliate member of the Institute for Quantum Computing. His research was supported by the Simons Foundation Grant 851334. \bigskip

\noindent{\bf Conflict of Interest Statement}

The authors certify that they have no affiliations with or involvement in any organization or entity with any financial interest or nonfinancial interest in the subject matter or materials discussed in this manuscript.

\newpage
\section*{Appendix 1. The circuit decomposition in Figure 2 does not produce the unitary matrix in Figure \ref{oldU}.}

We can construct the simple gates corresponding to 
$Y_\theta, Y_{\pi/4}, I_2\otimes I_2\otimes \sigma_z,$ and the three control gates,
as $U_1, U_2, \dots, U_6$ as follows.
\begin{verbatim}
a1 = [1 -sqrt(2)]'/sqrt(3); a2 = [sqrt(2), 1]'/sqrt(3);
b1 = [1 -1]'/sqrt(2); b2 = [1 1]'/sqrt(2);
e0 = [1 0]'; e1 = [0 1]';

U1 = [kron(e0,kron(e0,e0)),kron(e0,kron(e0,e1)),kron(a1,kron(e1,e0)),kron(a1,kron(e1,e1)),
kron(e1,kron(e0,e0)), kron(e1,kron(e0,e1)), kron(a2,kron(e1,e0)), kron(a2,kron(e1,e1))];

U2 = [kron(e0,kron(b1,e0)),kron(e0,kron(b1,e1)),kron(e0,kron(b2,e0)),kron(e0,kron(b2,e1)),
kron(e1,kron(e0,e0)), kron(e1,kron(e0,e1)), kron(e1,kron(e1,e0)), kron(e1,kron(e1,e1))];
        
Z = [1 0;0 -1];
U3 = kron(eye(2), kron(eye(2), Z) );

U4 = [kron(e0,kron(e0,e0)),kron(e1,kron(e0,e1)),kron(e0,kron(e1,e0)),kron(e1,kron(e1,e1)),
kron(e1,kron(e0,e0)), kron(e0,kron(e0,e1)), kron(e1,kron(e1,e0)), kron(e0,kron(e1,e1))];   

U5 = [kron(e0,kron(e0,e1)),kron(e0,kron(e0,e0)),kron(e0,kron(e1,e0)),kron(e0,kron(e1,e1)),
kron(e1,kron(e0,e1)), kron(e1,kron(e0,e0)), kron(e1,kron(e1,e0)), kron(e1,kron(e1,e1))];

U6 = [kron(e0,kron(e0,e0)),kron(e0,kron(e0,e1)),kron(e0,kron(e1,e0)),kron(e0,kron(e1,e1)),
kron(e1,kron(e1,e0)), kron(e1,kron(e1,e1)), kron(e1,kron(e0,e0)), kron(e1,kron(e0,e1))];

U6*U5*U4*U3*U2*U1 = 

         0         0         0         0         0   -1.0000         0         0
    0.7071         0    0.4082         0         0         0    0.5774         0
   -0.7071         0    0.4082         0         0         0    0.5774         0
         0         0         0    0.8165         0         0         0   -0.5774
         0         0   -0.8165         0         0         0    0.5774         0
         0    0.7071         0   -0.4082         0         0         0   -0.5774
         0   -0.7071         0   -0.4082         0         0         0   -0.5774
         0         0         0         0    1.0000         0         0         0

[sqrt(2/3), sqrt(1/3), sqrt(1/6), 1/sqrt(2)] =  [0.8165, 0.5774, 0.4082, 0.7071]
\end{verbatim}
\bigskip

\noindent So, we see that $U \ne U_6U_5U_4U_3U_2U_1$.

\newpage

\section*{Appendix 2. Matlab scripts to verify the circuit decompositions of the matrix $U$.}
\subsection*{Decomposition in Figure \ref{newUdec}}

\begin{verbatim}
 U=[0,0,0,0,0,0,0,-1; sqrt(2/3),0,0,0,sqrt(1/3),0,0,0;
    -sqrt(1/6),0,sqrt(1/2),0,sqrt(1/3),0,0,0; 0,sqrt(1/6),0,sqrt(1/2),0,-sqrt(1/3),0,0;
    -sqrt(1/6),0,-sqrt(1/2),0,sqrt(1/3),0,0,0; 0,sqrt(1/6),0,-sqrt(1/2),0,-sqrt(1/3),0,0;
    0,-sqrt(2/3),0,0,0,-sqrt(1/3),0,0; 0,0,0,0,0,0,1,0];

E0=[1,0;0,0];	E1=[0,0;0,1];	Z=[1,0;0,-1];	X=[0,1;1,0];

P1=kron(eye(2),kron(E1,eye(2)))+kron(X,kron(E0,eye(2)));
P2=kron(E1,kron(eye(2),X))+kron(E0,eye(4));
P3=kron(eye(2),kron(X,E1))+kron(eye(2),kron(eye(2),E0));
Q1=kron(eye(4),Z);
A2=sqrt(1/2)*[1,-1;1,1];
Q2=kron(E0,kron(A2,eye(2)))+kron(E1,eye(4));
A3=[-sqrt(1/3),sqrt(2/3);sqrt(2/3),sqrt(1/3)];
Q3=kron(A3,kron(E0,eye(2)))+kron(eye(2),kron(E1,eye(2)));

P1*P2*P3*Q1*Q2*Q3-U %must close to zero matrix
\end{verbatim}

\subsection*{Decomposition in Figure 5}
\begin{verbatim}
a=pi/8;
t=asin(sqrt(1/3))/2; %alpha/4 
A=[cos(a),-sin(a);sin(a),cos(a)]; %Ry(pi/4)
B=[cos(t),-sin(t);sin(t),cos(t)]; %Ry(alpha/2)

S1=kron(B',eye(4));
S2=kron(X*B,eye(4));
S3=kron(eye(2),kron(A,eye(2)));
S4=kron(eye(2),kron(A',eye(2)));
X1=kron(eye(2),kron(X,eye(2)));
X2=kron(X,eye(4));
Z0=kron(eye(4),Z);

%C-control-target
C12=kron(X,kron(E1,eye(2)))+kron(eye(2),kron(E0,eye(2)));
C21=kron(E1,kron(X,eye(2)))+kron(E0,kron(eye(2),eye(2)));
C01=kron(eye(2),kron(X,E1))+kron(eye(2),kron(eye(2),E0));
C20=kron(E1,kron(eye(2),X))+kron(E0,kron(eye(2),eye(2)));

X1*C12*C20*C01*Z0*S4*X2*C21*S3*C21*S2*C12*S1*X1-U %must close to zero matrix
 \end{verbatim}
\newpage 
\section*{Appendix 3. Circuits Generated by IBMQ}

Here we demonstrate how the IBM quantum machines may process the same user-input circuit differently for two separate runs. 

\begin{figure}[!ht]
\begin{subfigure}{1\textwidth}
\begin{center}
\includegraphics[scale=0.16]{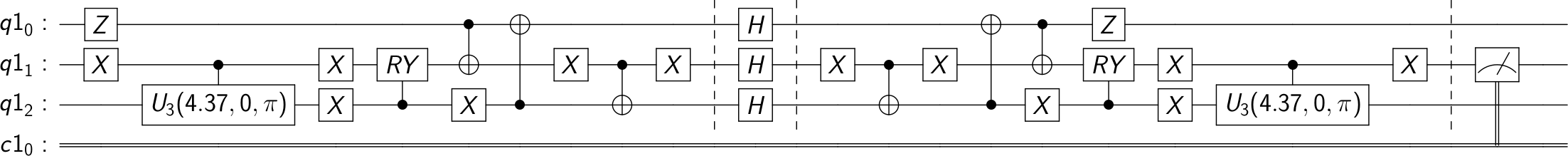}
\end{center}
\caption{user-input circuit diagram to implement QECC scheme}
\label{stdcircfig}
\end{subfigure}\medskip\\
\begin{subfigure}{1\textwidth}
\begin{center}
\includegraphics[scale=0.11]{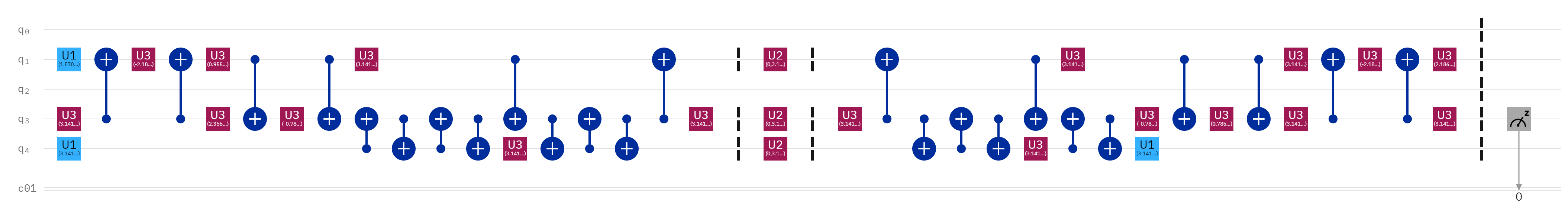}\\
\includegraphics[scale=0.11]{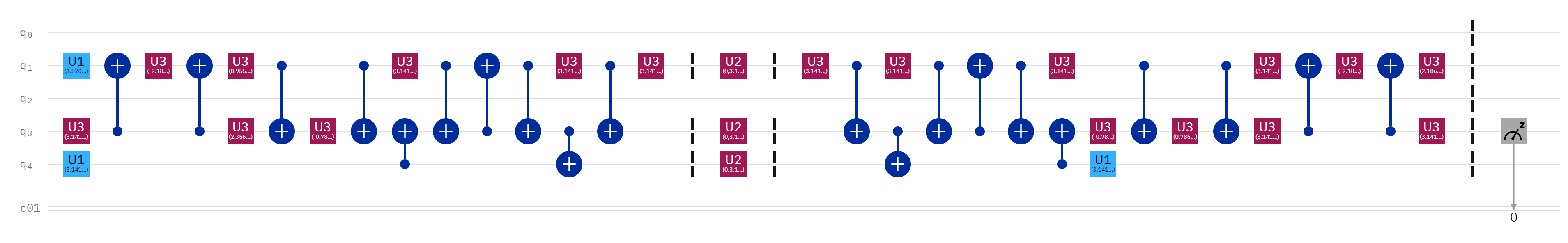}
\end{center}
\caption{Two different circuits generated by the transpiler for \texttt{ibmq\_valencia} given the input circuit in (a). }
\label{IBMCQ1sh}
\end{subfigure}\vspace{1cm}\\
\begin{subfigure}{1\textwidth}
\begin{center}
\includegraphics[scale=0.16]{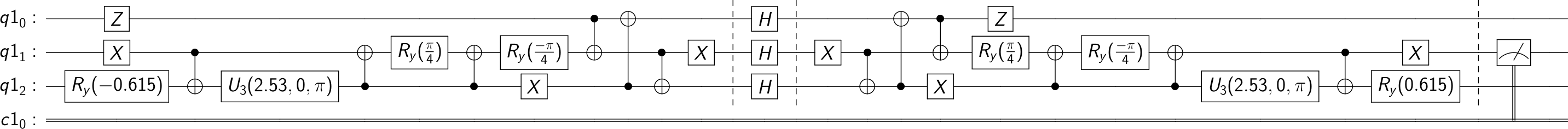}
\end{center}
\caption{user-input circuit diagram to implement QECC scheme}
\label{basiccircfig}
\end{subfigure}\medskip\\
\begin{subfigure}{1\textwidth}
\begin{center}
\includegraphics[scale=0.11]{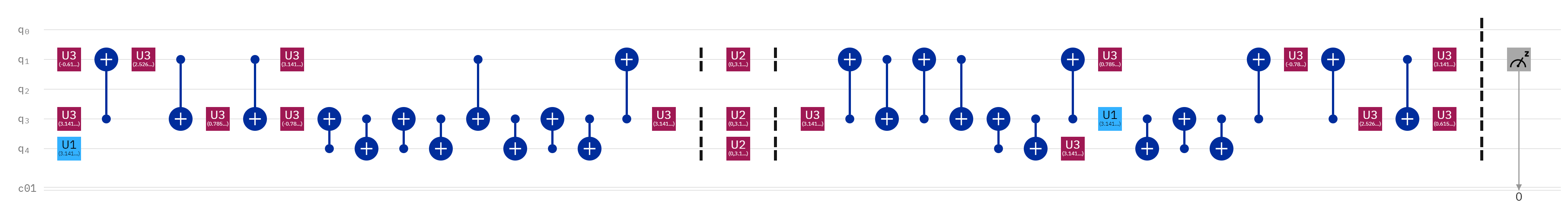}\\
\includegraphics[scale=0.11]{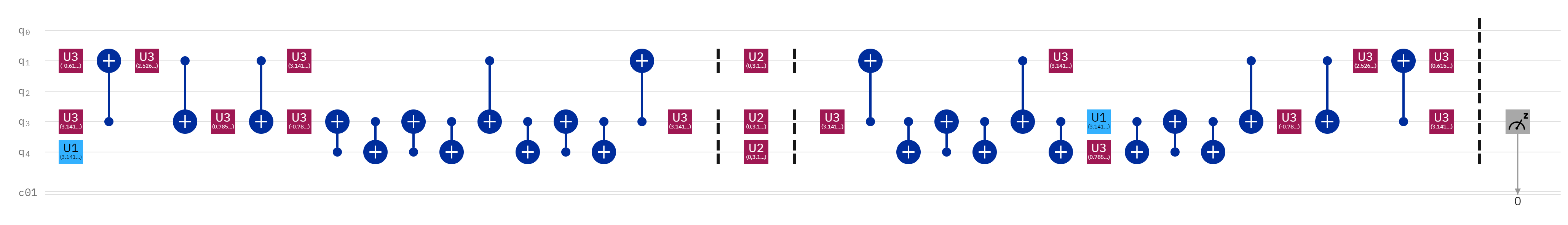}
\end{center}
\caption{Two different circuits generated by the transpiler for \texttt{ibmq\_valencia} given the input circuit in (c). }
\label{IBMCQ1bh}
\end{subfigure}
\caption{}
\label{app3}
\end{figure}

\newpage
\section*{Appendix 4. Results from the 5-qubit QECC implementation on the IBM Quantum Computers}
In this appendix, we present more experimental results for the implementation of the 5-qubit QECC presented in Section \ref{5qWWW}. Each experiment is run three times in the IBM quantum computers  \texttt{ibmq\_valencia}, \texttt{ibmq\_santiago},  \texttt{ibmq\_vigo}, \texttt{ibmq\_5\_yorktown},   \texttt{ibmq\_ourense} and \texttt{ibmq\_athens}. The leftmost histograms show the best (least error or highest probability for $|*0*0*\rangle$) of the three runs, while the rightmost histogram shows the worst of the three runs.

\begin{figure}[!ht]
\includegraphics[scale=0.085]{F7_SH1_all.png}
\includegraphics[scale=0.085]{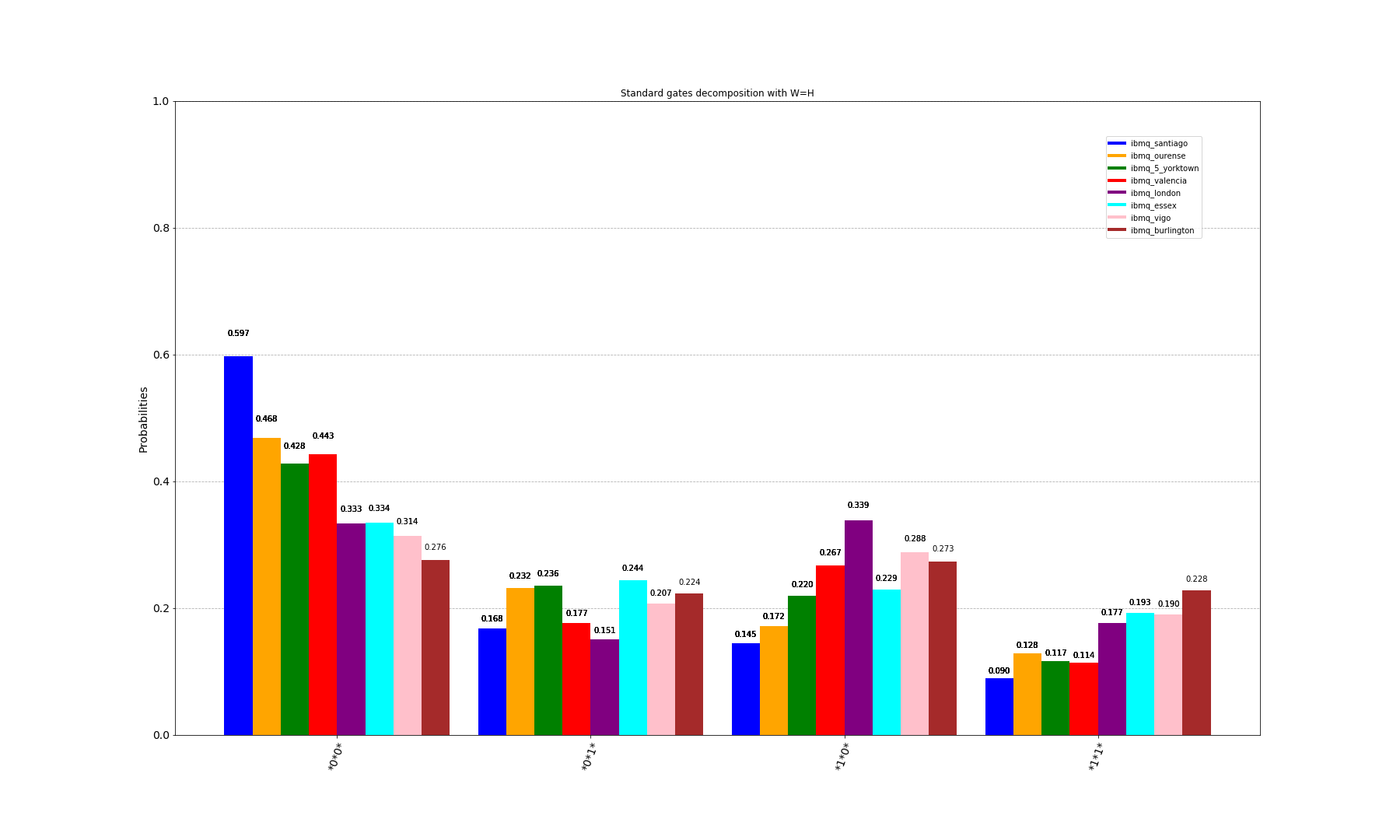}
\includegraphics[scale=0.085]{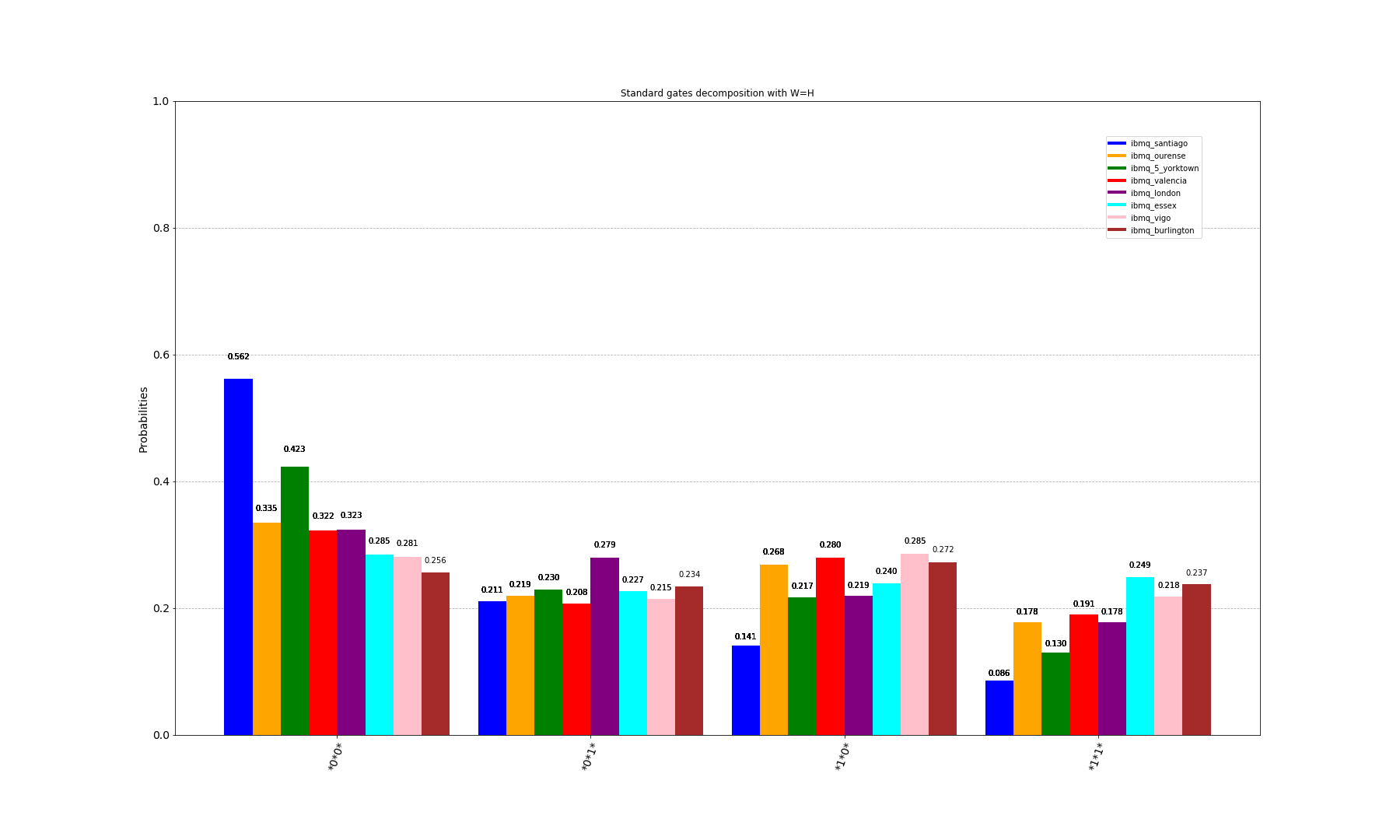}
\caption{using the standard gate decomposition of $U$ and $W=H$}
\end{figure}

\begin{figure}[!ht]
\includegraphics[scale=0.085]{F8_BH1_all.png}
\includegraphics[scale=0.085]{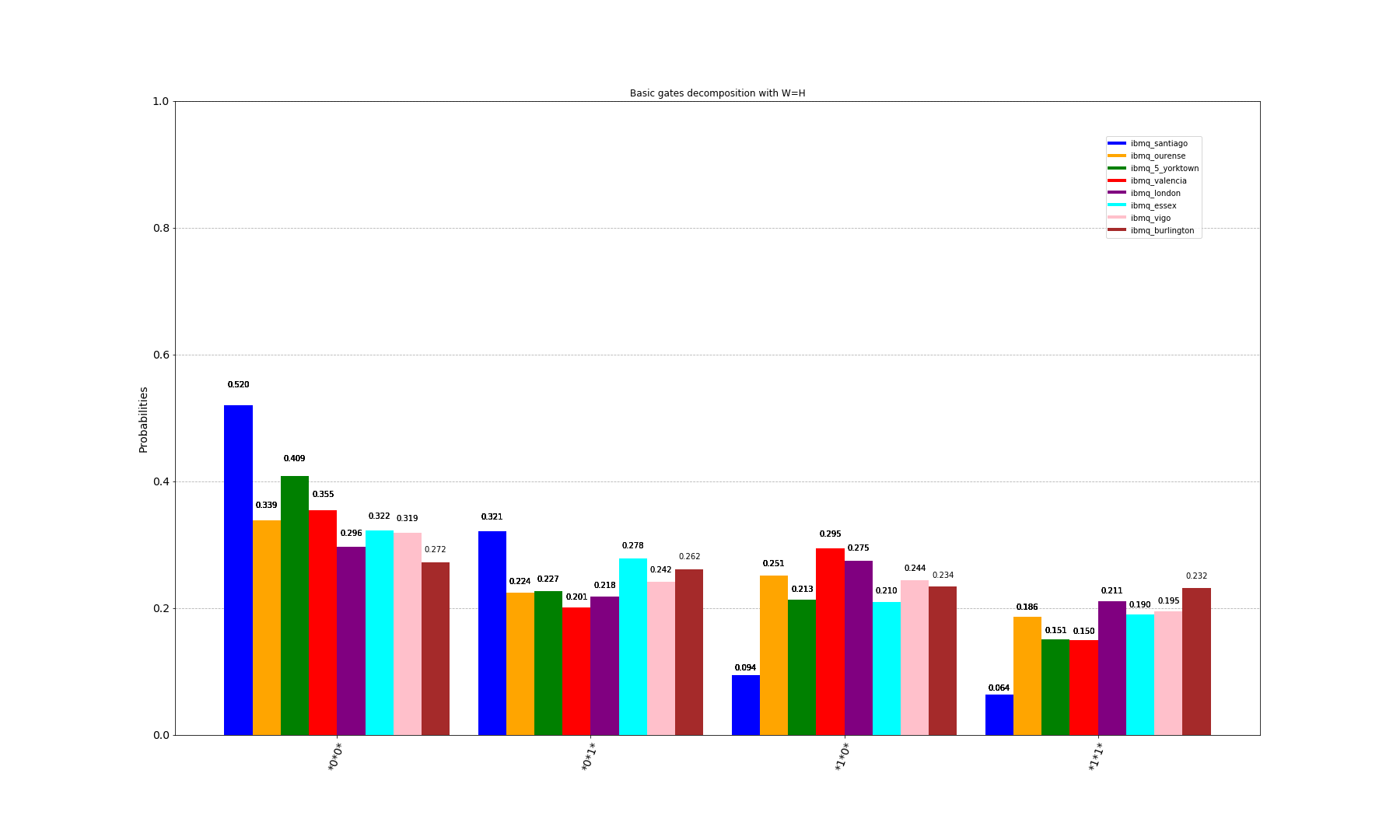}
\includegraphics[scale=0.085]{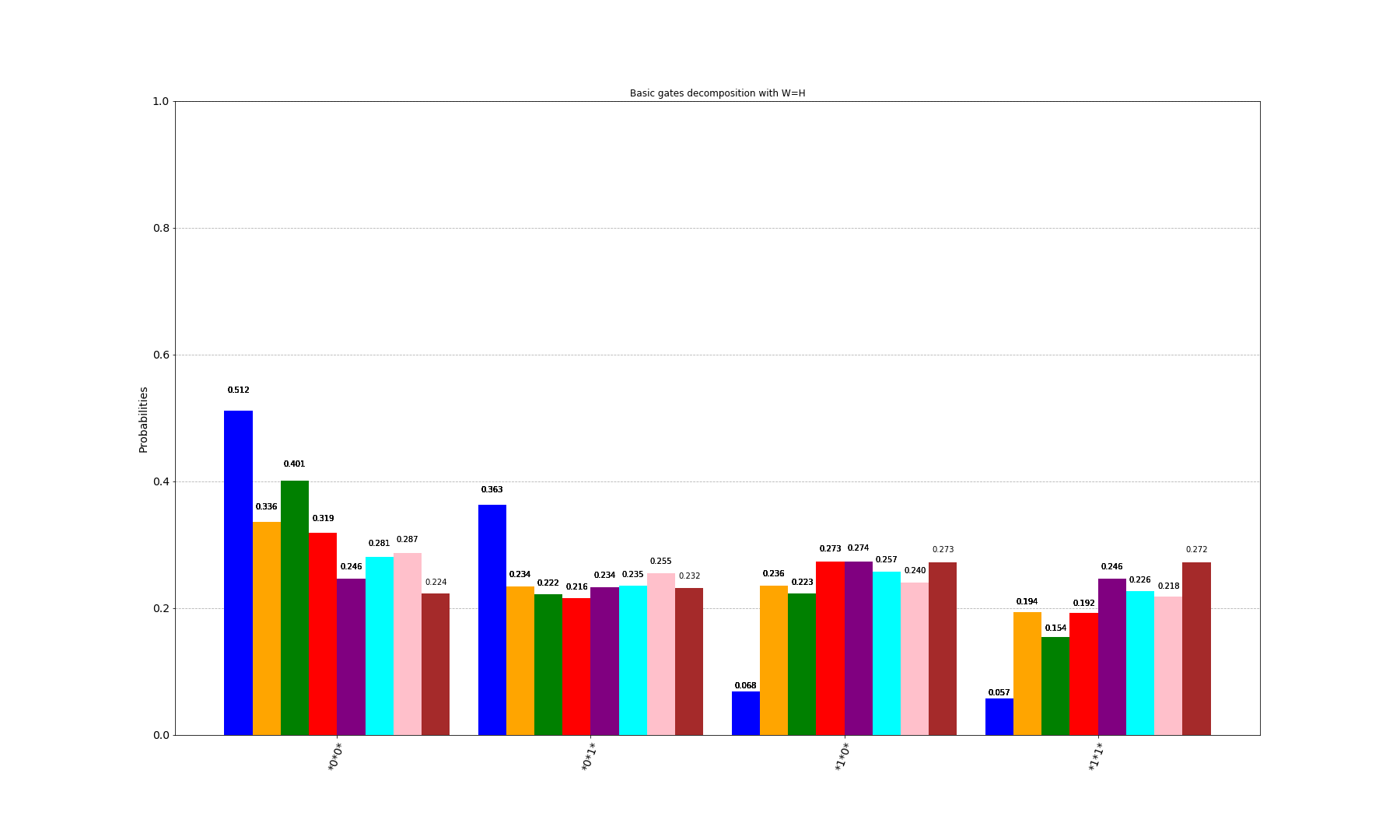}
\caption{using the basic gate decomposition of $U$ and $W=H$}
\end{figure}

\begin{figure}[!ht]
\includegraphics[scale=0.085]{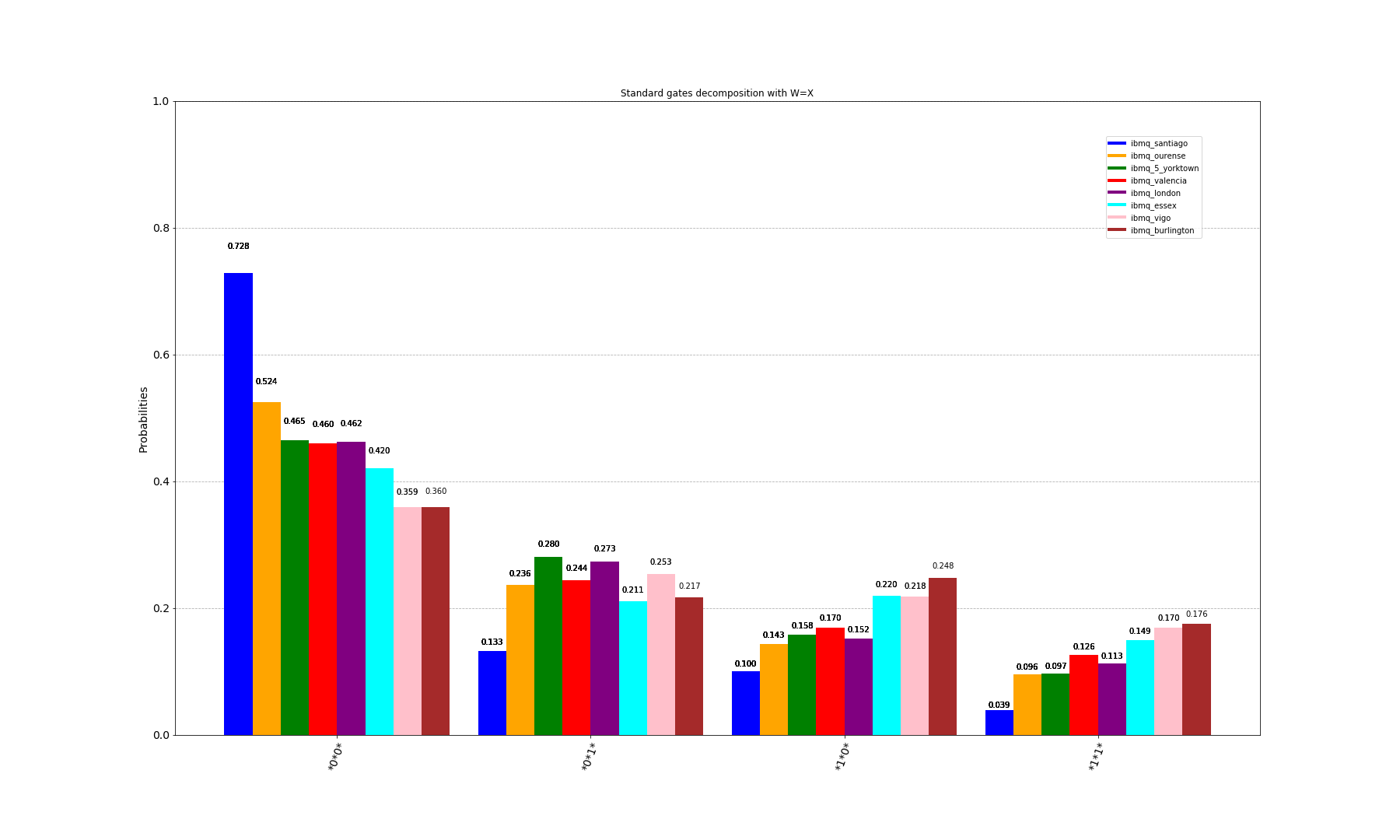}
\includegraphics[scale=0.085]{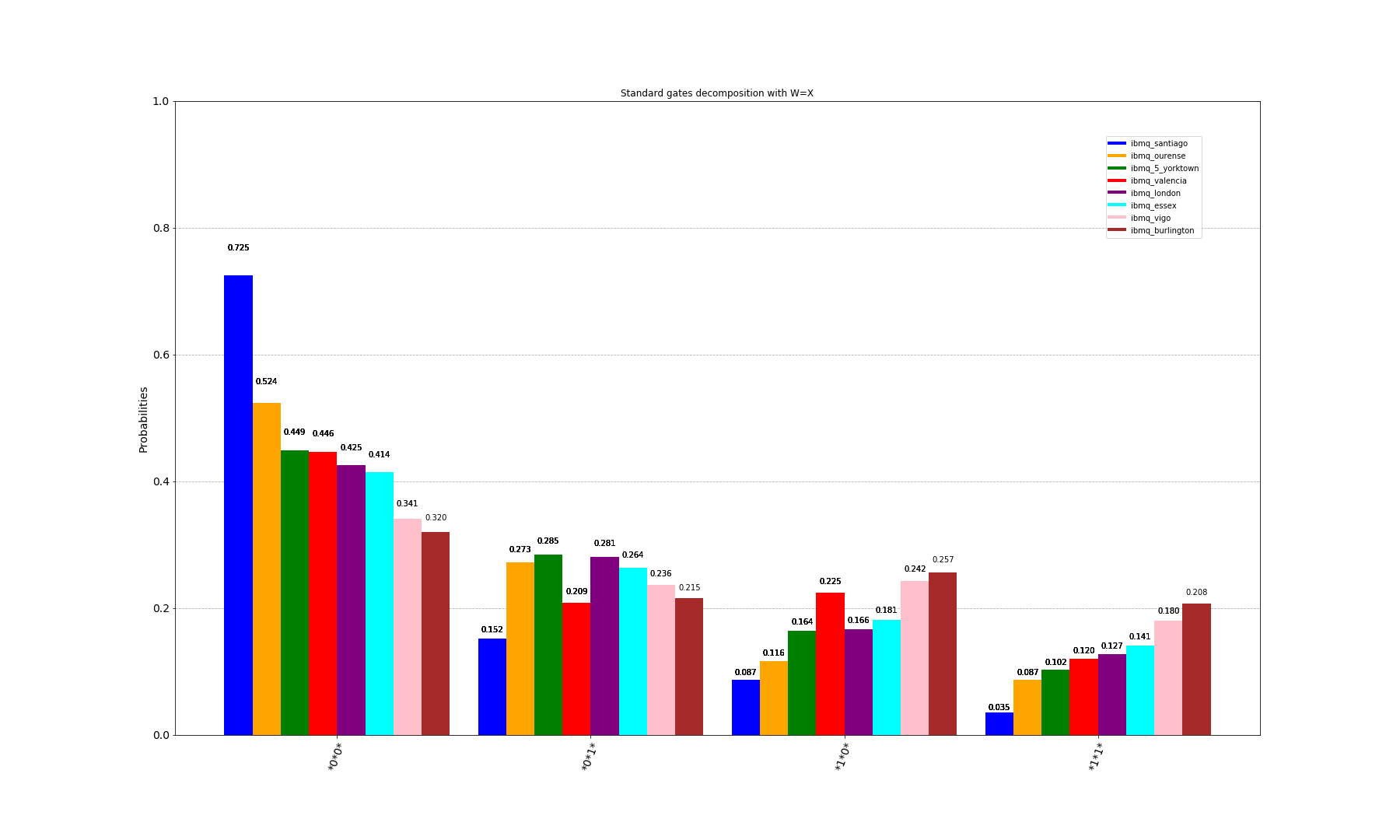}
\includegraphics[scale=0.085]{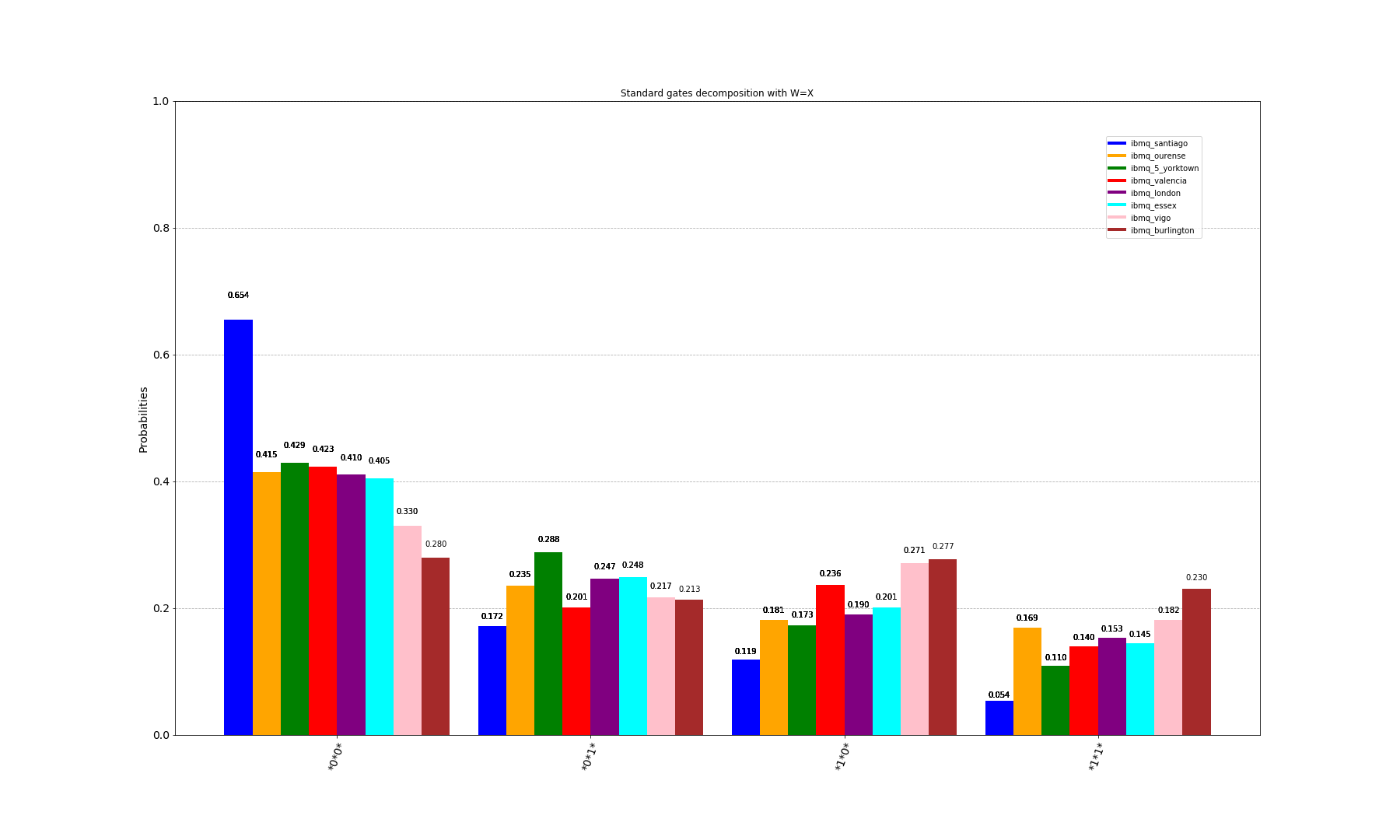}
\caption{using the standard gate decomposition of $U$ and $W=X$}
\end{figure}

\begin{figure}[!ht]
\includegraphics[scale=0.085]{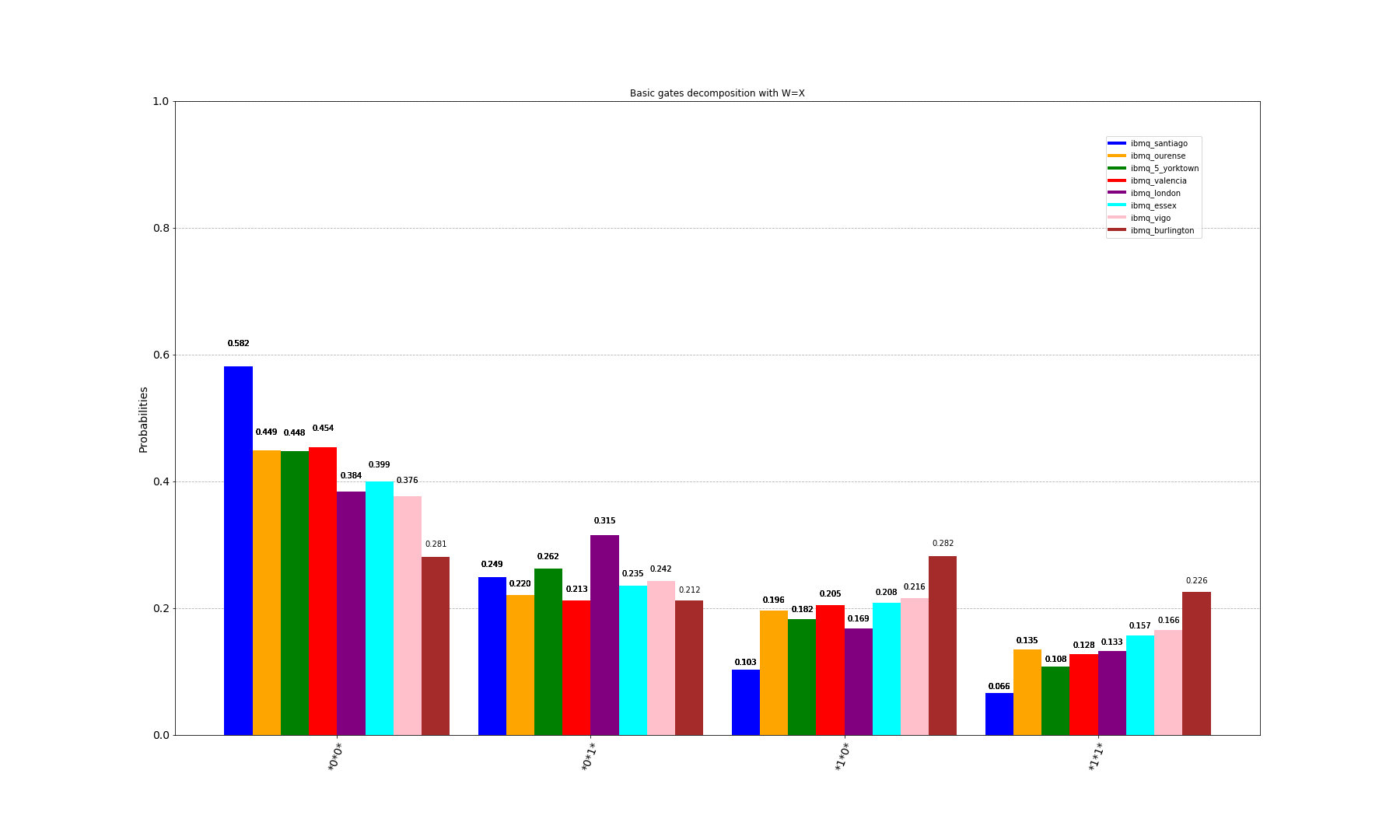}
\includegraphics[scale=0.085]{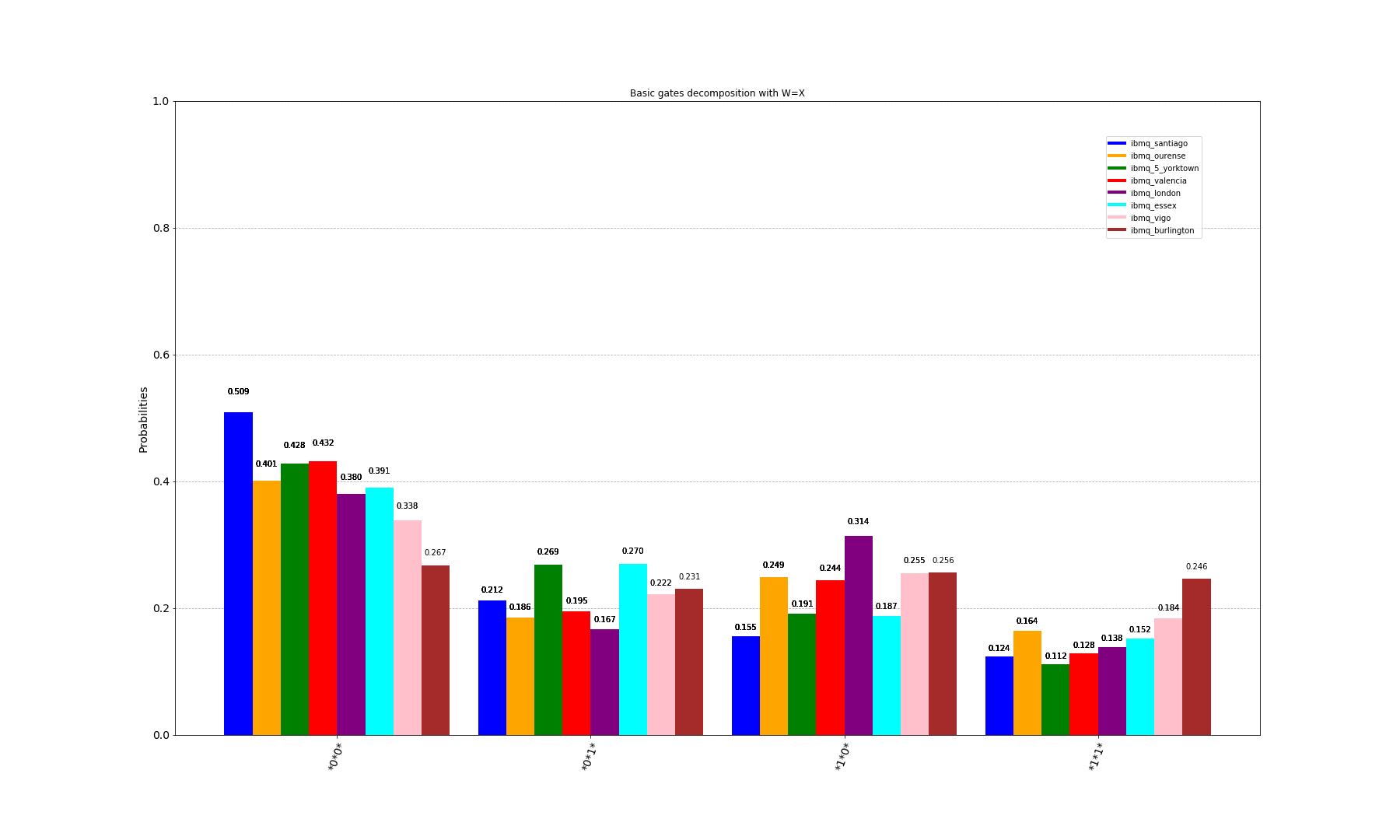}
\includegraphics[scale=0.085]{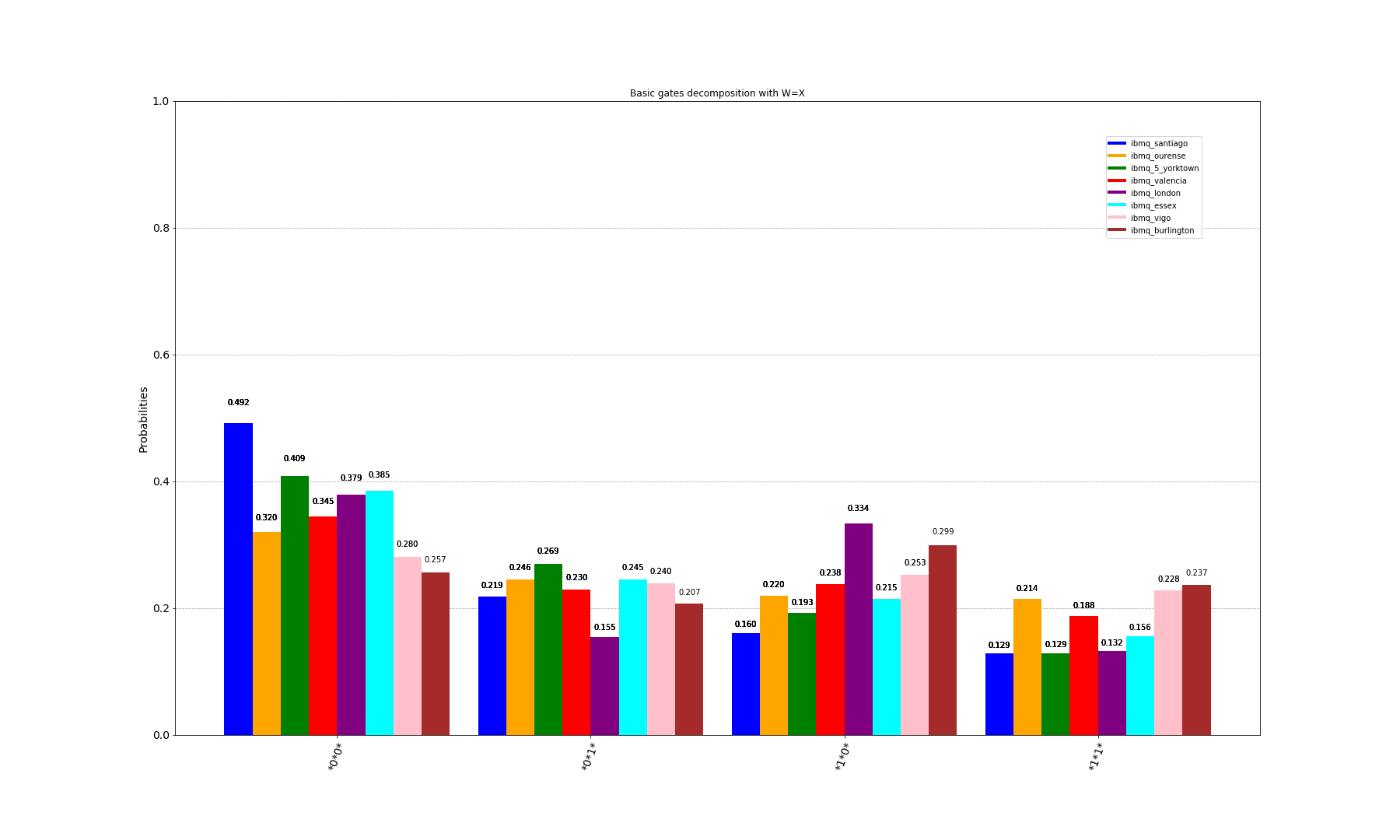}
\caption{using the basic gate decomposition of $U$ and $W=X$}
\end{figure}

\begin{figure}[!ht]
\includegraphics[scale=0.085]{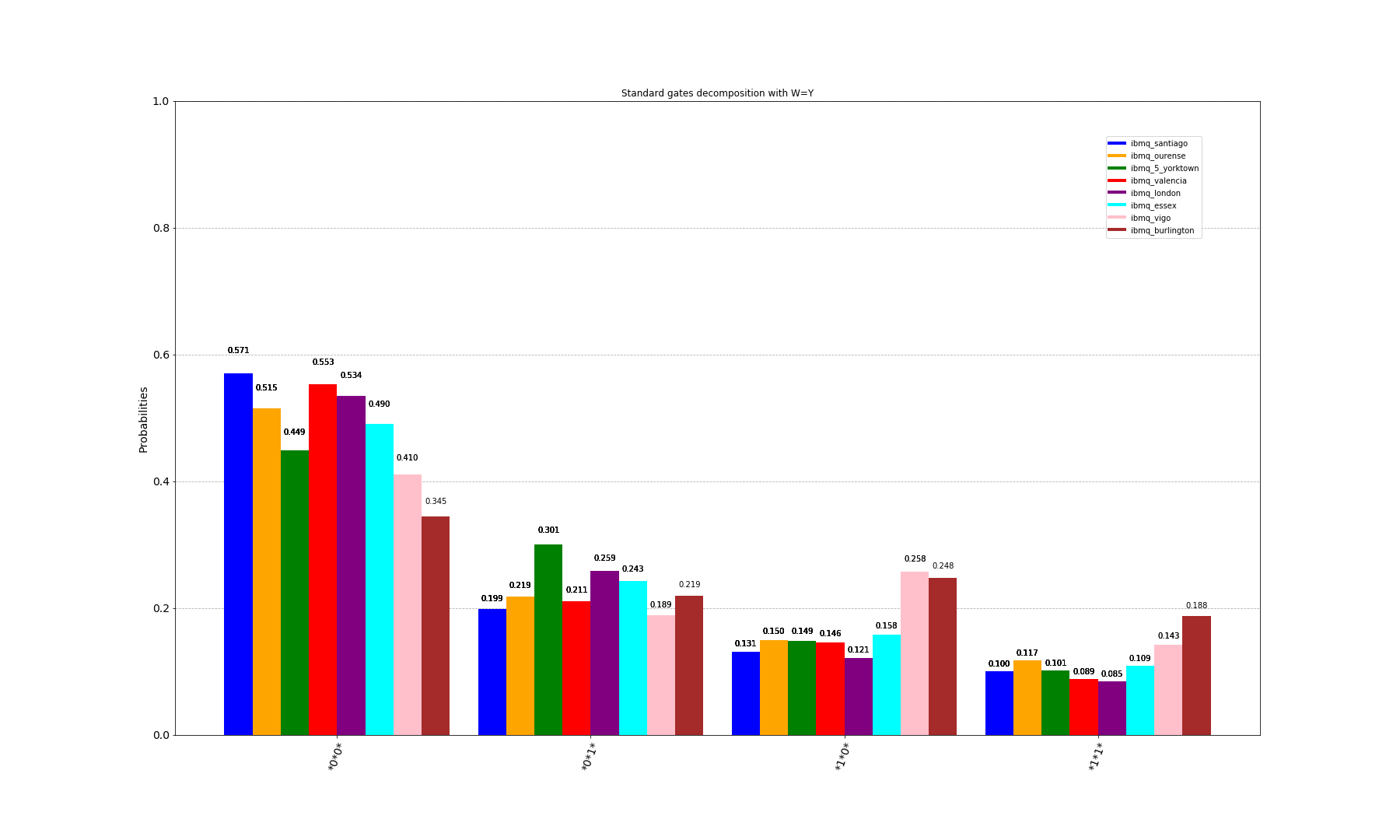}
\includegraphics[scale=0.085]{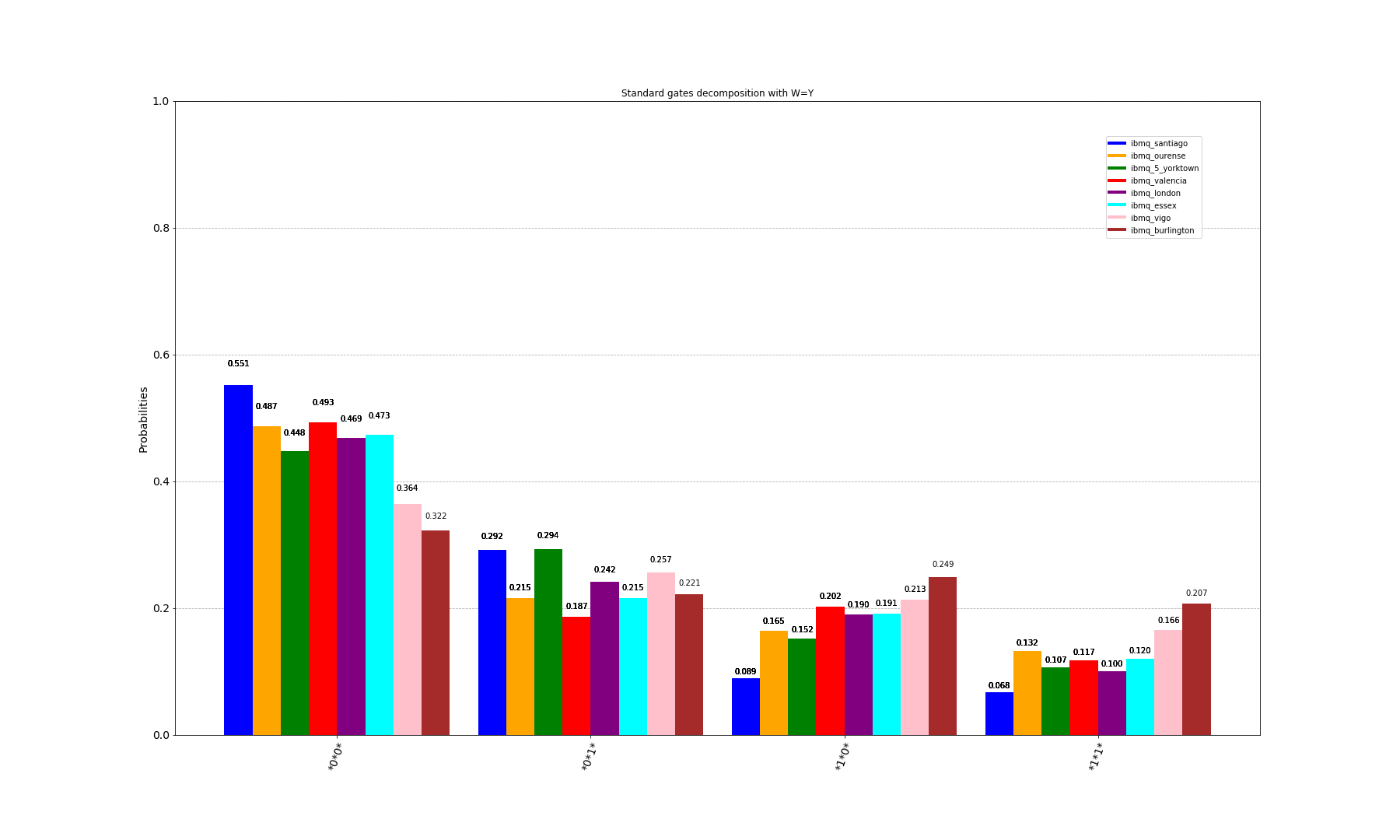}
\includegraphics[scale=0.085]{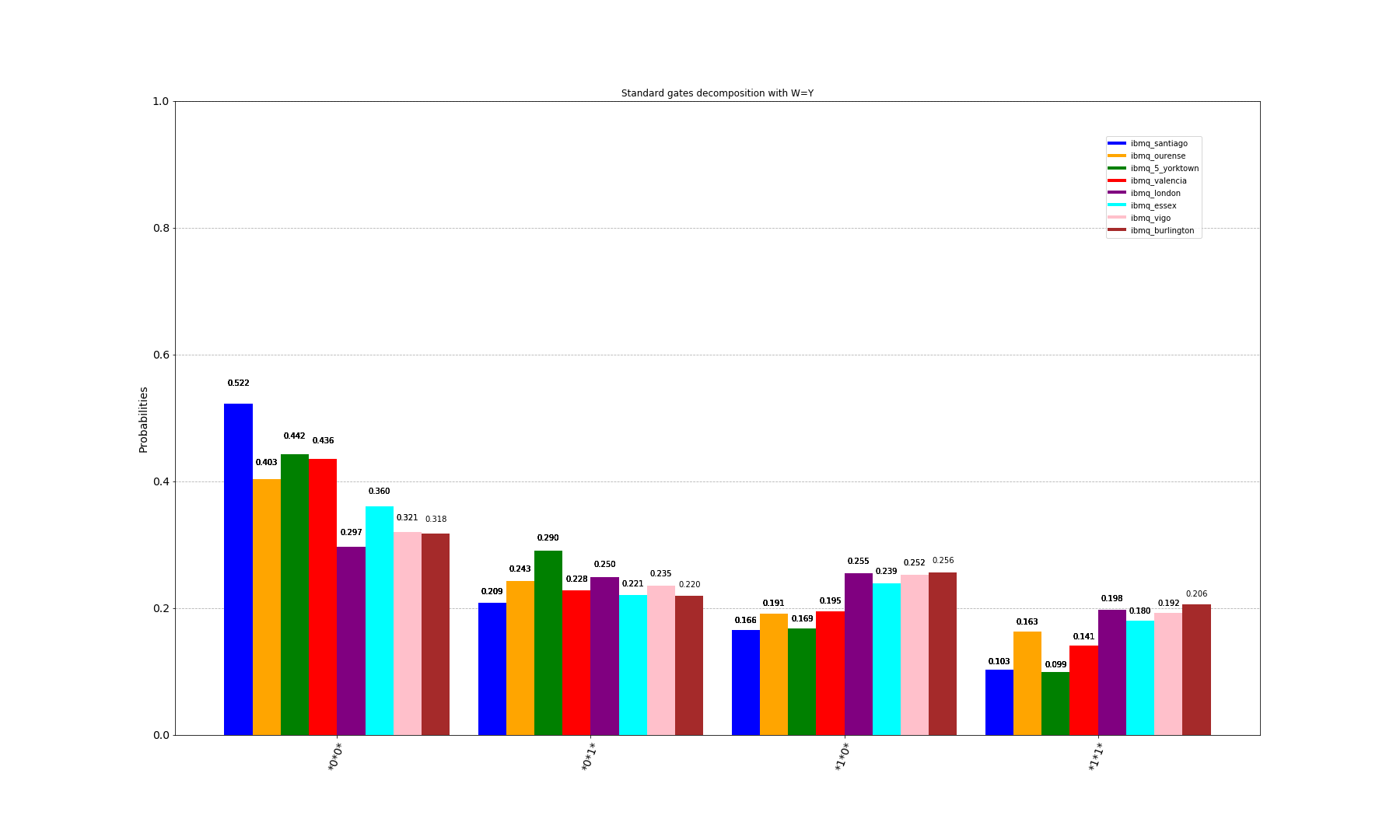}
\caption{using the standard gate decomposition of $U$ and $W=Y$}
\end{figure}

\begin{figure}[!ht]
\includegraphics[scale=0.085]{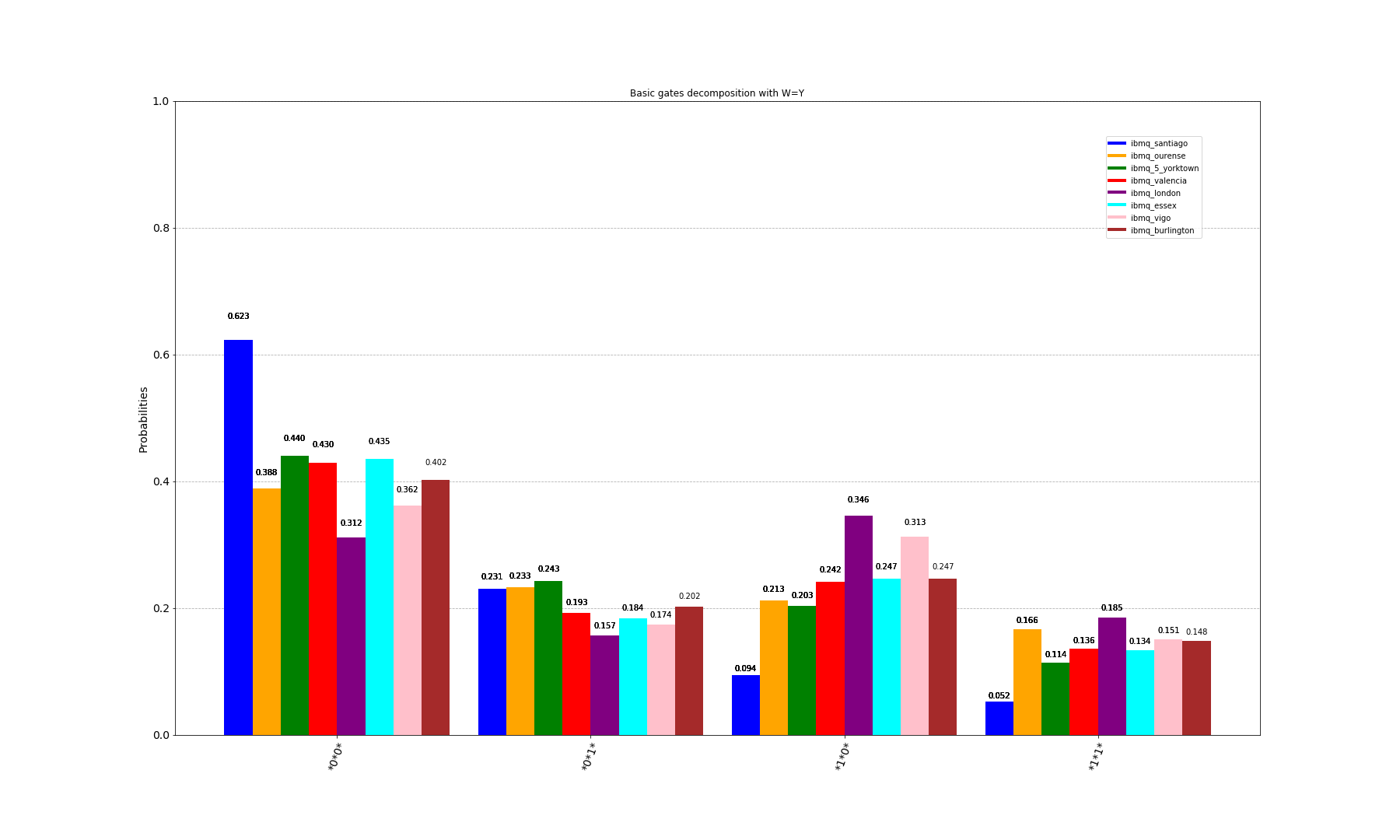}
\includegraphics[scale=0.085]{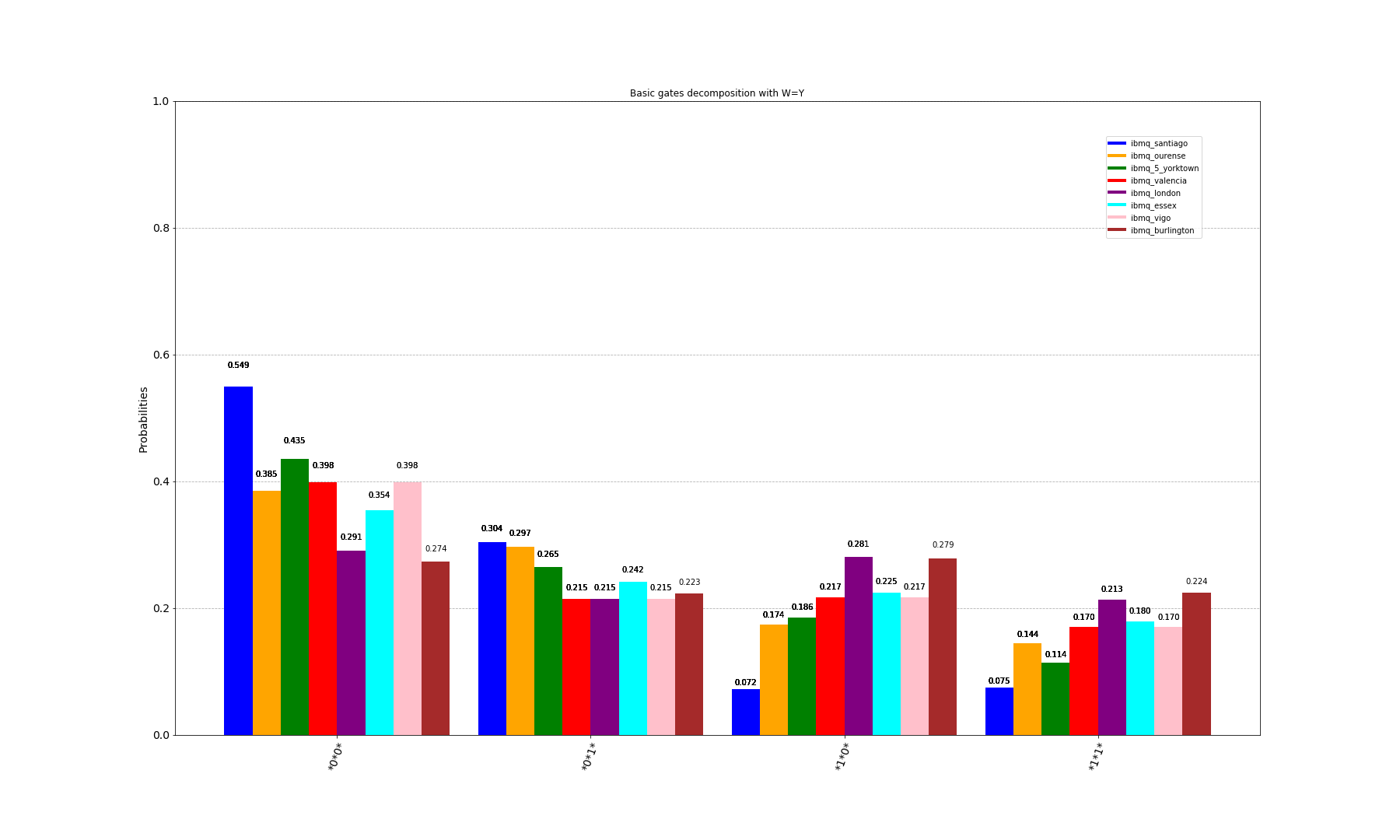}
\includegraphics[scale=0.085]{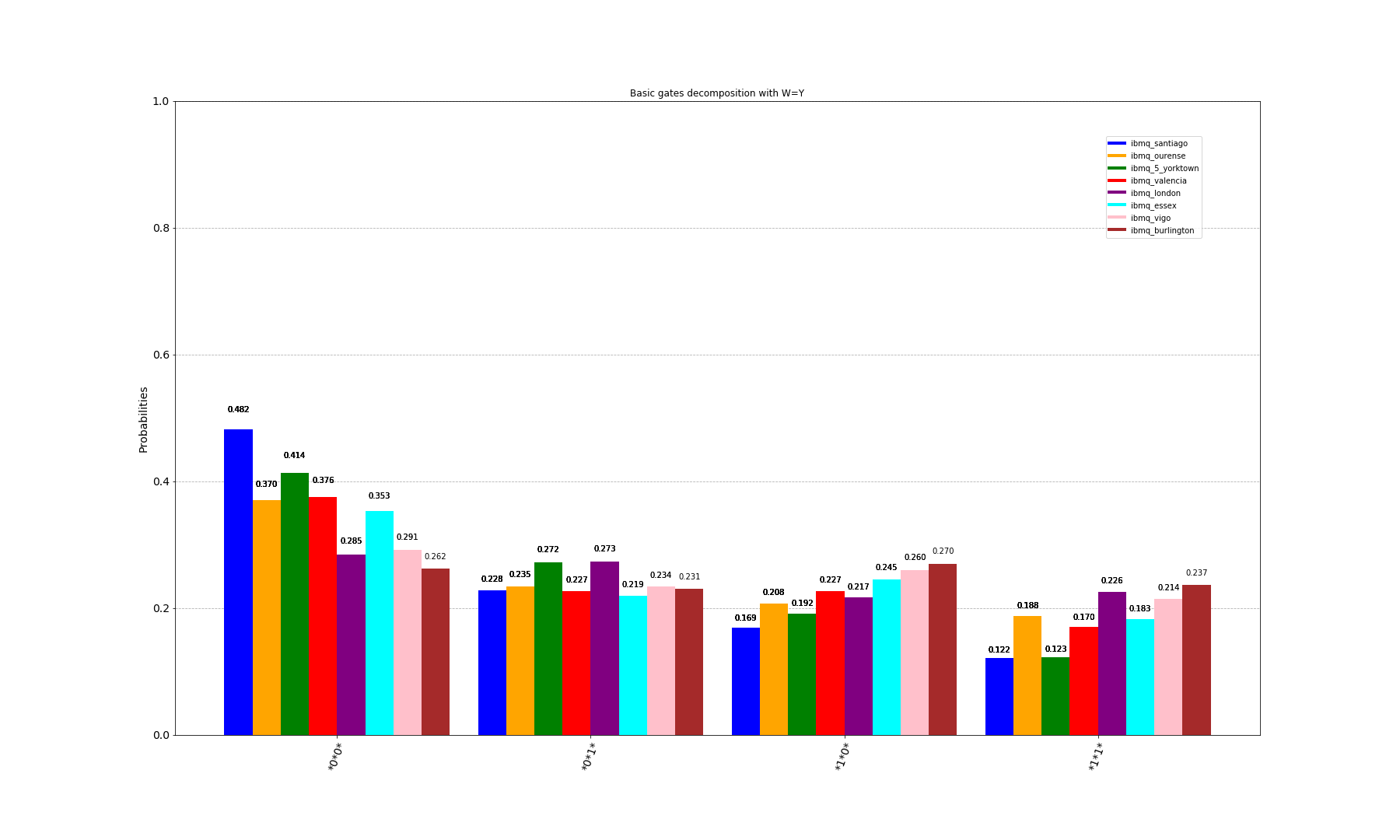}
\caption{using the basic gate decomposition of $U$ and $W=Y$}
\end{figure}

\begin{figure}[!ht]
\includegraphics[scale=0.085]{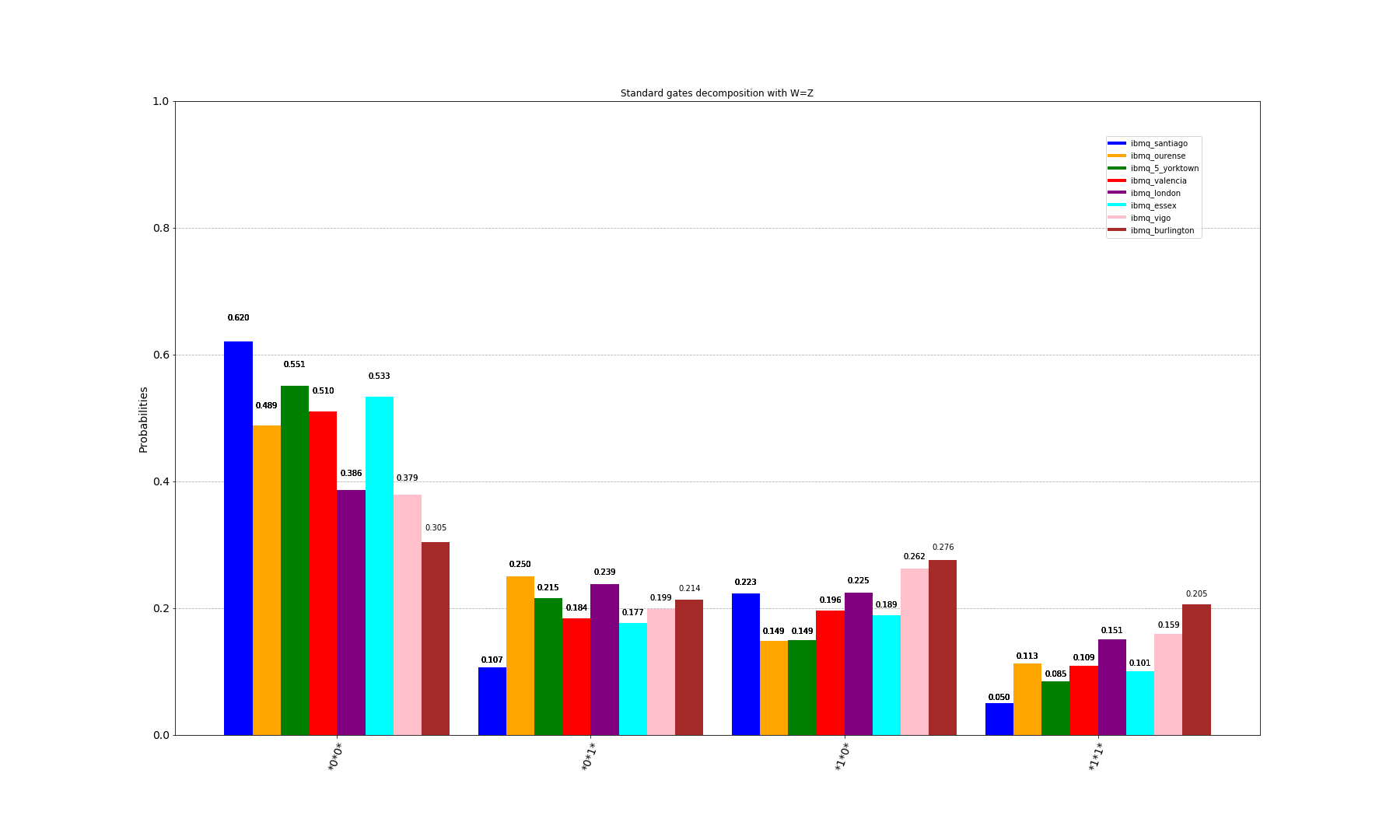}
\includegraphics[scale=0.085]{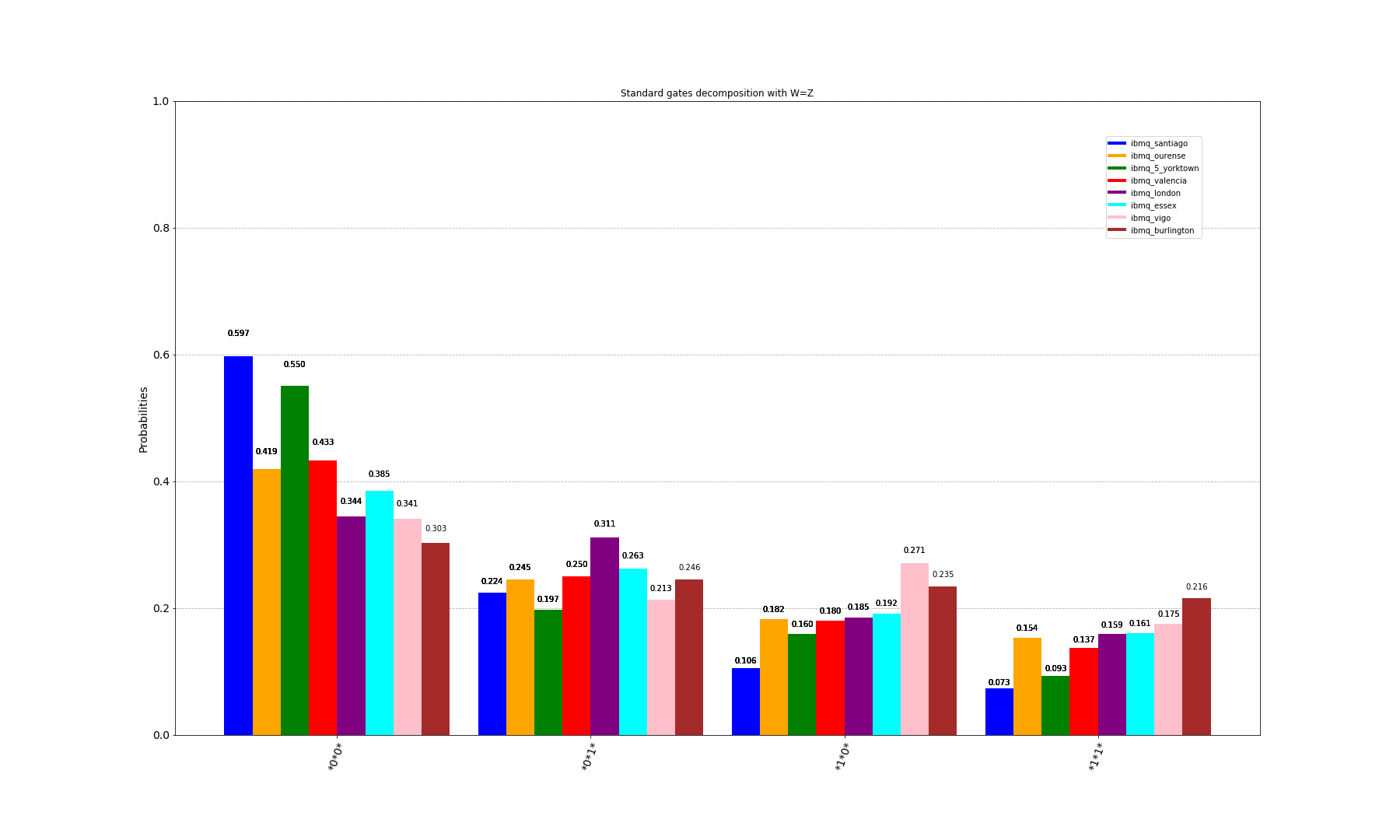}
\includegraphics[scale=0.085]{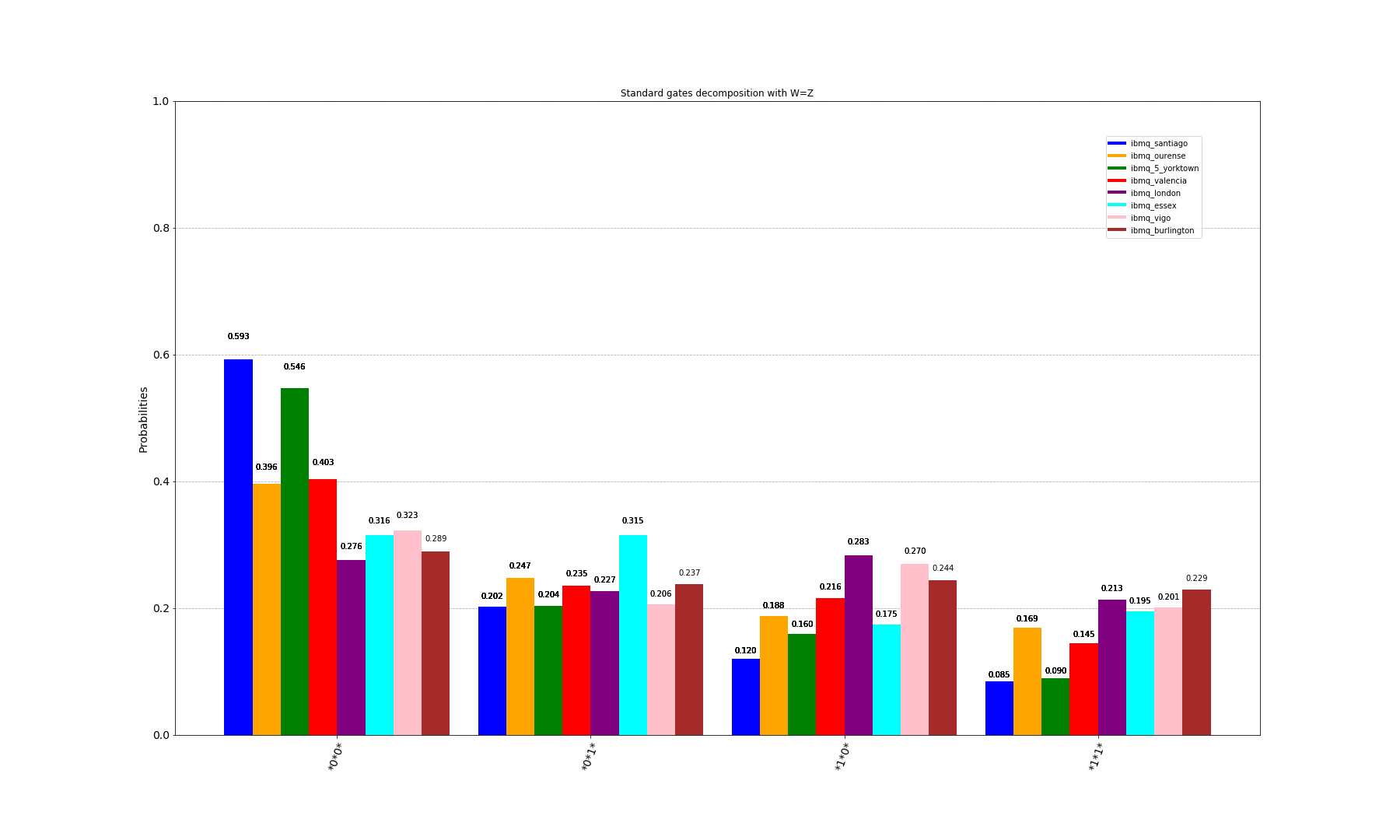}
\caption{using the standard gate decomposition of $U$ and $W=Z$}
\end{figure}

\begin{figure}[!ht]
\includegraphics[scale=0.085]{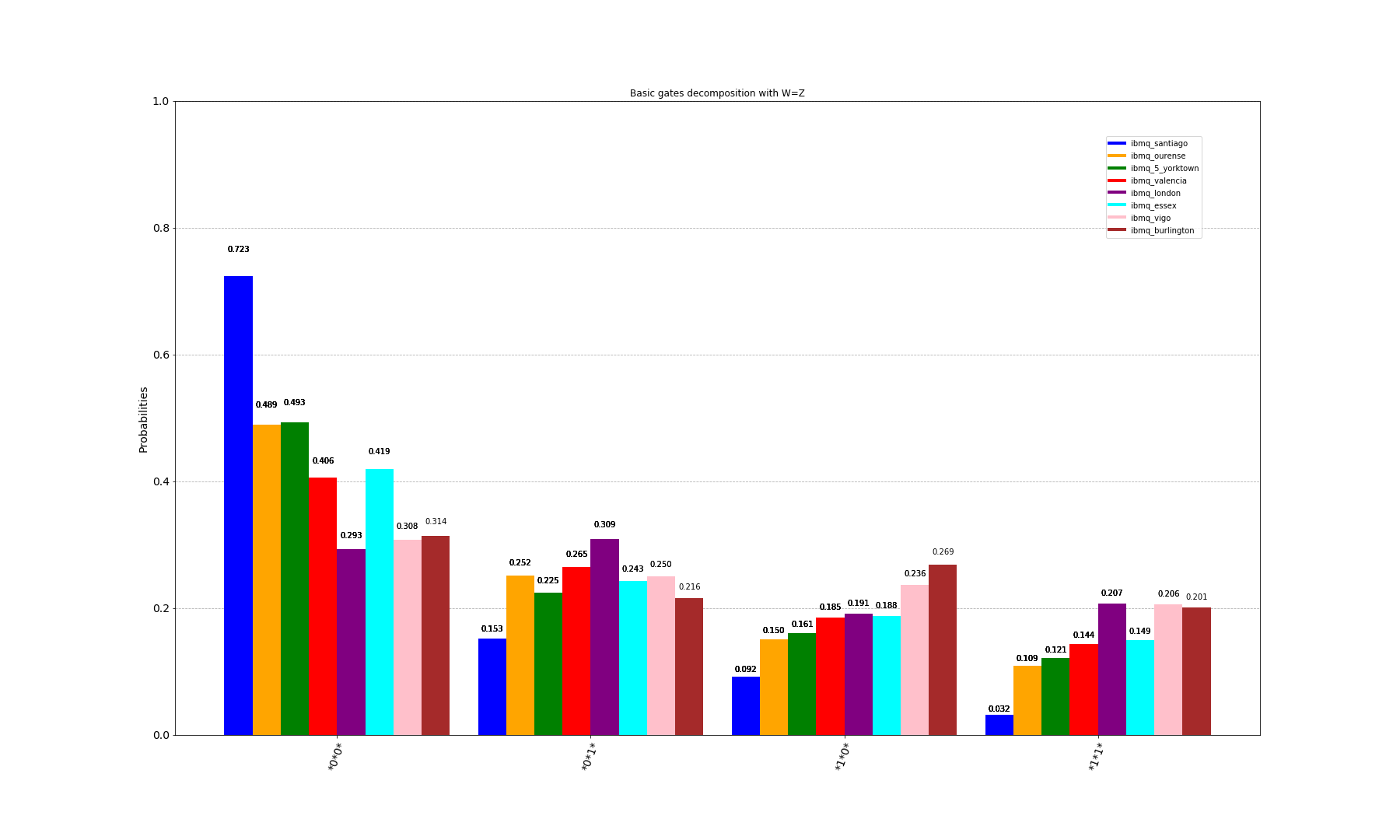}
\includegraphics[scale=0.085]{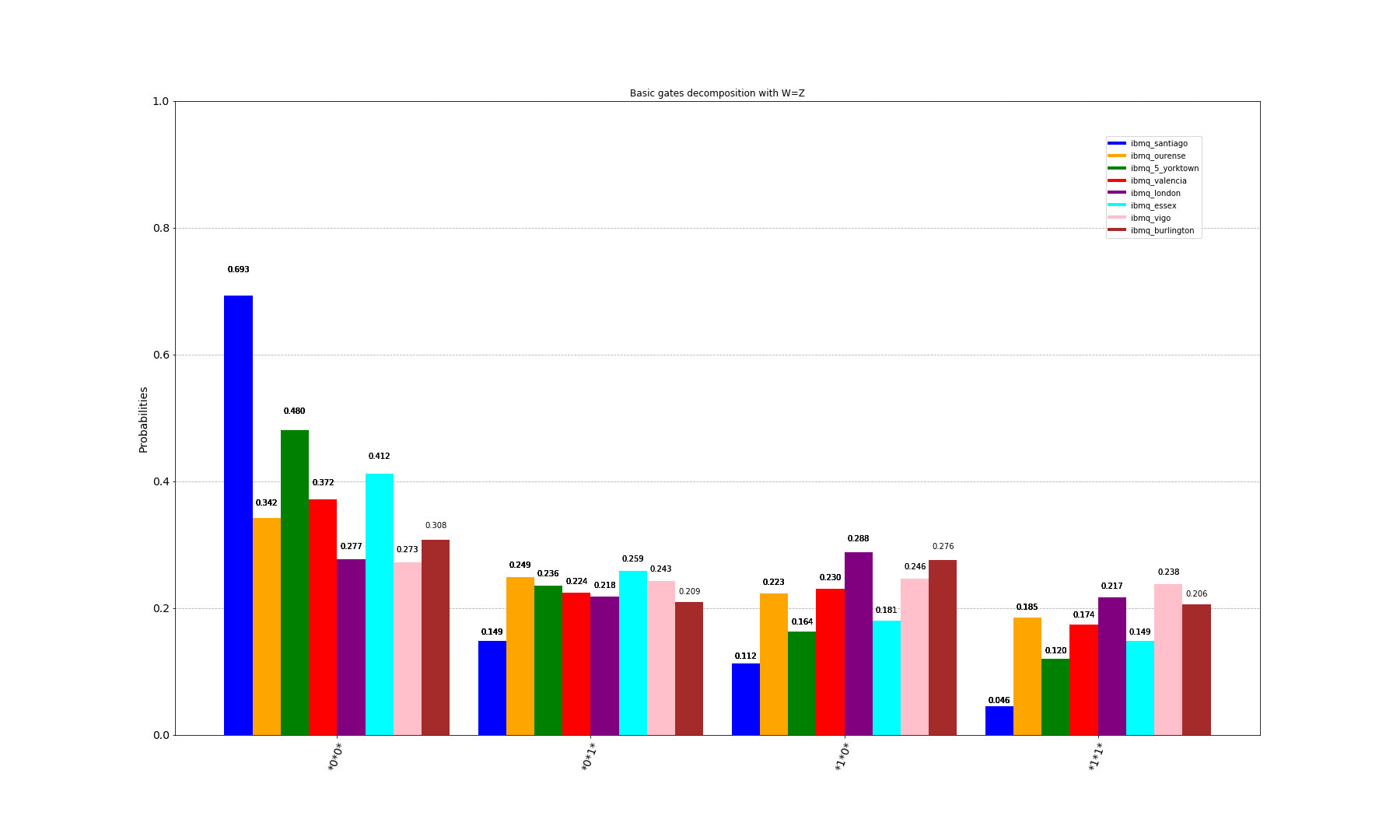}
\includegraphics[scale=0.085]{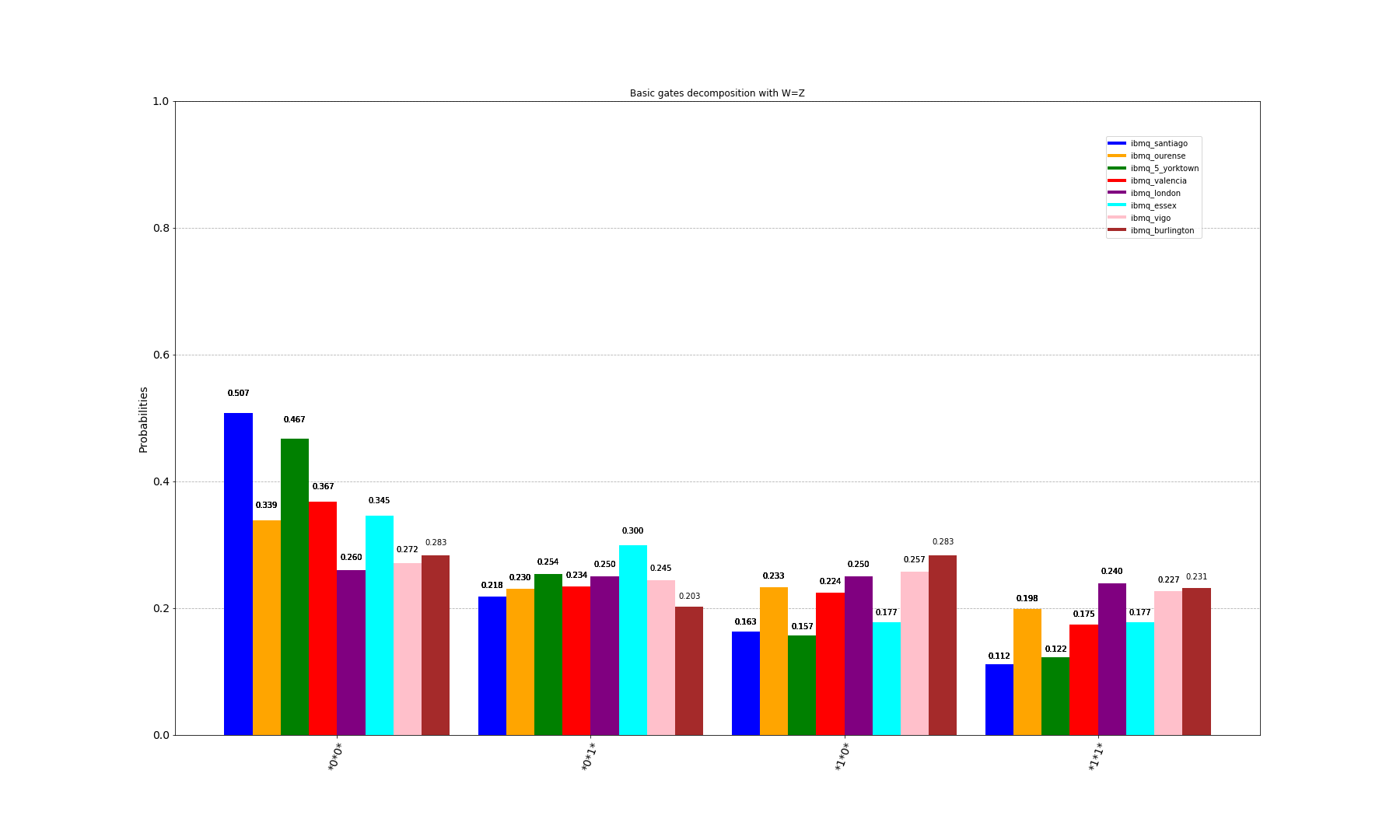}
\caption{using the basic gate decomposition of $U$ and $W=Z$}
\end{figure}

\begin{figure}[!ht]
\includegraphics[scale=0.085]{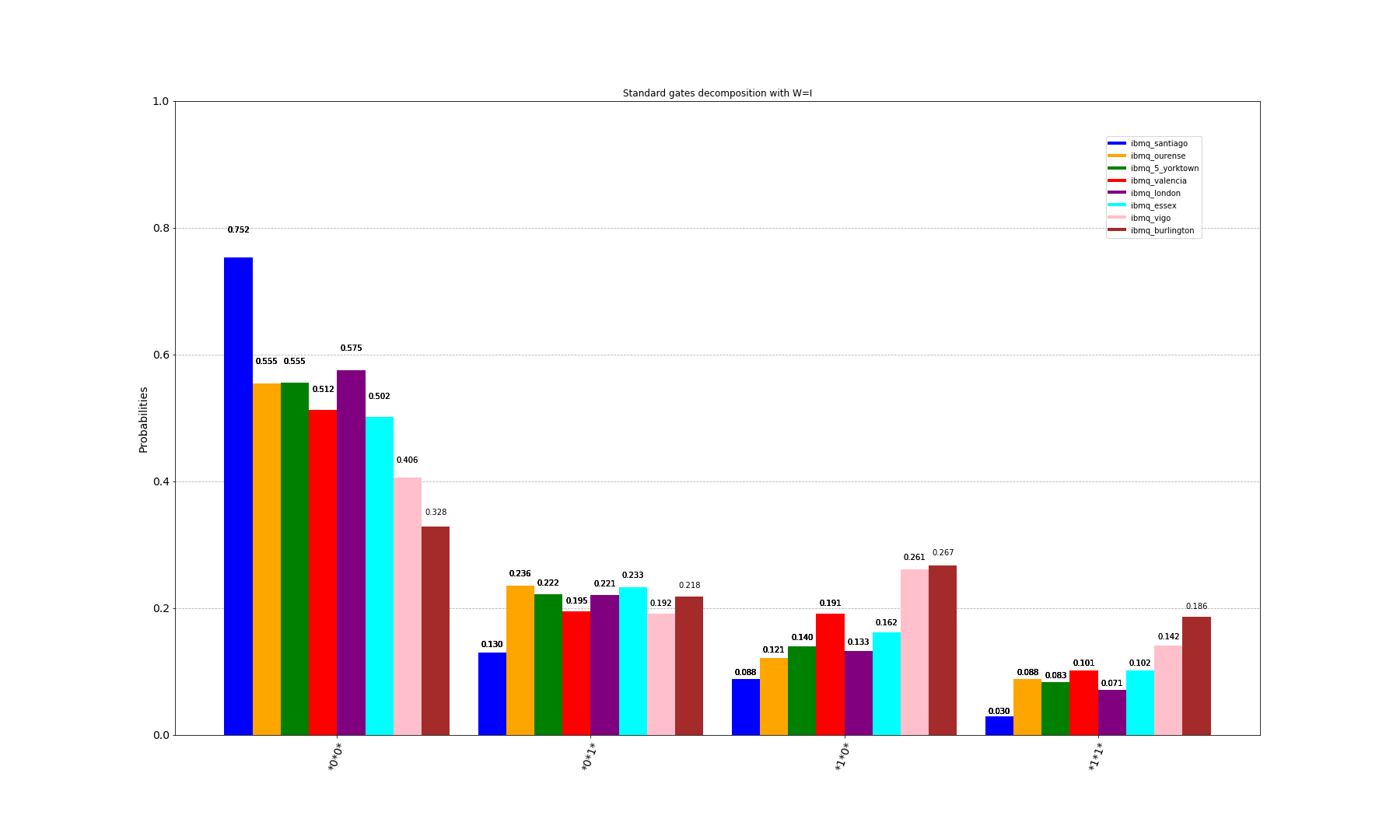}
\includegraphics[scale=0.085]{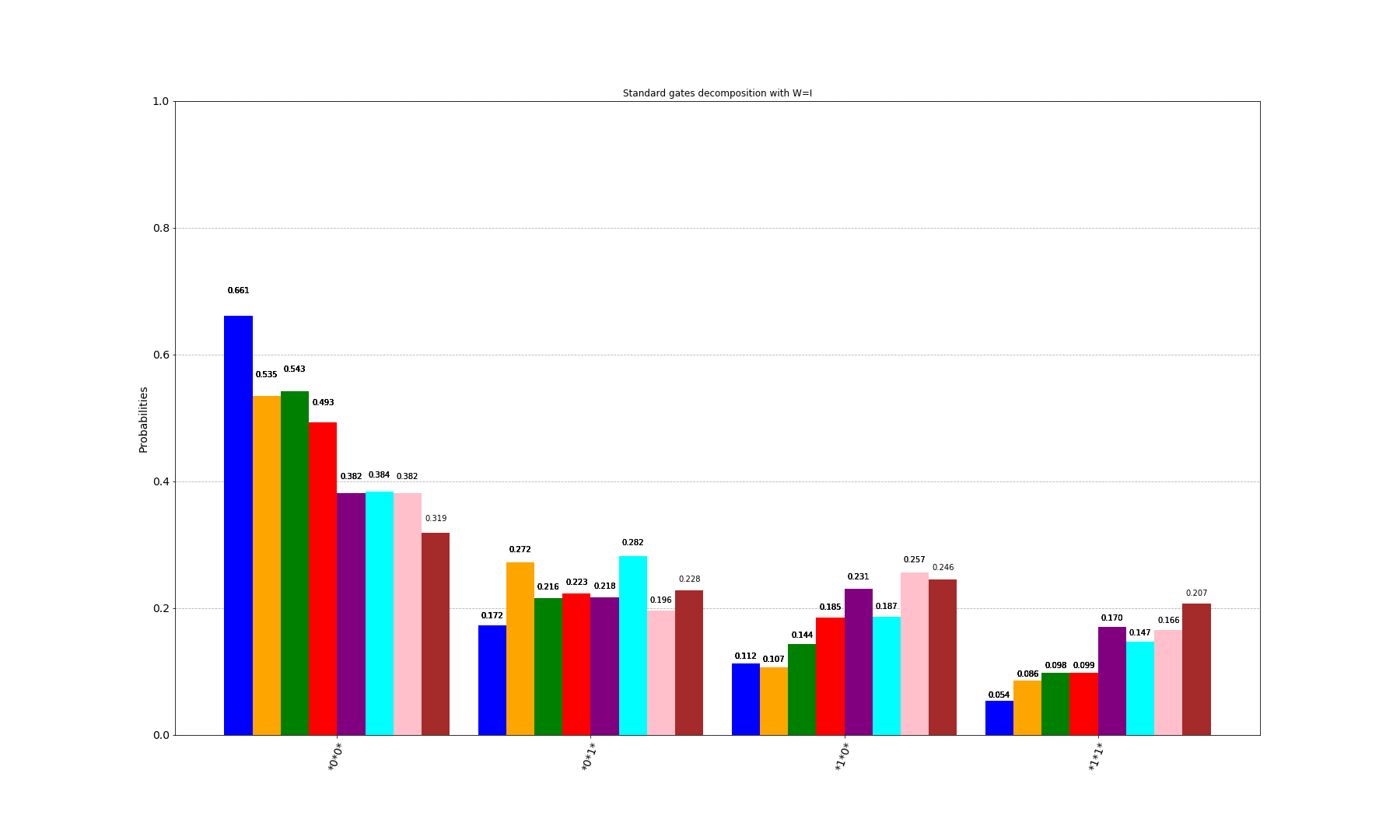}
\includegraphics[scale=0.085]{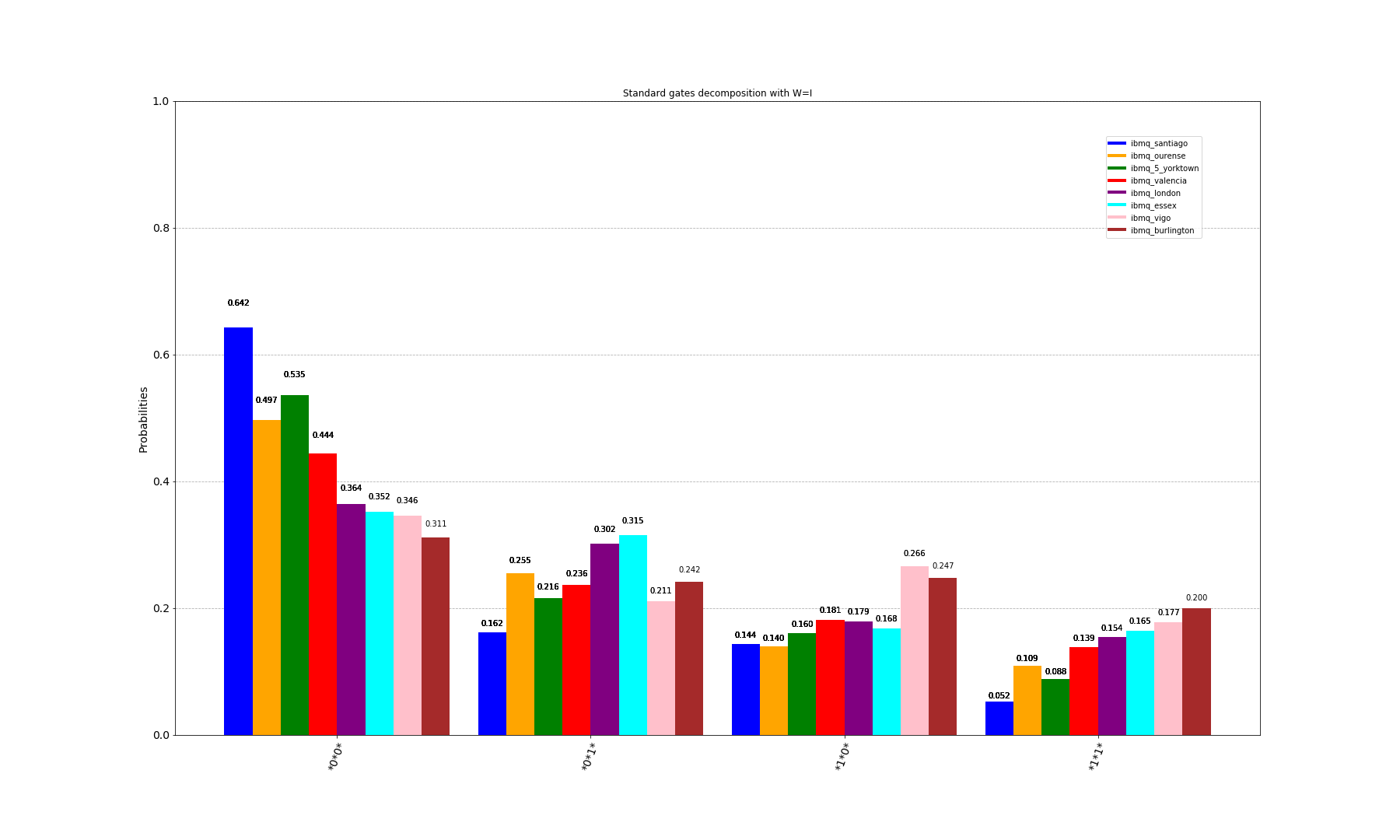}
\caption{using the standard gate decomposition of $U$ and $W=I$}
\end{figure}

\begin{figure}[!ht]
\includegraphics[scale=0.085]{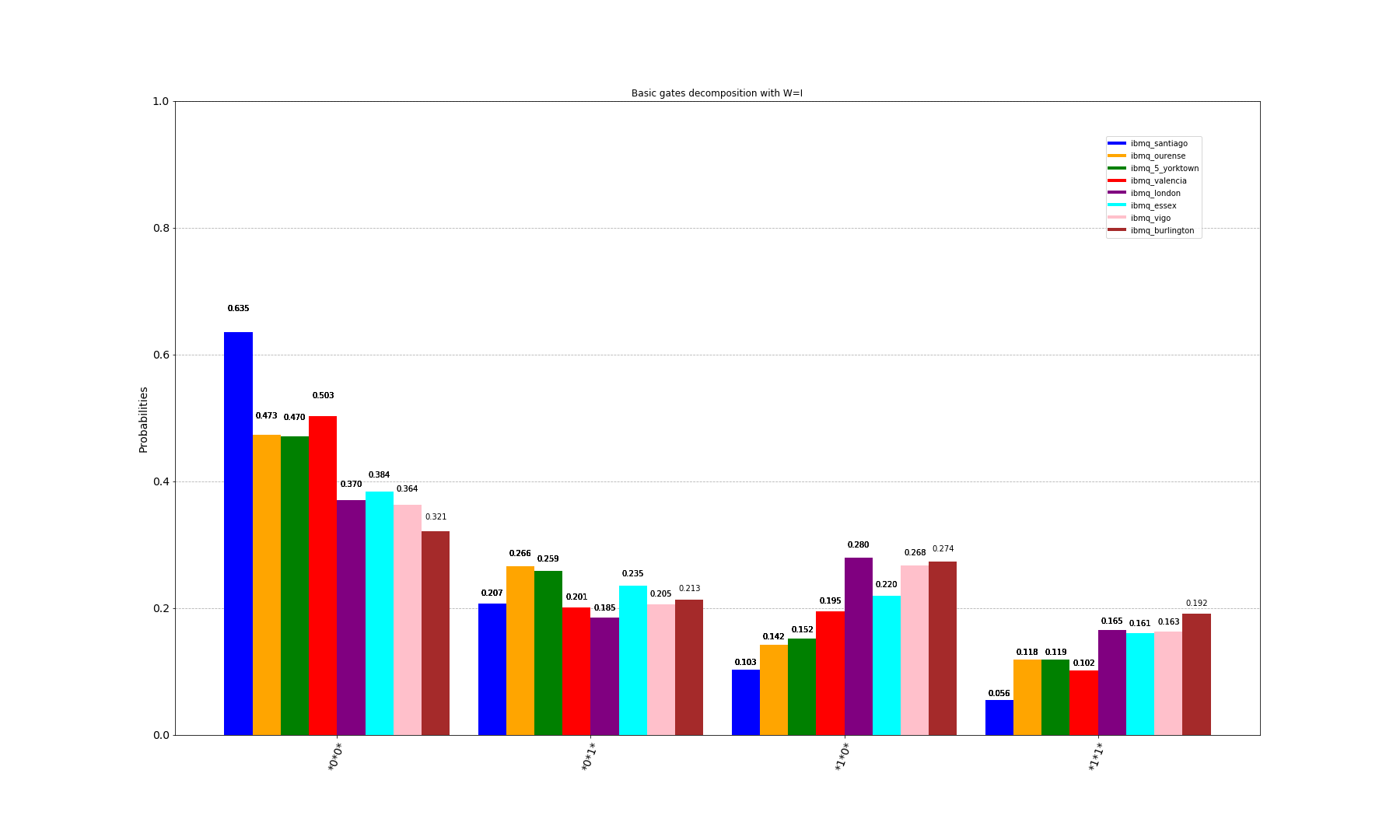}
\includegraphics[scale=0.085]{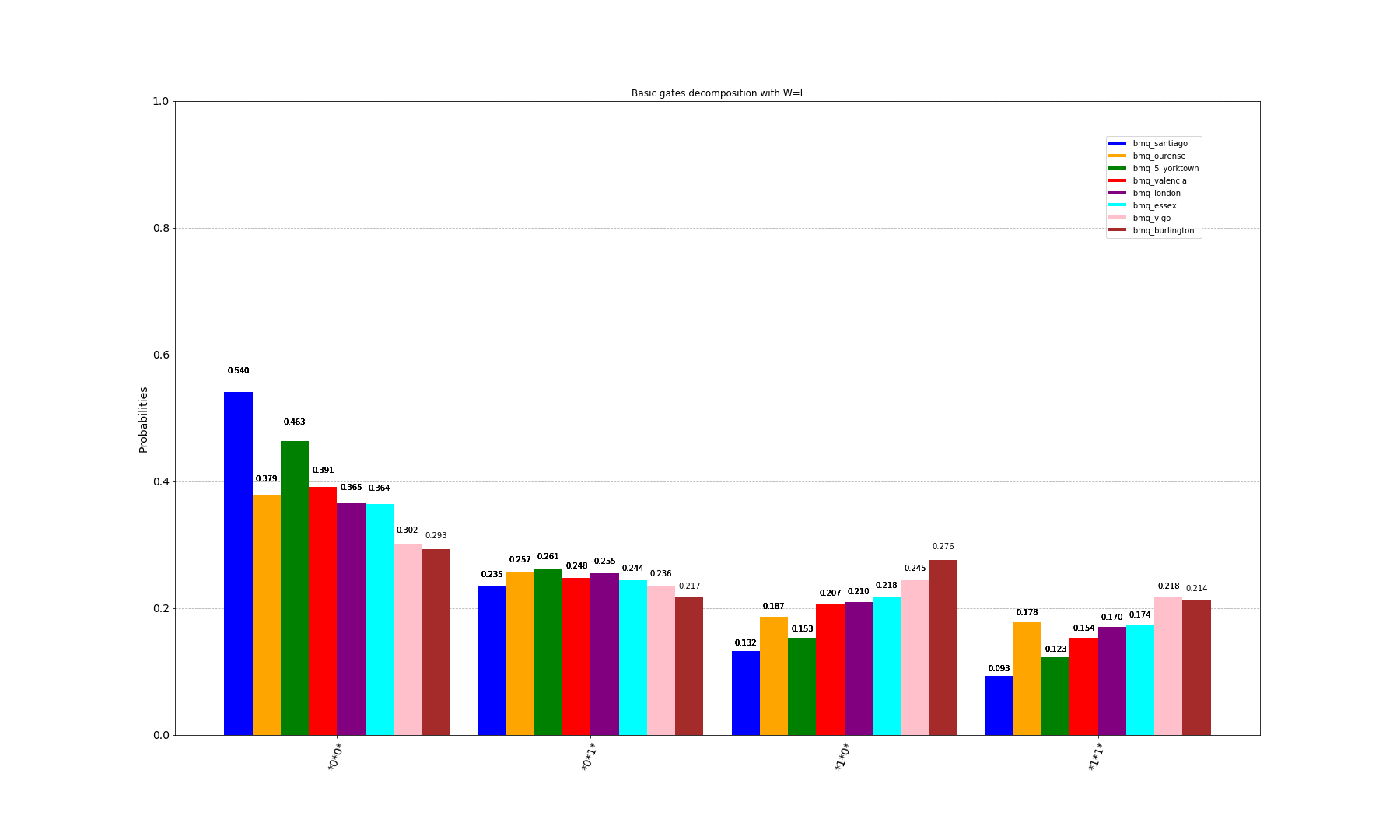}
\includegraphics[scale=0.085]{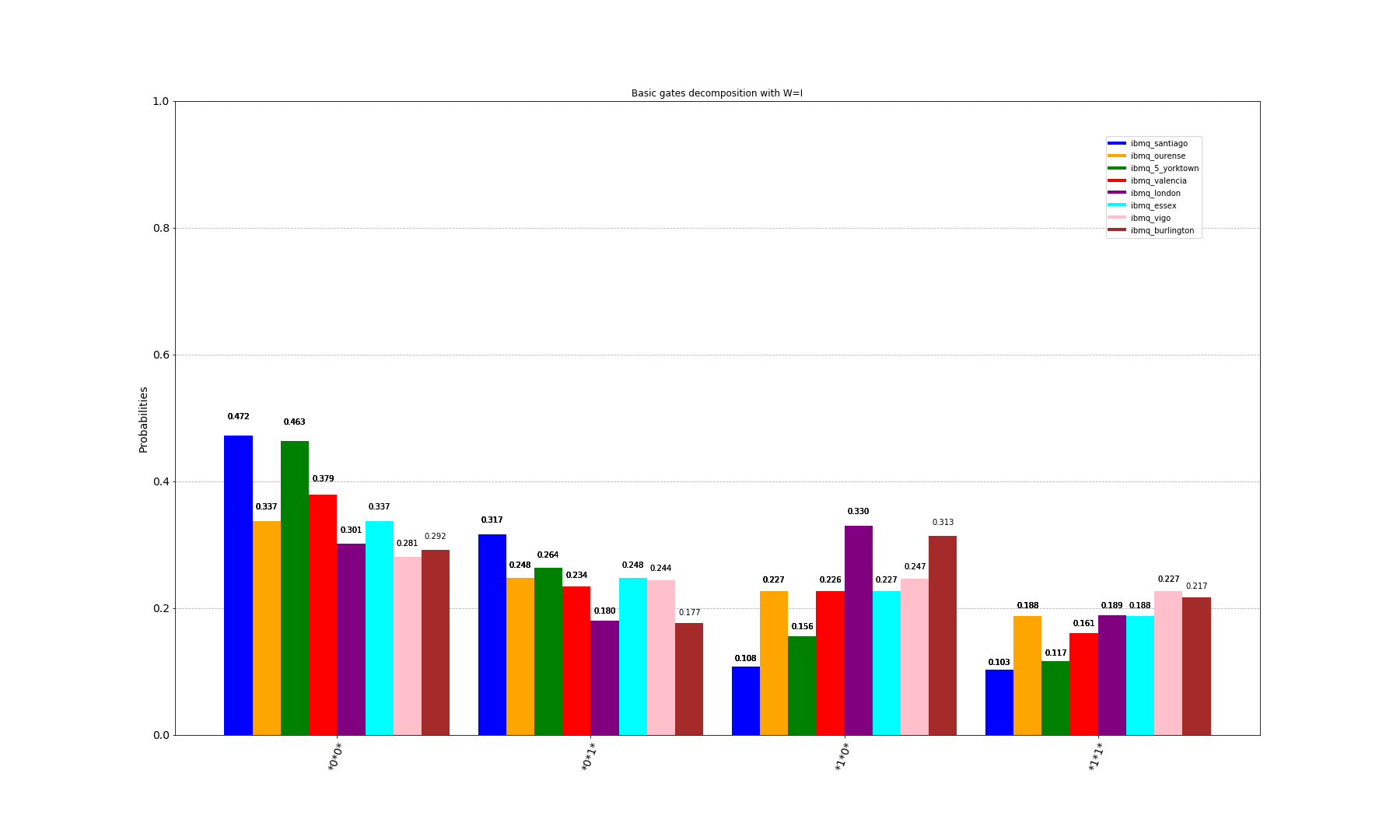}
\caption{using the basic gate decomposition of $U$ and $W=I$}
\end{figure}

\newpage
\section*{Appendix 5. }

The following illustrate the results obtained in implementing the 4-qubit QECC illustrated in equation )\ref{xyzeven}) using $|q_3q_2\rangle\in \{01,10,11\}$ and the IBM machines \texttt{ibmq\_santiago} and \texttt{ibmq\_athens}. 
\begin{figure}[!ht]
\begin{subfigure}{1\textwidth}
\includegraphics[width=1\linewidth]{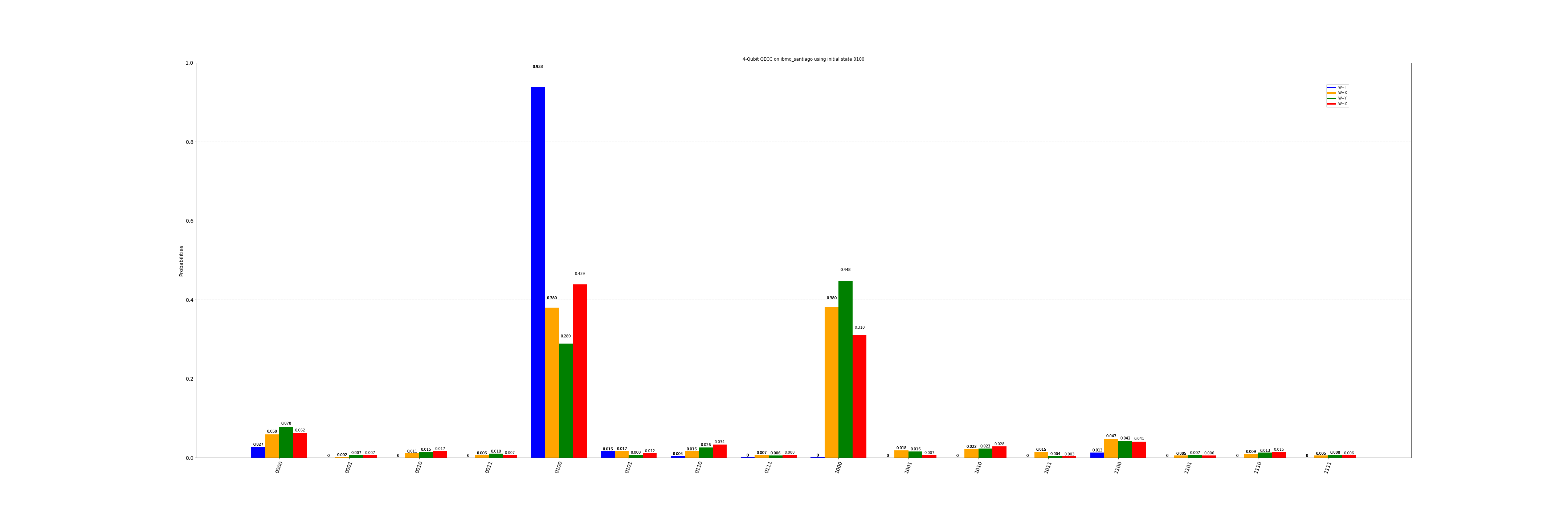}
\caption{using \texttt{ibmq\_santiago} and $|q_3q_2q_1q_0\rangle=|0100\rangle$}
\end{subfigure}
\begin{subfigure}{1\textwidth}
\includegraphics[width=1\linewidth]{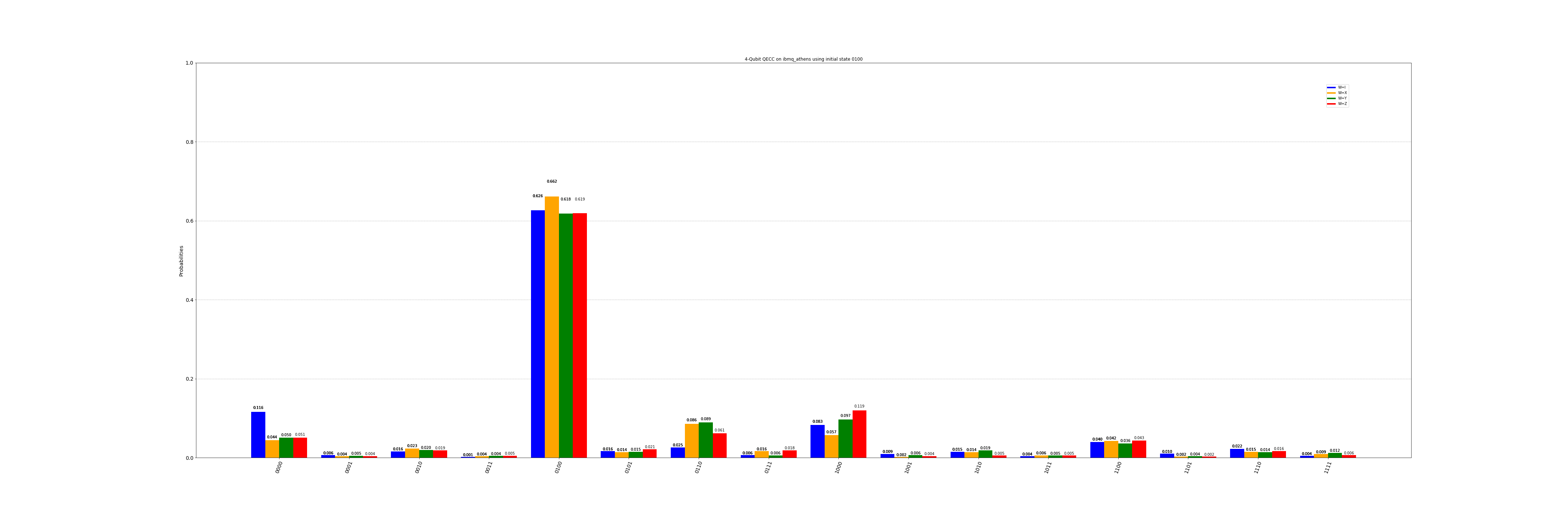}
\caption{using \texttt{ibmq\_athens} and $|q_3q_2q_1q_0\rangle=|0100\rangle$}
\end{subfigure}
\caption{}
\end{figure}

\begin{figure}[!ht]
\begin{subfigure}{1\textwidth}
\includegraphics[width=1\linewidth]{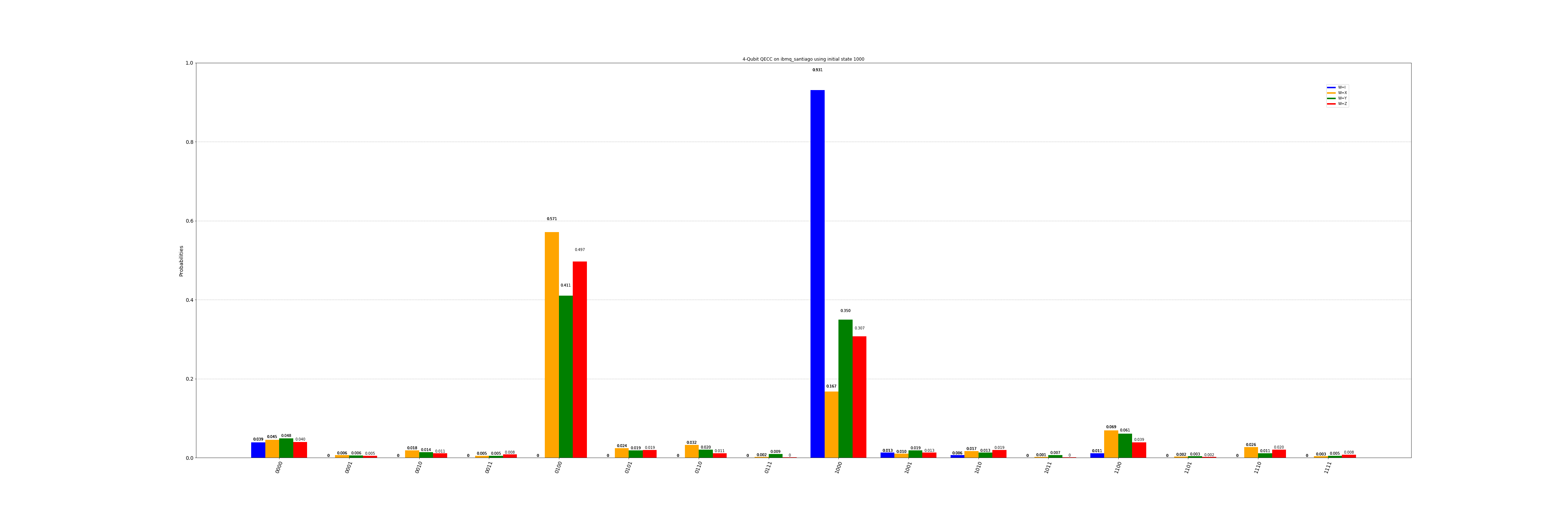}
\caption{using \texttt{ibmq\_santiago} and $|q_3q_2q_1q_0\rangle=|1000\rangle$}
\end{subfigure}
\begin{subfigure}{1\textwidth}
\includegraphics[width=1\linewidth]{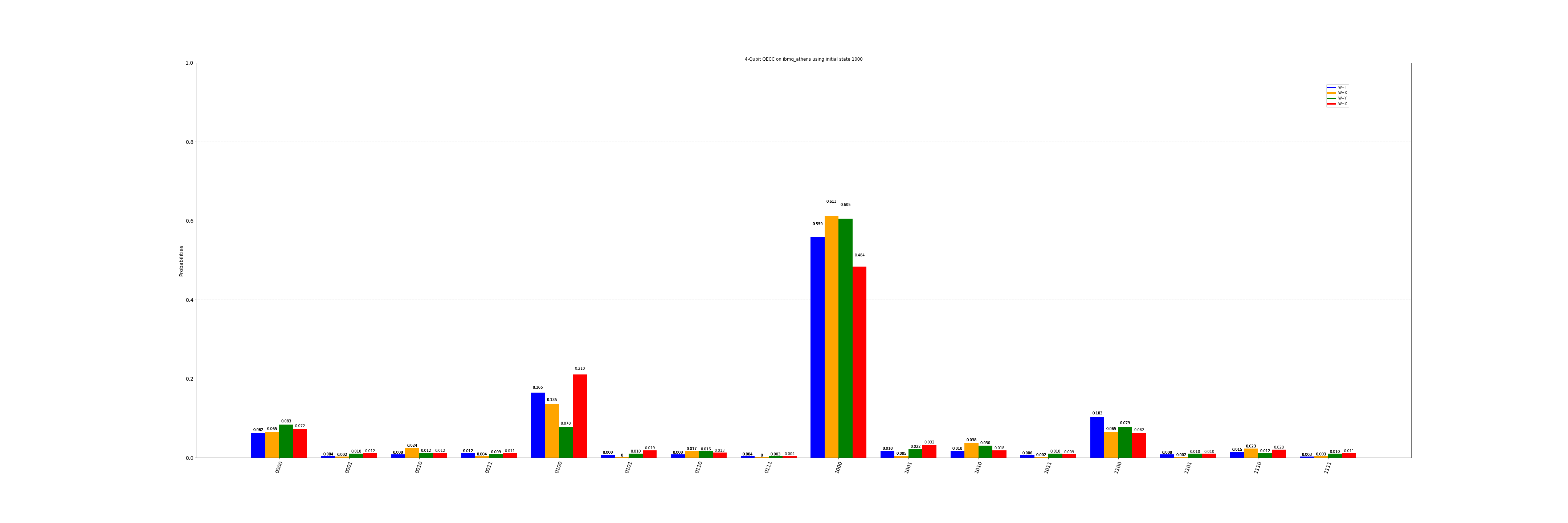}
\caption{using \texttt{ibmq\_athens} and $|q_3q_2q_1q_0\rangle=|1000\rangle$}
\end{subfigure}
\caption{}
\end{figure}

\begin{figure}[!ht]
\begin{subfigure}{1\textwidth}
\includegraphics[width=1\linewidth]{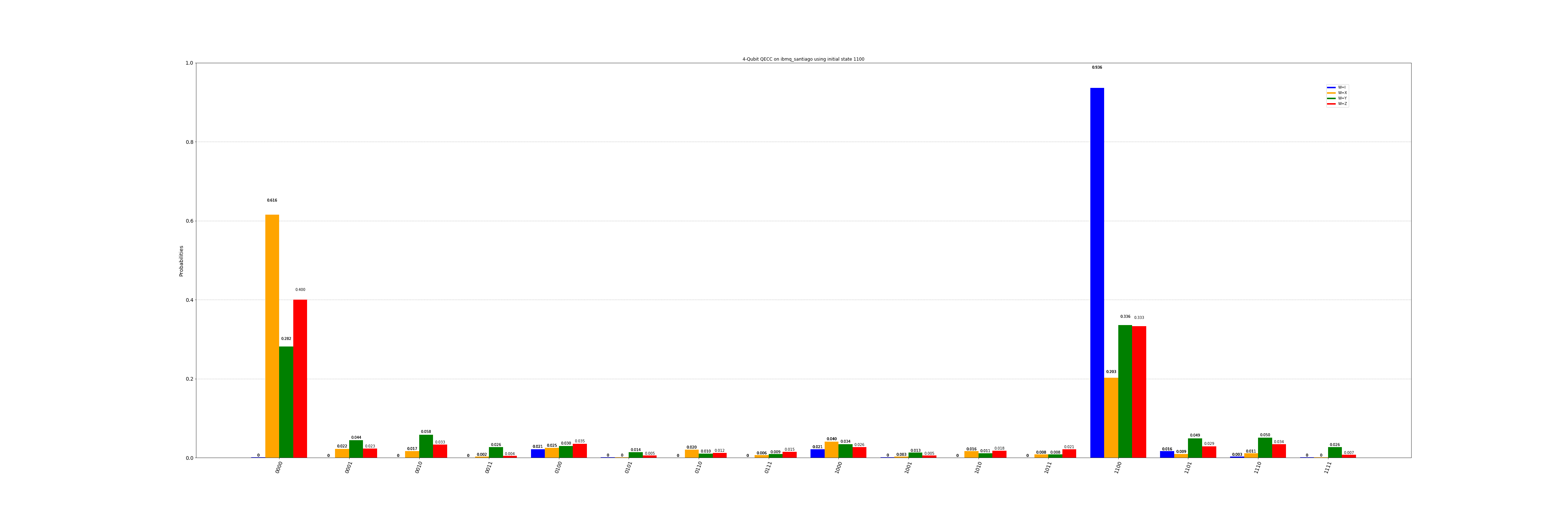}
\caption{using \texttt{ibmq\_santiago} and $|q_3q_2q_1q_0\rangle=|1100\rangle$}
\end{subfigure}
\begin{subfigure}{1\textwidth}
\includegraphics[width=1\linewidth]{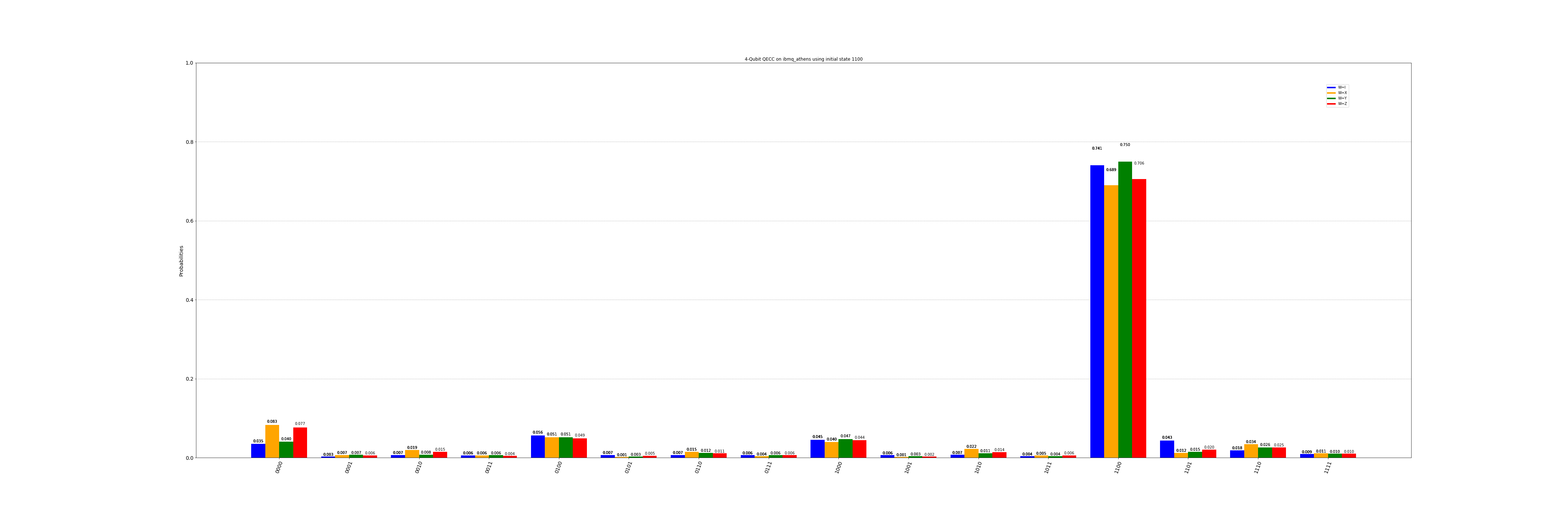}
\caption{using \texttt{ibmq\_athens} and $|q_3q_2q_1q_0\rangle=|1100\rangle$}
\end{subfigure}
\caption{}
\end{figure}


\begin{thebibliography}{WWW}
\bibitem{[1]}
P. Zanardi and M. Rasetti, Noiseless Quantum Codes, 
Physical Review Letters 79,3306 (1997).

\bibitem{[2]} P. Zanardi and M. Rasetti,,
Error Avoiding Quantum Codes Modern Physics Letters B 11,
1085 (1997).

\bibitem{[3]} P. Zanardi,
Dissipation and decoherence in a quantum register,
Physical Review A 57, 3276 (1998).

\bibitem{[4]} D. A. Lidar, I. L. Chuang, and K. B. Whaley,
Decoherence-Free Subspaces for Quantum Computation, 
Physical Review Letters 81, 2594 (1998).

\bibitem{[5]} J. Kempe, D. Bacon, D. A. Lidar, and K. B. Whaley,
Theory of decoherence-free fault-tolerant universal quantum 
computation, Physical Review A 63, 042307 (2001).

\bibitem{[6]} D. A. Lidar, D. Bacon, J. Kempe, and K. B. Whaley,
Protecting quantum information encoded in decoherence-free states 
against exchange errors,
Physical Review A 61, 052307 (2000).

\bibitem{[7]} D. A. Lidar, D. Bacon, J. Kempe, and K. B. Whaley,
Decoherence-free subspaces for multiple-qubit errors. I. 
Characterization, Physical Review A 63, 022306 (2001).


\bibitem{[8]} M.-D. Choi and D. W. Kribs, 
Method to Find Quantum Noiseless Subsystems
Phys. Rev. Lett. 96, 050501 (2006).

\bibitem{[9]} E. Knill, R. Laflamme, and L. Viola, ,
Theory of Quantum Error Correction for General Noise,
Physical Review Letters 84, 2525 (2000).

\bibitem{[10]} E. M. Fortunato, L. Viola, M. A. Pravia, E. Knill,
R. Laflamme, T. F. Havel, and D. G. Cory,
Exploring noiseless subsystems via nuclear magnetic resonance,
Phys. Rev. A 67, 062303 (2003).

\bibitem{[11]} L. Viola, E. M. Fortunato, M. a. Pravia, E. Knill,
R. Laflamme, and D. G. Cory, 
Experimental Realization of Noiseless Subsystems for Quantum 
Information Processing,
Science 293, 2059 (2001).

\bibitem{[12]} D. A. Lidar and T. A. Brun, 
Quantum Error Correction
Cambridge University Press, (2013).

\bibitem{[13]} Y. Kondo, C. Bagnasco, 
and M. Nakahara, 
Implementation of a simple operator-quantum-error-correction scheme,
Physical Review A 88, 022314 (2013).

\bibitem{[14]} M. S. Byrd, 
Implications of qudit superselection rules for the theory of 
decoherence-free subsystems,
Physical Review A 73, 032330 (2006).

\bibitem{[15]} C.K. Li,
M. Nakahara, Y.T. Poon, N.K. Sze, H. Tomita,
 Efficient Quantum Error Correction for Fully Correlated Noise, Phys. Lett. A, 375:3255-3258 (2011). 

\bibitem{[17]} C.-K. Li, M. Nakahara, Y.-T. Poon, and N.-S. Sze, 
Maximal error correction rates for collective rotation channels on qudits, Quantum Information Processing  14, 4039-4055 (2015).

\bibitem{LLP} 
C.K. Li, S. Lyles and Y.T. Poon,
Error correction schemes for fully correlated 
quantum channels protecting 
both quantum and classical information, Quantum Information 
Processing  19, no. 5, Paper No. 153 (2020).

\bibitem{Utkan} U. Gungordu, C.K. Li, M. Nakahara, 
Y.T. Poon, and N.S.  Sze,
Recursive encoding and decoding of the noiseless 
subsystem for qudits, 
Physical Review A 89, 042301 (2014). 
 
\bibitem{Tomita} C.K. Li,  M. Nakahara, Y.T. Poon, 
N.S. Sze and H. Tomita,
Recursive Encoding and Decoding of 
Noiseless Subsystem and Decoherence Free Subspace, 
Physical Review A 84, 044301 (2011). 

\bibitem{Yang}
C.-P. Yang and J. Gea-Banacloche,
Three-qubit quantum error-correction scheme for collective decoherence,
Physical Review A 63, 022311 (2001).

\end{thebibliography}
\end{document}